\RequirePackage{silence}
\documentclass[fleqn,usenatbib,useAMS]{mnras} 
\usepackage{graphicx}
\usepackage{amsmath}
\usepackage{substitutefont}
\substitutefont{TS1}{aer}{cmr}

\DeclareMathOperator{\sinc}{sinc}
\usepackage{amsfonts}
\usepackage{float}
\usepackage{bm}
\usepackage{cancel}
\usepackage{ae,aecompl}
\usepackage{array}
\usepackage{soul}
\usepackage{mathtools}
\usepackage{multirow}
\usepackage[utf8]{inputenc}
\usepackage{booktabs}
\usepackage{graphicx}
\usepackage{makecell}

\DeclareRobustCommand{\appropto}{\mathrel{\vcenter{
		\offinterlineskip\halign{\hfil$##$\cr 
			\propto\cr\noalign{\kern2pt}\sim\cr\noalign{\kern-2pt}}}}}

\defcitealias{Banik_2018_Centauri}{BZ18}
\defcitealias{Pittordis_2019}{PS19}
\defcitealias{Pittordis_2023}{PS23}

\title[Wide binary orbital velocities challenge MOND]{Strong constraints on the gravitational law from \emph{Gaia} DR3 wide binaries} 

\author[I. Banik et al.]{Indranil Banik$^{1}$\thanks{Email: \href{mailto:ib45@st-andrews.ac.uk}{ib45@st-andrews.ac.uk} (Indranil Banik);\newline $~~~~~~$ \href{mailto:w.j.sutherland@qmul.ac.uk}{w.j.sutherland@qmul.ac.uk} (Will Sutherland)}, Charalambos Pittordis$^2$, Will Sutherland$^2$, Benoit Famaey$^3$,
\newauthor
Rodrigo Ibata$^3$, Steffen Mieske$^4$ and Hongsheng Zhao$^1$ \vspace{10pt} \\
$^{1}$Scottish Universities Physics Alliance, University of Saint Andrews, North Haugh, Saint Andrews, Fife, KY16 9SS, UK \\
$^2$The School of Physical and Chemical Sciences, Queen Mary University of London, Mile End Road, London, E1 4NS, UK \\
$^3$Universit\'{e} de Strasbourg, CNRS UMR 7550, Observatoire astronomique de Strasbourg, 11 rue de l'Universit\'{e}, 67000 Strasbourg, France \\
$^4$European Southern Observatory, Alonso de Cordova 3107, Vitacura, Santiago, Chile}

\pubyear{2023}
\begin{document}
\label{firstpage}

\pagerange{\pageref{firstpage}--\pageref{lastpage}}

\maketitle

\begin{abstract} 
We test Milgromian dynamics (MOND) using wide binary stars (WBs) with separations of $2-30$~kAU. Locally, the WB orbital velocity in MOND should exceed the Newtonian prediction by $\approx 20\%$ at asymptotically large separations given the Galactic external field effect (EFE). We investigate this with a detailed statistical analysis of \emph{Gaia} DR3 data on 8611 WBs within 250~pc of the Sun. Orbits are integrated in a rigorously calculated gravitational field that directly includes the EFE. We also allow line of sight contamination and undetected close binary companions to the stars in each WB. We interpolate between the Newtonian and Milgromian predictions using the parameter $\alpha_{\rm{grav}}$, with 0 indicating Newtonian gravity and 1 indicating MOND. Directly comparing the best Newtonian and Milgromian models reveals that Newtonian dynamics is preferred at $19\sigma$ confidence. Using a complementary Markov Chain Monte Carlo analysis, we find that $\alpha_{\rm{grav}} = -0.021^{+0.065}_{-0.045}$, which is fully consistent with Newtonian gravity but excludes MOND at $16\sigma$ confidence. This is in line with the similar result of Pittordis and Sutherland using a somewhat different sample selection and less thoroughly explored population model. We show that although our best-fitting model does not fully reproduce the observations, an overwhelmingly strong preference for Newtonian gravity remains in a considerable range of variations to our analysis. Adapting the MOND interpolating function to explain this result would cause tension with rotation curve constraints. We discuss the broader implications of our results in light of other works, concluding that MOND must be substantially modified on small scales to account for local WBs.

\end{abstract}


\section{Introduction}
\label{Introduction}

Our current best understanding of gravity is encapsulated by the theory of General Relativity \citep[GR;][]{Einstein_1915}. This reduces to Newtonian dynamics in the weak-field non-relativistic limit \citep*{Rowland_2015, Almeida_2016, Ciotti_2022}. According to this theory, the gravity from a point mass follows an inverse square law, which historically was inferred from the fact that the rotation velocity $v_c$ of a planet around the Sun declines with its heliocentric distance $r$ as $v_c \propto 1/\sqrt{r}$. In principle, this Keplerian decline should also apply to galaxies as a whole if one considers tracers of their potential far beyond the bulk of their detectable baryonic mass (stars and gas). However, it is well known that galaxy rotation curves are flat \citep*[][and references therein]{Bosma_1978, Rubin_1978, Faber_1979} $-$ this is nicely illustrated in figure~15 of \citet{Famaey_McGaugh_2012}.

Several explanations have been put forward for this missing gravity problem, which is also apparent in a number of other ways like dwarf galaxy velocity dispersions \citep[figure 11 of][]{McConnachie_2012} and weak gravitational lensing \citep{Brouwer_2021}. The most popular idea is that the extra gravity needed to bind galaxies and enhance their effect on passing photons comes from an otherwise undetected halo of particles beyond the well-tested standard model of particle physics \citep[for a review, see][]{Peebles_2017_DM_review}. This forms the basis for the currently prevailing Lambda cold dark matter ($\Lambda$CDM) standard model of cosmology \citep*{Efstathiou_1990, Ostriker_Steinhardt_1995}. However, DM particles have not been found despite decades of highly sensitive searches for collisions with terrestrial detectors \citep{LUX_2017, LUX_2023}. Similarly, there is no sign of $\gamma$-rays from DM annihilation or decay processes in supposedly DM-dominated Galactic satellites, providing stringent limits on the allowed properties of the hypothetical particles \citep*{Hoof_2020}.

This has led some workers to consider that the missing gravity problem might not be due to seemingly undetectable particles. The most developed alternative is Milgromian dynamics \citep[MOND;][]{Milgrom_1983}. It proposes that the gravity $g$ in an isolated spherically symmetric system is asymptotically related to the Newtonian gravity $g_{_N}$ of the baryons alone according to
\begin{eqnarray}
    g ~\to~ \begin{cases}
    g_{_N} \, , & \textrm{if} ~g_{_N} \gg a_{_0} \, , \\
    \sqrt{a_{_0} g_{_N}} \, , & \textrm{if} ~g_{_N} \ll a_{_0} \, .
    \end{cases}
    \label{g_cases}
\end{eqnarray}
MOND introduces a fundamental acceleration scale $a_{_0}$ below which the gravitational field deviates substantially from Newtonian expectations.\footnote{This is true in percentage terms, but the difference in absolute terms is $\la a_{_0}$.} Just like with the Newtonian constant $G$, we must determine $a_{_0}$ empirically. Using the rotation curves of a handful of disc galaxies with properties similar to our own, it has been found that $a_{_0} = 1.2 \times 10^{-10}$~m/s\textsuperscript{2} \citep*{Begeman_1991}. This value has remained very stable over many decades \citep*{Gentile_2011, McGaugh_Lelli_2016, Chae_2022}.

MOND works remarkably well at predicting the dynamics of galaxies across a huge range in baryonic mass, surface brightness, and gas fraction \citep{Famaey_McGaugh_2012, McGaugh_Lelli_2016, Lelli_2017, Kroupa_2018, Li_2018}. Observations show that galaxies follow a remarkably tight `radial acceleration relation' (RAR) between the radial components of $\bm{g}$ and $\bm{g}_{_N}$ as deduced from rotation curves and the baryonic mass distribution, respectively \citep{McGaugh_2020}. This conclusion is not much dependent on assumptions about the stellar mass-to-light ratio because one can restrict attention to gas-rich galaxies \citep{McGaugh_2012}. The intrinsic scatter in the RAR is only 0.034~dex, which given the uncertainties is consistent with zero \citep{Desmond_2023}. The RAR also extends to elliptical galaxies and even to weak gravitational lensing down to $g_{_N} \approx 10^{-5} \, a_{_0}$ \citep{Brouwer_2021}. The gravitational forces in virialized galaxies are thus very well described by MOND \citep[for a review, see section~3 of][]{Banik_Zhao_2022}.

These successes are complemented by recent numerical MOND simulations of galaxies, which allow an exploration of non-axisymmetric dynamical features like bars and spiral arms \citep{Banik_2020_M33, Roshan_2021_disc_stability}. MOND has also been shown to get about the right star formation rates in disc galaxies with an appropriate size and gas fraction for their stellar mass \citep{Nagesh_2023}. While these works typically target isolated galaxies, MOND has also been applied to interacting galaxies, in particular to understand the Antennae \citep*{Renaud_2016}, the tidal stability of dwarf spheroidal galaxies in the gravitational field of the Fornax Cluster \citep{Asencio_2022}, the asymmetric tidal tails of the globular cluster Palomar 5 \citep{Thomas_2018} and open star clusters in the Solar neighbourhood \citep{Kroupa_2022}, and the formation of the Local Group (LG) satellite planes around the Milky Way (MW) and M31 from tidal debris expelled by a past flyby encounter between their discs around 8~Gyr ago \citep{Zhao_2013, Bilek_2018, Banik_2022_satellite_plane}.

Many of these results were motivated by tensions and inconsistencies within the $\Lambda$CDM paradigm. For instance, it is known to face severe difficulties explaining the LG satellite galaxy planes \citep[][and references therein]{Pawlowski_2021}. Hydrodynamical MOND simulations of M33 \citep{Banik_2020_M33} were motivated by difficulties in understanding why it is weakly barred, bulgeless, and has a two-armed spiral $-$ behaviour which is in many ways the opposite of what arises in Newtonian simulations with a live halo \citep*{Sellwood_2019}. Problems understanding galaxy bars in $\Lambda$CDM are certainly not confined to M33: the model faces a $13\sigma$ tension when confronted with the observed ratios between bar lengths and corotation radii \citep{Roshan_2021_bar_speed}. This fast bar tension has been rather directly linked to dynamical friction between the rotating bar and the hypothetical CDM halo \citep[figure~12 of][]{Roshan_2021_disc_stability}. Those authors demonstrated that the problem is alleviated in MOND (see their figure~20), with a similar result found in subsequent more detailed simulations that include hydrodynamics and star formation \citep[see figure~16 of][]{Nagesh_2023}. The presence of bars in galaxies with a low surface brightness is also very problematic for $\Lambda$CDM because these galaxies are supposedly embedded in a dominant dark halo, which reduces the role of disc self-gravity \citep*{Kashfi_2023}. Milgromian disc galaxies are always completely self-gravitating, making it easier to understand why even galaxies with a large enhancement to their Newtonian baryonic rotation curve commonly have bars and spiral arms \citep*{McGaugh_1995_image, McGaugh_1998_MOND, McGaugh_1998_LCDM}.

MOND has also been applied on galaxy cluster and cosmological scales with the neutrino hot dark matter ($\nu$HDM) paradigm developed by \citet{Angus_2009} and the aether scalar tensor (AEST) model \citep{Skordis_2019, Skordis_2021, Skordis_2022} $-$ for a review, we refer the reader to section~9 of \citet{Banik_Zhao_2022}. Here again some of the successes relate to situations which are severely problematic for $\Lambda$CDM, in particular the KBC void and Hubble tension \citep*{Keenan_2013, Haslbauer_2020} and the El Gordo massive galaxy cluster collision at redshift $z = 0.87$ \citep*{Asencio_2021, Asencio_2023}. Thus, the successes of MOND extend well beyond its initial motivation forty years ago from the rotation curves of high surface brightness disc galaxies. At the same time, the motivation for considering alternatives to $\Lambda$CDM extends well beyond a mere failure to detect the hypothetical particles \citep[for an extensive review, see][]{Banik_Zhao_2022}.

\subsection{The importance of wide binaries (WBs)}
\label{WB_importance}

In this contribution, we seek to test the validity of MOND on very small length scales by galactic standards, greatly diminishing the role of DM \citep{Acedo_2020}. One might think that the role of MOND would also be diminished, but since MOND does not introduce a fundamental new length scale, it should remain valid in systems with a much smaller mass and size than a typical galaxy provided the accelerations are low enough. Equation~\ref{g_cases} indicates that Newtonian and Milgromian gravity diverge when $g \la a_{_0}$. Around an isolated point mass $M$, this occurs beyond its MOND radius
\begin{eqnarray}
    r_{_M} ~\equiv~ \sqrt{\frac{GM}{a_{_0}}} \, .
    \label{MOND_radius}
\end{eqnarray}
The MOND radius of the Sun is 7000~AU (7~kAU) or 0.03~pc. This is much smaller than the typical separation between stars in the Solar neighbourhood, where the Newtonian Jacobi or tidal radius for a system with $M = 2 \, M_\odot$ is 350~kAU \citep[equation 43 of][]{Jiang_2010}. It is thus possible to test MOND at kAU distances from a star provided suitable tracers can be found. The properties of long-period comets that reach kAU distances have been used to argue in favour of the Milgromian enhancement to the Solar potential \citep{Penner_2020_comets}, while its predicted anisotropy may explain the apparent orbital alignment of Kuiper belt objects with a large pericentre \citep{Brown_2023, Migaszewski_2023}. However, this evidence is rather indirect. More direct evidence could be obtained using interstellar precursor missions travelling at a few percent of the speed of light $c$ \citep{Banik_2019_spacecraft}, but this remains well beyond current technology.

These difficulties can be overcome by using stars in WBs, defined in this work as binary stars with orbital acceleration $\la a_{_0}$, implying kAU-scale separations. The nearest star to the Sun is in such a WB \citep{Beech_2009, Beech_2011}. MOND can thus be tested using the orbital acceleration of Proxima Centauri around $\alpha$ Centauri A and B, which have a very small separation and can be treated as one object \citep*{Kervella_2017}. Unfortunately, the $\approx 0.5 \, \mu$as astrometric precision required is roughly two orders of magnitude beyond the reach of current instruments \citep{Banik_2019_Proxima}. Instead of using the acceleration, it is much more promising to use the relative velocity, but then the wide binary test (WBT) of gravity has to be done statistically by considering a large number of WBs to average over orbital phases and projection effects. This was first suggested by \citet*{Hernandez_2012}, with \citet{Pittordis_2018} later exploring the behaviour of WBs in several gravity theories. Since WBs come in a range of masses and separations, the WBT is usually phrased as a statistical test involving the distribution of the dimensionless parameter
\begin{eqnarray}
    \widetilde{v} ~\equiv~ v_{\rm{rel}} \div \overbrace{\sqrt{\frac{GM}{r_{\rm{sky}}}}}^{\rm{Newtonian} \, v_c} ,
    \label{v_tilde_definition}
\end{eqnarray}
where $M$ is the total mass of a WB with sky-projected separation $r_{\rm{sky}}$ and relative velocity $\bm{v}_{\rm{rel}}$, with $v \equiv \left| \bm{v} \right|$ for any vector $\bm{v}$. Thus, the relative velocity is normalized to the Newtonian circular velocity at the projected separation. Since the actual separation is larger, $\widetilde{v}$ is smaller than if the Newtonian $v_c$ had been calculated using the 3D separation. Moreover, we only consider the sky-projected part of $\bm{v}_{\rm{rel}}$ (which we denote $\bm{v}_{\rm{sky}}$) as the data is not yet accurate enough for a 3D version of the WBT (though see Section~\ref{Future_prospects}). This is mainly due to the parallaxes being too inaccurate to reliably constrain the 3D separation (Section~\ref{rp_trick}), though there is also some additional uncertainty from the required correction for the gravitational redshift of each star \citep{Loeb_2022}. Considering only sky-projected quantities further reduces $\widetilde{v}$, which in this contribution is based on using $\bm{v}_{\rm{sky}}$ in Equation~\ref{v_tilde_definition} $-$ we do not use the full 3D $\widetilde{v}$. Note that even though we will mostly be dealing with the magnitudes of $\bm{r}_{\rm{sky}}$ and $\bm{v}_{\rm{sky}}$ in this work, these are actually 2D vectors within the sky plane. This is important to the work of \citet*{Hwang_2022}, who used the angle $\psi$ between $\bm{r}_{\rm{sky}}$ and $\bm{v}_{\rm{sky}}$ to obtain interesting constraints on the WB orbital eccentricity distribution (Section~\ref{WB_orbital_parameter_distribution}).

The first detailed MOND calculations of WB orbits showed that the Milgromian circular velocity would be enhanced by 20\% over the Newtonian expectation in the asymptotic regime of large separations, a result which can be understood analytically \citep[][hereafter \citetalias{Banik_2018_Centauri}]{Banik_2018_Centauri}.\footnote{$\widetilde{v}$ in this contribution has the same meaning as $\widetilde{v}_{\rm{sky}}$ in \citetalias{Banik_2018_Centauri}.} Numerical MOND simulations of nearby star clusters also show a 20\% enhanced velocity dispersion over their Newtonian counterparts for essentially the same reason \citep{Kroupa_2022}. The enhancement is limited by the Galactic gravity on the Solar neighbourhood, which is important in MOND because it is a non-linear theory (Equation~\ref{g_cases}). As a result, the internal behaviour of a self-gravitating system is influenced by an external source of gravity, even in the absence of tidal effects. This non-standard external field effect (EFE) has been an important aspect of MOND since the beginning, not only because it is theoretically inevitable in a non-linear theory, but also because the EFE is necessary to suppress the MOND enhancement to the velocity dispersions of the Pleiades and Praesepe open star clusters \citep[see section~3 of][]{Milgrom_1983}.

Many further arguments for the EFE can be made nowadays \citep[for a review, see section~3.3 of][]{Banik_Zhao_2022}. It significantly enhances the tidal susceptibility of dwarf galaxies in a cluster environment by weakening their self-gravity, thereby helping to explain the observed signs of tidal disturbance in the Fornax Cluster's dwarf galaxies and the lack of low surface brightness dwarfs towards its centre \citep{Asencio_2022}. As explained in that work, tidal radii are necessarily in the EFE-dominated regime, making their results particularly sensitive to the EFE. Its subtle imprint on the outer parts of disc galaxy rotation curves has recently been detected by comparing galaxies in isolated and more crowded environments \citep{Chae_2020_EFE, Chae_2021}. The EFE is also crucial when calculating the escape velocity from a mass distribution, since the Milgromian potential of an isolated mass is infinitely deep \citep*{Famaey_2007, Zhao_2010_two_body, Banik_2018_escape}. This is not true once the EFE is included because at large distances where the EFE dominates, the gravitational field of a point mass returns to the Newtonian inverse square law, albeit with a higher normalization and some angular dependence \citep[see figure~1 of][]{Banik_Zhao_2022}. This EFE-dominated regime is sometimes known as the quasi-Newtonian regime, though in this contribution we simply refer to it as the asymptotic regime because local WBs lack an extended region in which they are isolated and at low acceleration. This is due to the Galactic external field $g_e$ being slightly stronger than $a_{_0}$, as evident from the Galactic rotation curve \citep[e.g.][]{Zhu_2023}.

Given the importance of $g_e$ for the WBT, it is fortunate that we know $\bm{g}_e$ from a direct measurement of the acceleration of the Solar System relative to distant quasars using their changing aberration angle, which directly tells us that $\bm{g}_e/c = 5.05 \pm 0.35 \, \mu$as/yr in a direction very close to that of the Galactic centre \citep{Klioner_2021}. This precludes substantial deviations of $g_e$ from the value deduced kinematically from the Galactic rotation curve (Section~\ref{WB_population}). Even so, the predicted 20\% enhancement to the orbital velocities of local WBs is related to the shape of the MOND interpolating function in the regime close to the critical acceleration scale $a_{_0}$. It is thus somewhat model-dependent as it does not directly relate to the weak-field asymptotic limit at the heart of MOND. Nevertheless, recent detailed rotation curve studies have mostly converged on the form of the interpolating function in this regime, though of course slight variations are still possible. We will return to this point in Section~\ref{MOND_interpolating_function}.

Prospects for the WBT were discussed extensively in \citetalias{Banik_2018_Centauri}, whose section~8 clarified the main systematic effects that would likely have to be considered. The main issue was expected to be undetected close binary (CB) companions to one or both of the stars in a WB. Indeed, the nearest star to the Sun is in just such a triple system: Proxima Centauri is on a wide orbit about $\alpha$~Centauri A and B, whose mutual orbital semi-major axis of 23~AU is negligible compared to the 13~kAU distance from their barycentre to Proxima Centauri \citep{Kervella_2016, Kervella_2017}. At larger distances from the Sun, it is possible that a similar CB would not be resolved, but it would still have significant effects on the observed kinematics that would not arise if Proxima Centauri were orbiting a single star (Section~\ref{CB_impact}). \citet[][hereafter \citetalias{Pittordis_2019}]{Pittordis_2019} considered the WBT in light of actual data from \emph{Gaia} Data Release 2 \citep[DR2;][]{Gaia_2018}, highlighting that there is indeed an extended tail towards much higher $\widetilde{v}$ than could plausibly arise from genuine WB orbital motion, which cannot realistically yield $\widetilde{v} \ga 2$ for any reasonable modification to gravity. The possible presence of CBs was strongly suggested by the analysis of \citet{Clarke_2020} and later by \citet{Belokurov_2020}, who argued that some of the WBs with unexpectedly large relative velocities also have a \emph{Gaia} astrometric solution that poorly fits the observations. Since the astrometric solution only includes parallax and proper motion, it was argued that the excess `noise' could be due to astrometric acceleration induced by a CB, which was previously shown to typically induce a much larger orbital acceleration than the WB orbital motion \citepalias[see section~8.2 of][]{Banik_2018_Centauri}. Further follow-up observations should be able to confirm the existence of CB companions around some WBs \citep*{Manchanda_2023}, especially once the astrometric time series are published in \emph{Gaia}~DR4 (see Section~\ref{Future_prospects}).

Some line of sight (LOS) contamination is also inevitable, though this is expected to be rather modest thanks to the excellent quality of the \emph{Gaia} data \citepalias[see section~3 of][]{Pittordis_2019}. Their work shows a clear main peak to the $\widetilde{v}$ distribution due to WB orbital motion. Investigation of its properties already decisively rejects theories where local WBs behave like isolated MOND systems without the weakening of their self-gravity due to the Galactic EFE (see their figure~11). This is because the gravity binding WBs can be deduced from the main peak to the $\widetilde{v}$ distribution at $\widetilde{v} \la 2$, with CB and LOS contamination leading to an extended low amplitude tail out to much higher $\widetilde{v}$. If the orbital velocity becomes asymptotically flat with increasing separation, we would expect the mode of the $\widetilde{v}$ distribution to be significantly higher in systems with large $r_{\rm{sky}}$. The predicted peak shift is in catastrophic disagreement with the observations, which show at most only a modest peak shift. This still leaves open the possibility that WBs obey MOND with its inevitable EFE, but it falsifies the quantized inertia proposal \citep*{McCulloch_2019} and repeated claims to have confirmed that WBs follow the isolated MOND prediction without considering the full $\widetilde{v}$ distribution \citep{Hernandez_2019_WB, Hernandez_2022}.

In this contribution, we conduct a detailed statistical analysis to test the MOND prediction for local WBs. We use a rigorous grid solution to the MOND field equation including the EFE \citepalias{Banik_2018_Centauri} and then add possible perturbations from undetected CBs around one or both of the stars in each WB. We jointly infer properties of the WB population, the undetected CB population, and the extent of LOS contamination, which becomes important towards large $r_{\rm{sky}}$ and $\widetilde{v}$. Both the model used and the exploration of its parameter space are much more detailed than in any previous attempt at the WBT. To mitigate possible biases, the plan was prepared in advance \citep*{Banik_2021_plan} and has barely been modified to deal with the actual \emph{Gaia}~DR3 \citep{Gaia_2023}.

In the following, we explain how we reduce the \emph{Gaia} dataset to a form suitable for our analysis and what quality cuts we employ (Section~\ref{Observed_WB_distribution}). We then introduce our detailed model for the WB dataset and explain how we compare it to observations (Section~\ref{Model}). Our results are presented in Section~\ref{Results} and discussed in Section~\ref{Discussion}, which includes a comparison with prior WBT results (Section~\ref{Other_WBT_results}). We conclude in Section~\ref{Conclusions}.

\section{The observed WB distribution}
\label{Observed_WB_distribution}

The primary dataset for the WBT comes from the precise results obtained by the \emph{Gaia} mission \citep{Gaia_2016}. Local WBs were extracted from the \emph{Gaia} dataset using the methods described in \citet[][hereafter \citetalias{Pittordis_2023}]{Pittordis_2023}, who slightly adapted the methods in an earlier work \citepalias[section~2 of][]{Pittordis_2019} based on \emph{Gaia}~DR2 \citep{Gaia_2018} due to the improved quality in \emph{Gaia} Early Data Release~3 \citep[EDR3;][]{Gaia_2021}. We supplement EDR3 with radial velocities (RVs) from the full \emph{Gaia}~DR3 \citep{Gaia_2023}.

In Section~\ref{Gaia_DR3_sample}, we describe the basic quality cuts that we impose. We then conduct a Monte Carlo propagation of the uncertainties (Section~\ref{Monte_Carlo_error_propagation}). This leads us to impose further quality cuts, which we briefly summarize in Section~\ref{Refined_quality_cuts} alongside our choice of restrictions on the parameter ranges.

\subsection{The \emph{Gaia} DR3 sample \& basic quality cuts}
\label{Gaia_DR3_sample}

The details of the sample selection are given in \citetalias{Pittordis_2023}, whose main points we summarize below. We begin by selecting all stars from \emph{Gaia}~EDR3 at a Galactic latitude $\lvert b \rvert > 15^\circ$ with an apparent \emph{Gaia}-band ($G$-band) magnitude $m_G < 17$ and measured parallax $\varpi > 4$~mas (estimated distance $<250$~pc) uncorrected for parallax bias \citep{Lindegren_2021}. From this sample of 2.1~million single stars, candidate WBs are selected by requiring:   
\begin{enumerate}
    \item Projected separation $r_{\rm{sky}} \le 50$~kAU;
    \item Star distances consistent with each other within the lesser of 8~pc or $4\times$ the combined distance uncertainty, i.e. $\lvert d_A - d_B \rvert \leq \min \left( 4\sigma_d, 8 \, \rm{pc} \right)$, where $A$ and $B$ label the stars in each candidate WB; and
    \item Projected velocity difference between the stars of $v_{\rm{sky}} < 3$~km/s, as inferred from the difference in proper motions but assuming both stars are at the mean estimated distance $\bar{d} \equiv \left( d_A + d_B \right)/2$.
\end{enumerate}
From this preliminary list of binaries (WB-EDR3), sky regions are removed around four known open clusters \citepalias[see table~1 of][]{Pittordis_2023}.

To remove some probable triple systems or groups, we reject all WBs in which either star is common to more than one candidate binary in WB-EDR3. We also search for comoving companion stars to a fainter limit using a `faint star' sample constructed from \emph{Gaia}~EDR3 stars with $m_G < 20$ and parallax $\varpi > 10/3$~mas. For each star in each candidate WB, we search for companions in the faint star sample with the following criteria:
\begin{enumerate}
    \item Parallax distance consistent with the main star at $4\sigma$;
    \item Angular separation less than 2/3 of the main binary separation (since hierarchical triples are expected to be unstable if the inner orbit separation $\ga 0.4\times$ the outer separation);
    \item Angular separation $>0.5\arcsec$ to avoid barely-resolved companions; and
    \item Projected velocity difference from the main star $\leq 5$~km/s.
\end{enumerate}
If any such `third star' is found, the candidate binary is rejected. Note however that this will not reject triples with an inner orbit closer than about 100~AU or where the third star is so faint that $m_G > 20$, so it remains crucial to model the CB population. We return to this point in Section~\ref{aint_distribution}.

\citetalias{Pittordis_2023} also applied a per-star quality cut based on the \emph{Gaia} parameters \citep[see equation~1 of][]{Arenou_2018}. WBs were only considered if both stars satisfy
\begin{eqnarray} 
    \label{Arenou_cut}
    \sqrt{\frac{\chi^2}{\nu}} &\le& 1.2 \max \left( 1, \, \exp \left[ -0.2 \left( m_G - 19.5 \right) \right] \right) \, , \\
    \chi^2 &\equiv& {\tt astrometric\_chi2\_al}  \, , \nonumber \\
    \nu &\equiv& {\tt astrometric\_n\_good\_obs\_al} - 5 \, . \nonumber
\end{eqnarray}
This cut is based on \emph{Gaia}~DR2, so it should be fairly conservative in \emph{Gaia}~DR3 given the greater number of observations per star and the longer time baseline. Throughout this study, any quality cuts applied at the individual star level are implemented by excluding the whole WB if either of its two stars fails to pass the relevant quality cut.

Applying all the above quality cuts yields a sample of 73k candidate WBs. The majority have $r_{\rm{sky}} \le 1$~kAU, which is not very useful for the WBT. We therefore publicly release a smaller version of this sample consisting of 19786 WBs that satisfy $1.5 < r_{\rm{sky}}/\rm{kAU} < 40$, which is somewhat wider than the parameter range we use for the WBT.

From the \emph{Gaia}~DR3 WBs that pass the above quality cuts, we restrict ourselves to the range $r_{\rm{sky}} = 2-30$~kAU and $\widetilde{v} \leq 5$. These choices were fixed in advance to limit the possibility of biasing the results \citep{Banik_2021_plan}. The upper limit to $\widetilde{v}$ is much larger than can plausibly arise from WB orbital motion. This is necessary to properly sample the extended tail to the $\widetilde{v}$ distribution, which we need in order to properly constrain the CB and LOS contamination causing the tail. In particular, LOS contamination is expected to become more significant at high $\widetilde{v}$ because $\widetilde{\bm{v}}$ is a 2D quantity, so we can get much better leverage on the extent of LOS contamination if we have data at high $\widetilde{v}$. Even so, we do not want to include data at very high $\widetilde{v}$ because the likelihood of systematics increases the wider the `aperture' used for the WBT.

We could increase the sample size substantially by going down to even lower separations, but then there would be a risk of swamping the analysis with data on Newtonian systems. In this case, small improvements to the fit in the Newtonian regime would become more important to the statistical analysis than checking whether the $\widetilde{v}$ distribution differs between systems within their MOND radius and those with larger separations. On the other hand, we need to go down to rather low $r_{\rm{sky}}$ to provide a `Newtonian anchor' population that cannot be affected by MOND, especially since the RAR requires a fairly smooth MOND interpolating function such that significant MOND effects persist even when the accelerations slightly exceed $a_{_0}$ (Section~\ref{MOND_interpolating_function}). Model parameters besides the gravity law can be constrained much better if we have a significant number of WBs which cannot be much affected by MOND, since then the parameters inferred here will not be degenerate with the gravity law. In particular, a Newtonian anchor population should be extremely valuable in constraining the CB population parameters $-$ WBs are expected to have similar CB populations regardless of whether their WB orbital acceleration is below or above $a_{_0}$. Our analysis does not explicitly consider some subset of our sample to be completely immune to MOND \citep[unlike][]{CHAE_2023}, but instead relies on the predicted MOND effects being different at low and high $r_{\rm{sky}}$. We will see later that since our sample goes up to $M = 4 \, M_\odot$ and thus has systems with MOND radii of up to 14~kAU (Equation~\ref{MOND_radius}), we have plenty of WBs separated by a small fraction of their MOND radius. This is especially true given the steeply declining $r_{\rm{sky}}$ distribution of WBs \citepalias{Banik_2018_Centauri, Pittordis_2019, Pittordis_2023}, which also implies that the 3D separation is likely to only slightly exceed $r_{\rm{sky}}$ (Equation~\ref{Deprojection}).

\subsection{The mass-luminosity relation}
\label{ML_relation}

We estimate the mass of each star in our WB sample from its absolute $G$-band magnitude $M_G$ using a similar technique to \citetalias{Pittordis_2019}, which we briefly describe below. We use the $M_V$ and $\left( V - I \right)$ colour at different stellar masses $M$ as tabulated in \citet{Pecaut_2013}.\footnote{\url{https://www.pas.rochester.edu/~emamajek/EEM_dwarf_UBVIJHK_colors_Teff.txt} [3.3.2021]} We relate this to \emph{Gaia} photometry using the relation between $\left( V - I \right)$ and $\left( G - V \right)$ colours from the first Johnson-Cousins relation in table~C2 of \citet{Riello_2021}, which states that
\begin{eqnarray}
    \left( G - V \right) \!\!\! &=& \!\!\! -0.01597 - 0.02809 \left( V - I \right) - 0.2483 \left( V - I \right)^2 \nonumber \\
    && \!\!\! + 0.03656 \left( V - I \right)^3 - 0.002939 \left( V - I \right)^4 .
\end{eqnarray}

\begin{figure}
    \centering
    \includegraphics[width=0.47\textwidth]{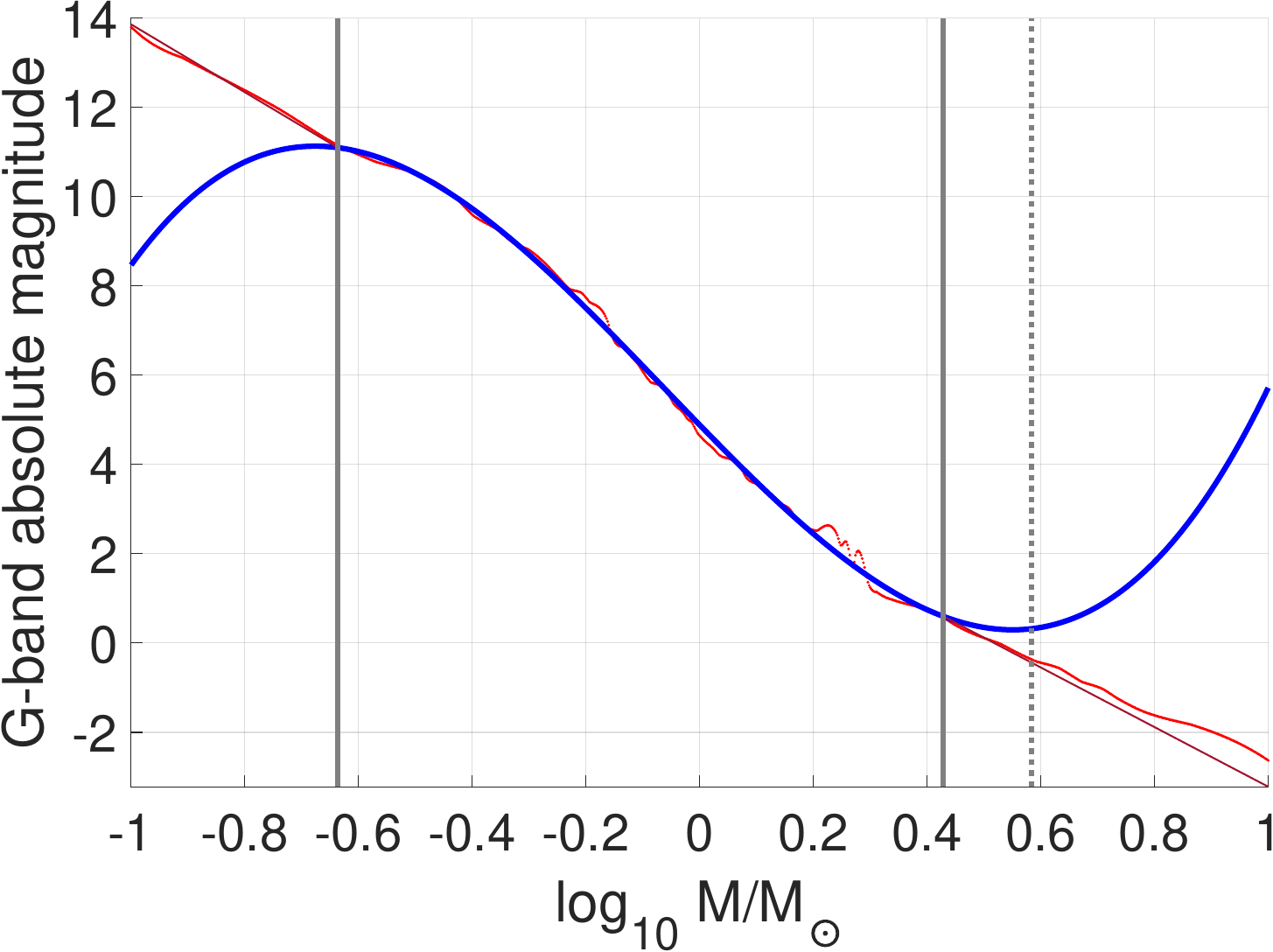}
    \caption{The relation between the mass of a star and its absolute $G$-band magnitude (red line). Results are based on the absolute $V$-band magnitude and $\left( V - I \right)$ colour tabulated in \citet{Pecaut_2013} and the relation between $\left( V - I \right)$ and $\left( G - V \right)$ colours from the first Johnson-Cousins relation in table~C2 of \citet{Riello_2021}. Our cubic fit to the data (blue line; Equation~\ref{Cubic_mass_luminosity_relation}) is used over the mass range between the vertical solid grey lines. Outside this range, a linear relation is assumed, as indicated by the solid purple lines. The dotted vertical line shows the maximum mass of any star in our WB sample.}
    \label{Mass_luminosity_relation}
\end{figure}

We now have a complete set of equations to obtain $M_G$ for any $M$. However, this is difficult to invert because of small-scale irregularities in the relation arising from the complexities of stellar astrophysics (Figure~\ref{Mass_luminosity_relation}). We therefore fit the relation using the cubic
\begin{eqnarray}
    M_G ~=~ 4.887 - 5.693x + 0.4164x^2 + 0.9611x^3 \, ,
    \label{Cubic_mass_luminosity_relation}
\end{eqnarray}
where $x \equiv \ln M/M_\odot$. Since each unit decrease in $M_G$ corresponds to slightly less than an $\mathrm{e}$-fold increase in luminosity $L$, Equation~\ref{Cubic_mass_luminosity_relation} tells us that $L \appropto M^5$ when $M = M_\odot$ ($x = 0$), representing a steep mass-luminosity relation. We use our cubic fit over the range $0.6 < M_G < 11.1$. At the low (high) mass end, we assume the relation becomes linear and has a slope of $-3.3$ ($-2.9$). The fitted relation between $M_G$ and $x$ is shown as the solid blue line, confirming that the above cubic provides a fairly good fit over the range we use it, i.e. between the vertical solid grey lines. The purple lines outside this range show that a linear relation here also provides a reasonably good fit, at least up to the dotted vertical grey line showing the maximum mass of any star in our sample. A cubic is the lowest order polynomial which seems to provide a good match to the tabulated data in \citet{Pecaut_2013}, mainly because the slope is similar at low and high masses but there is a steepening at intermediate masses. 

We invert our piece-wise analytic relation between $\ln M/M_\odot$ and $M_G$ to obtain the mass of any star from its $M_G$, which we obtain from its apparent magnitude using its trigonometric parallax. The linear relations assumed at very low and very high masses can be readily inverted. We invert our cubic fit over most of the mass and $M_G$ range using the Newton-Raphson method. To speed this up, we obtain an initial guess for the mass using a linear fit of the form
\begin{eqnarray}
    M_G ~\approx~ 5.023 - 5.102x \, .
    \label{Linear_mass_luminosity_relation}
\end{eqnarray}
This can be inverted analytically without iterative techniques. Our estimate that $-\frac{d\ln M}{dM_G} = 0.20$ is similar to previous estimates of $0.074 \ln 10 = 0.17$ \citepalias[equation~3 of][]{Pittordis_2023} and $0.0725 \ln 10 = 0.17$ \citepalias[equation~3 of][]{Pittordis_2019}.

We consider uncertainties in $M_G$ arising from those in the trigonometric parallax and the \emph{Gaia} photometric uncertainty in the apparent magnitude, which we convert from a fractional uncertainty to a magnitude uncertainty. We also allow a small uncertainty in the conversion of luminosity to mass. This must play some role, but the $\approx 42\%$ asymptotic enhancement to the radial gravity predicted in MOND combined with our estimate that $\frac{dM_G}{d\ln M} = -5.1$ (Equation~\ref{Linear_mass_luminosity_relation}) implies that the MOND signal corresponds to a difference in $M_G$ of about $5.1\ln1.4 = 1.7$ magnitudes, which is very large compared to the slight fluctuations about our cubic mass-luminosity relation evident in Figure~\ref{Mass_luminosity_relation}. In other words, the steep mass-luminosity relation implies that mass uncertainties should have a very small impact compared to uncertainties in the relative velocity arising from uncertain proper motions, especially when the key quantity entering $\widetilde{\bm{v}}$ is essentially the relative proper motion between the two stars in each WB. The mass of a star with known absolute magnitude can be constrained to within $\approx6\%$ \citep{Eker_2015, Mann_2015}, but due to the square root term in Equation~\ref{v_tilde_definition}, this would have only a 3\% impact on $\widetilde{v}$. The impact on the WBT is further reduced because it is mostly concerned with the width of the $\widetilde{v}$ distribution, which is only affected at second order by measurement uncertainties. Even so, we do assign a modest 0.024~dex or 5.5\% uncertainty to our stellar mass estimates at fixed luminosity for reasons discussed in Section~\ref{FLAME_calibration}. We will see later that our main conclusion is the same as a different study which used a linear relation between $M_G$ and $x$ over the full mass range relevant for the WBT, thus leading to rather different errors in the conversion of absolute magnitudes to masses \citepalias{Pittordis_2023}.

\subsection{Monte Carlo error propagation}
\label{Monte_Carlo_error_propagation}

We use the full $5 \times 5$ \emph{Gaia} covariance matrix to propagate uncertainties in the sky position, parallax, and proper motion of each star in our WB sample into an uncertainty in $\widetilde{v}$. We will see later that this uncertainty is very small for a large number of WBs, so our approach is to consider only those systems and obtain their $\left( r_{\rm{sky}}, \widetilde{v} \right)$ directly from the raw observables. The resulting $\left( r_{\rm{sky}}, \widetilde{v} \right)$ distribution is then used in our main analysis, which therefore does not directly include measurement uncertainties $-$ these are included indirectly in the sample selection. Our procedure avoids adding measurement errors to values which already contain measurement errors, as that would lead to adding measurement errors twice to the latent values.

For our Monte Carlo error propagation, we diagonalize the correlation matrix and generate Gaussian random numbers to propagate uncertainties along each of the eigenvector directions, bearing in mind also the eigenvalues. The correlation matrix is dimensionless, so we relate it to the covariance matrix using the \emph{Gaia} catalogue uncertainty of each parameter. Uncertainties in the parallax and proper motion are important, but we do not propagate the uncertainty in the sky position of each star at the reference epoch, instead taking the published position at face value. This simplifies our analysis greatly because it enables us to work with a fixed definition of the sky plane for each WB, which we take to be the plane whose normal vector is the angular bisector of the directions toward each of the stars in the WB. Since the sky positions of the stars in our WB catalogue are expected to be particularly precise, there should be negligible uncertainty associated with this definition. Apart from this minor change, we follow the same method as that described in \citet{Banik_2019_line} using mock parallax and proper motion data which samples the \emph{Gaia} error distribution as described above. We assume that the projected separation of each WB has only a negligible uncertainty, so we obtain $r_{\rm{sky}}$ from the raw data.

\subsubsection{The systemic RV}
\label{Systemic_RV}

Compared to the detailed plan we set up earlier \citep{Banik_2021_plan}, a major simplification has become possible with regards to the systemic RV of each WB because we can generally obtain this directly from \emph{Gaia}~DR3 \citep{Gaia_2023} without having to guess it from the proper motion and distance to the Galactic disc plane. The RV of each star enters the calculation of $\bm{v}_{\rm{sky}}$ through perspective effects \citep{Shaya_2011, Banik_2019_line, Badry_2019_geometry} $-$ a receding WB with no internal motion will have a shrinking apparent separation on the sky. The impact is suppressed by the angular separation of the WB on our sky plane, but the systemic RV still cannot be completely neglected: for a WB with $r_{\rm{sky}} = 10$~kAU, a typical systemic RV of 20~km/s translates into an apparent proper motion of 10~m/s at a heliocentric distance of 100~pc. Since the Newtonian $v_c$ of two Sun-like stars with a 10~kAU separation is 420~m/s, the impact on $\widetilde{v}$ would be $\la 0.03$ depending on the geometry. While this is small, we opt to try and include the impact of recessional motion because WBs can be even closer and/or have a larger RV.

We can see from the above that the at most km/s level difference in RV between the stars in each WB would have only a negligible impact on $\widetilde{v}$. This is because relative motions along the LOS and within the sky plane should be comparable, but the impact of the former are suppressed by the angular WB separation on the sky. Consequently, we limit ourselves to including only the systemic RV of each WB. We find the systemic RV by taking an inverse variance weighted mean of each star's RV when this is available for both stars, leading to an uncertainty smaller than either star's RV uncertainty.\footnote{We combine the RV measurements when these are known for both stars because the typical RV uncertainty is around 1~km/s, which is much more than the expected WB relative velocity. Thus, differences in the RV are best understood as due to random errors rather than intrinsic WB orbital motion.} If the RV is only available for one star, we adopt this as the systemic value. We then assign both stars a recessional velocity equal to the systemic RV, errors in which are propagated as part of the Monte Carlo error propagation.

\subsubsection{Reducing uncertainty in the relative distance}
\label{rp_trick}

Another source of perspective effects \citep[discussed further in][]{Shaya_2011} is the difference in heliocentric distances to the stars in each WB. The relative LOS separation is important because proper motions must be multiplied by distances to convert them from angular to physical velocities. The \emph{Gaia} parallaxes typically do not provide very tight constraints on the LOS separation: while the astrometry is spectacular by historical standards, trigonometric distances to stars 100~pc away are not accurate at the kAU level \citepalias[the typical uncertainty is about 80~kAU; see section~6.2 of][]{Banik_2018_Centauri}. Using the same arguments as in Section~\ref{Systemic_RV}, a system with a tangential motion of 20~km/s but whose LOS separation is uncertain by 80~kAU would have an almost 80~m/s uncertainty in $\bm{v}_{\rm{sky}}$. This translates to an uncertainty in $\widetilde{v}$ of up to 0.2 depending on the geometry, which could hamper the WBT.

We intuitively expect that a WB's 3D separation is not much larger than its $r_{\rm{sky}}$, as suggested by the referee to \citet{Banik_2019_line}. We exploit this insight using the technique discussed in its section~2.3, which we briefly summarize here. The mean of the heliocentric distances to the two stars is taken to be the inverse variance weighted mean of the parallax distances to each of the stars in the WB, causing the mean distance to vary slightly between trials and thereby affect the distance modulus. Assuming a power-law prior on the 3D separation $r$ of the form $P \left( r \right) \propto r^{-1.6}$ \citep{Lepine_2007, Andrews_2017, Badry_2018}, we obtain a posterior inference on the ratio $x \equiv r/r_{\rm{sky}}$. Following equation~18 of \citet{Banik_2019_line}, we obtain that
\begin{eqnarray}
    P \left( x \right) ~\propto~ \frac{x^{-2.6}}{\sqrt{x^2 - 1}} \, , \quad x \geq 1 .
    \label{Deprojection}
\end{eqnarray}
As part of the Monte Carlo error propagation, we then randomly sample this distribution to obtain the separation of each WB along the LOS. We then push one of the stars away from us and bring the other star towards us by half this amount, with the displacement of each star being purely along its sky direction at the reference epoch. We use a random coin toss in each Monte Carlo trial to decide which star is to be pushed away from us. The revised distance to each star alters its estimated absolute magnitude and thus its mass.

This technique drastically reduces the uncertainty on the relative LOS distance in systems with a very low $r_{\rm{sky}}$ \citep[compare the black and blue points in figure~3 of][]{Banik_2019_line}. One disadvantage is that it implicitly assumes that the WB is genuine, so our results are less robust against LOS contamination. However, we mitigate this in our statistical analysis by including an allowance for LOS contamination (Section~\ref{Chance_alignments}).

Our approach is similar to that used in \citet{Nelson_2021}, though they used a prior of $r^{-1.5}$ instead of $r^{-1.6}$. A more important difference is that those authors were only interested in the 3D separation between the stars, whereas we need to find the heliocentric velocity of each star, which requires knowledge of the individual LOS distances. This causes our calculated $\widetilde{v}$ parameter to depend on which star is closer to us. Since we do not have this information, we use Monte Carlo trials to quantify the resulting uncertainty. We stress that in our main analysis, the adopted value of $\widetilde{v}$ is based on assigning the same LOS distance to both stars. The purpose of our deprojection is to quantify the uncertainty and combine this with other sources of uncertainty so that we can remove systems with insufficiently precise $\widetilde{v}$.

The technique discussed above assigns the relative LOS distance a value of zero and a non-Gaussian uncertainty of $\approx r_{\rm{sky}}$. This would only reduce the uncertainties if the parallaxes were not accurate enough to directly constrain the relative distance. Thus, we only apply the above technique if $r_{\rm{sky}}$ is smaller than the uncertainty in the relative LOS distance given by the \emph{Gaia} trigonometric parallaxes. While we generally expect this to be the case, we do not enforce it. The 3D geometry of nearby WBs may well be clear from existing \emph{Gaia} data, with the definition of `nearby' extending to larger distances as the data improves. Our approach is thus to use whichever method is expected to be more accurate. Note that if \emph{Gaia} parallaxes are used directly, then the relative distance along the LOS may have a non-zero mean value when averaged across different Monte Carlo trials as these would propagate uncertainties in each parallax. In this case, the nominal values adopted for our main analysis do not rely on the deprojection algorithm outlined above because each star is assigned its observed parallax distance.

\subsubsection{Improved mass estimates}
\label{FLAME_calibration}

It is possible to slightly improve the accuracy of our mass estimates by using the Final Luminosity, Age and Mass Estimator \citep[FLAME;][]{Pichon_2007} work package in \emph{Gaia}~DR3. Masses obtained in this way are not available for all stars, but when they are available, they should be much more precise because they involve a detailed analysis of the spectrum. Figure~3 of \citet{Hernandez_2023} shows that FLAME masses can differ by $\approx 0.05 \, M_\odot$ from estimates using a linear relation between absolute magnitude and the logarithmic mass. Unfortunately, restricting to only WBs where both stars have a FLAME mass would reduce the sample size too much for the WBT to be feasible. We therefore apply a small correction to the masses estimated using our cubic fit to the \citet{Pecaut_2013} mass-luminosity relation (Equation~\ref{Cubic_mass_luminosity_relation}). Denoting these masses by $M_{PM}$ and the revised masses with the FLAME calibration as $M_F$, a good fit is given by
\begin{eqnarray}
    \label{FLAME_adjustment}
    M_F &=& M_{PM} - 0.07 \tanh \left( \frac{M_{PM} - 0.75 \, M_\odot}{0.16 \, M_\odot} \right) f \, ,\\
    f &=& \begin{cases}
        \exp \left( 2 - \frac{M_\odot}{M_{PM}} \right), \quad \quad \textrm{ if} ~M_{PM} \leq 0.5 \, M_\odot \, , \\
        1, \qquad \quad \quad \, \textrm{ if} ~0.5 \, M_\odot \leq M_{PM} \leq 0.75 \, M_\odot \, , \\
        \exp \left( \frac{0.75 \, M_\odot - M_{PM}}{2 \, M_\odot} \right), \textrm{ if} ~M_{PM} \geq 0.75 \, M_\odot \, .
    \end{cases} \nonumber
\end{eqnarray}
The adjustment is tapered using the factor $f$ so it rapidly decays for stars with mass $M < 0.5 \, M_\odot$ as there are no FLAME masses below this, so we assume that $M_{PM}$ becomes very reliable at the low-mass end. We will see later that our results are not much affected by restricting our WB sample to systems where both stars have $M > 0.5 \, M_\odot$, which has the advantage that FLAME results can serve as a calibration. $M_{PM}$ and FLAME masses gradually converge when $M \gg 0.75 \, M_\odot$.

Our small adjustment to $M_{PM}$ is designed to eliminate a small systematic discrepancy with the FLAME mass in the stars for which the latter is available. Even so, our calculated $M_F$ does not exactly coincide with the FLAME mass. Based on the typical difference between the two mass estimates and assuming that the FLAME masses are very accurate, we assign our estimated masses a random Gaussian uncertainty of 0.024~dex or 5.5\% at fixed absolute magnitude. A 5.5\% uncertainty is in line with the fact that masses estimated directly from luminosities were expected to have a 6\% uncertainty \citep{Eker_2015, Mann_2015}, as argued in section~6.3 of \citetalias{Banik_2018_Centauri}. Note that our mass estimates also include uncertainty in the absolute magnitude due to that in the apparent magnitude and the trigonometric parallax.

\subsection{Refined quality cuts for the WBT}
\label{Refined_quality_cuts}

The quality cuts discussed in Section~\ref{Gaia_DR3_sample} provide a reasonably carefully prepared sample of WBs that can be used in further analyses. Unfortunately, some of these WBs are unsuitable for our highly precise analysis. The additional quality cuts in this work beyond those mentioned there are to exclude:
\begin{enumerate}
    \item Stars in sky directions with a high total Galactic extinction towards an extragalactic source (Section~\ref{Dust_extinction});
    \item Stars which are substantially below the main sequence on a colour-luminosity diagram, which can make the mass estimate inaccurate (Section~\ref{CMD_cut});
    \item Stars which sometimes appear as multiple peaks in the \emph{Gaia} images, suggesting the presence of a CB companion that can sometimes be resolved (Section~\ref{ipd_cut});
    \item WBs where the RV is unknown for both stars (Section~\ref{Systemic_RV_requirement});
    \item WBs where the RV is known for both stars and there is a mismatch at $>3\sigma$ confidence (Section~\ref{Similar_RV_requirement}); and
    \item WBs whose $\widetilde{v}$ is too uncertain (Section~\ref{v_tilde_uncertainty}).
\end{enumerate}
We also restrict ourselves to WBs with $2 < r_{\rm{sky}}/\rm{kAU} < 30$ and $\widetilde{v} < 5$ for reasons discussed earlier. To reduce the computational cost, we only consider WBs with a total mass of $0.464 - 4.31 \, M_\odot$. These cuts are now described further.

\subsubsection{Dust extinction}
\label{Dust_extinction}

As discussed in Section~\ref{ML_relation}, our estimated mass for each star relies entirely on its absolute $G$-band magnitude, which in turn is inferred from its apparent magnitude and trigonometric parallax. The difference between apparent and absolute magnitudes is assumed to arise purely from the distance modulus, which implicitly assumes negligible dust extinction. This should be a good assumption for stars within 250~pc and $>15^\circ$ from the Galactic disc, especially as known star clusters are excluded \citepalias[table~1 of][]{Pittordis_2023}. Even so, we can further reduce the impact of dust by considering only WBs where both stars have a $V$-band extinction $A_V < 0.5$ based on the dust maps of \citet*{Schlegel_1998}, which reduces our sample size by 8.5\%. We note that their published values refer to the total Galactic extinction along the LOS towards an extragalactic source. Since the WBs in our sample are all within 250~pc, the extinction towards any of the stars in our sample should be far smaller, making our cut on $A_V$ quite conservative.

\subsubsection{Colour-magnitude diagram}
\label{CMD_cut}

\begin{figure}
    \centering
    \includegraphics[width=0.47\textwidth]{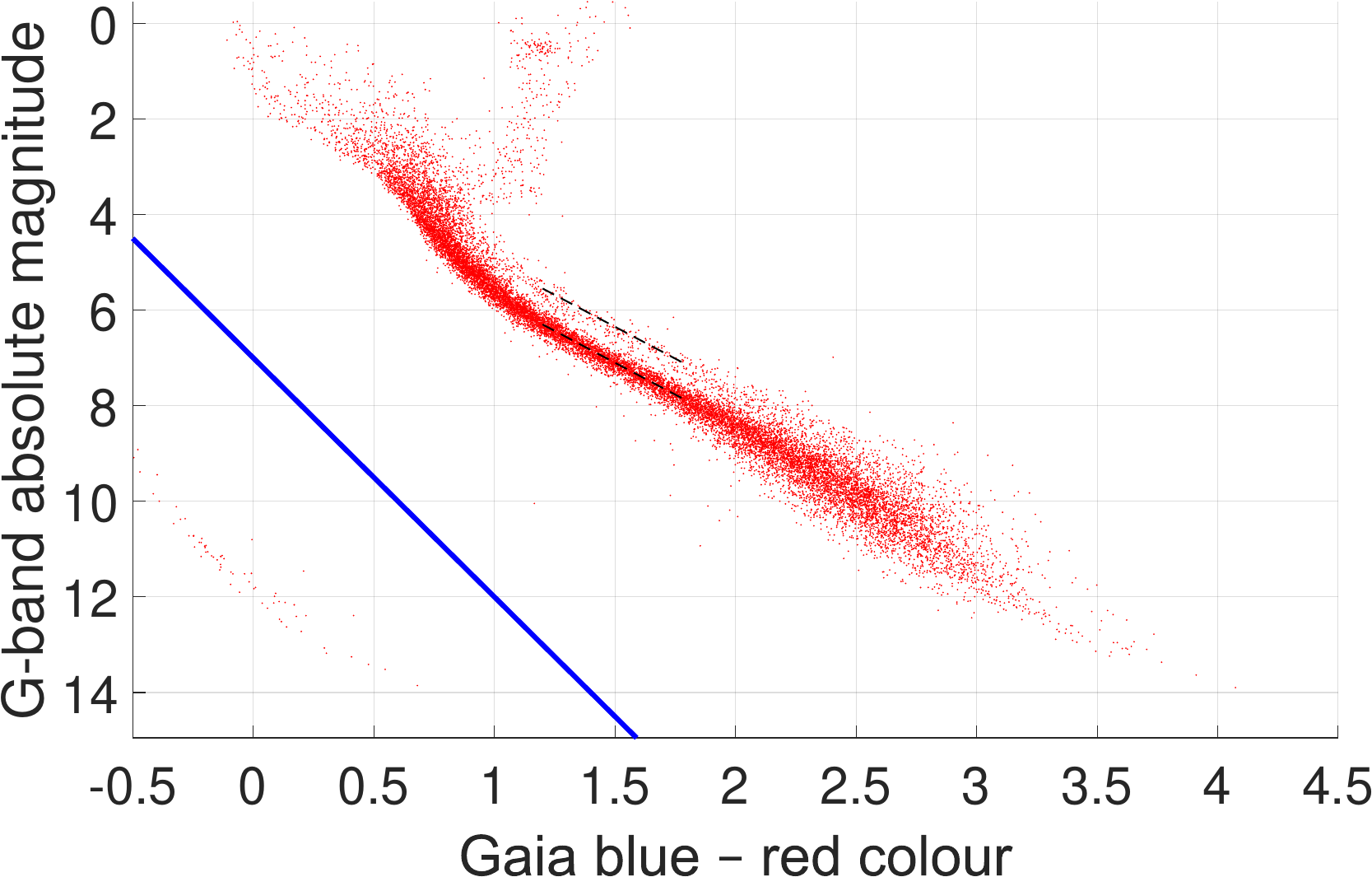}
    \caption{The colour-magnitude diagram of our sample just prior to removing 66 WBs where one star is below the solid blue line, which is designed to remove the white dwarfs evident at the lower left. The two dashed grey lines have an offset corresponding to a factor of two in luminosity. The lower line passes through the main sequence, while the upper line passes through the secondary track evident at higher luminosity. We attribute this double main sequence to unresolved binaries where the stars have an exactly equal mass (Appendix~\ref{Equal_mass_binaries}).}
    \label{Colour_magnitude_diagram}
\end{figure}

Our mass estimates also rely on a mass-luminosity relation designed for main sequence stars \citep{Pecaut_2013}. To check if this is a good assumption, we use Figure~\ref{Colour_magnitude_diagram} to plot the relation between the absolute $G$-band magnitude and the colour, which we define as the difference in apparent magnitude between the \emph{Gaia} blue and red passbands. As expected, the vast majority of the stars in our sample are on the main sequence. However, a small number of stars are very far below the main sequence $-$ these are probably white dwarfs (WDs). We remove these by excluding stars below the solid blue line, which achieves a good separation between the main sequence and the WD track. Only a very small number of WDs are removed by this cut and these are all well separated from the main sequence, so any remaining WDs should have a negligible impact on our results.

An interesting feature of our main WB sample's colour-magnitude diagram is a concentration of stars parallel to but brighter than the main sequence (Figure~\ref{Colour_magnitude_diagram}). This `double main sequence' is caused by unresolved CBs where both stars have an exactly equal mass, as demonstrated by the parallel dashed grey lines with an offset corresponding to a factor of two in luminosity. We do not exclude stars belonging to this double main sequence because our analysis already accounts for it (Appendix~\ref{Equal_mass_binaries}).

A major complication to the WBT is the presence of undetected CB companions (Section~\ref{Introduction}). We follow a `mitigate + simulate' strategy to minimize the uncertainties introduced by CBs. This entails reducing the presence of CB companions and modelling the distribution of those that remain. Since we do not aim to have a sample that is completely free of stars with a CB companion, we benefit from a much larger sample size than the study of \citet{Hernandez_2023}. We discuss below how we attempt to reduce the proportion of contaminated WBs. Details of our CB model will be presented later (Section~\ref{Undetected_companions}).

\subsubsection{Multiple peaks in \emph{Gaia} images}
\label{ipd_cut}

\begin{figure}
    \centering
    \includegraphics[width=0.47\textwidth]{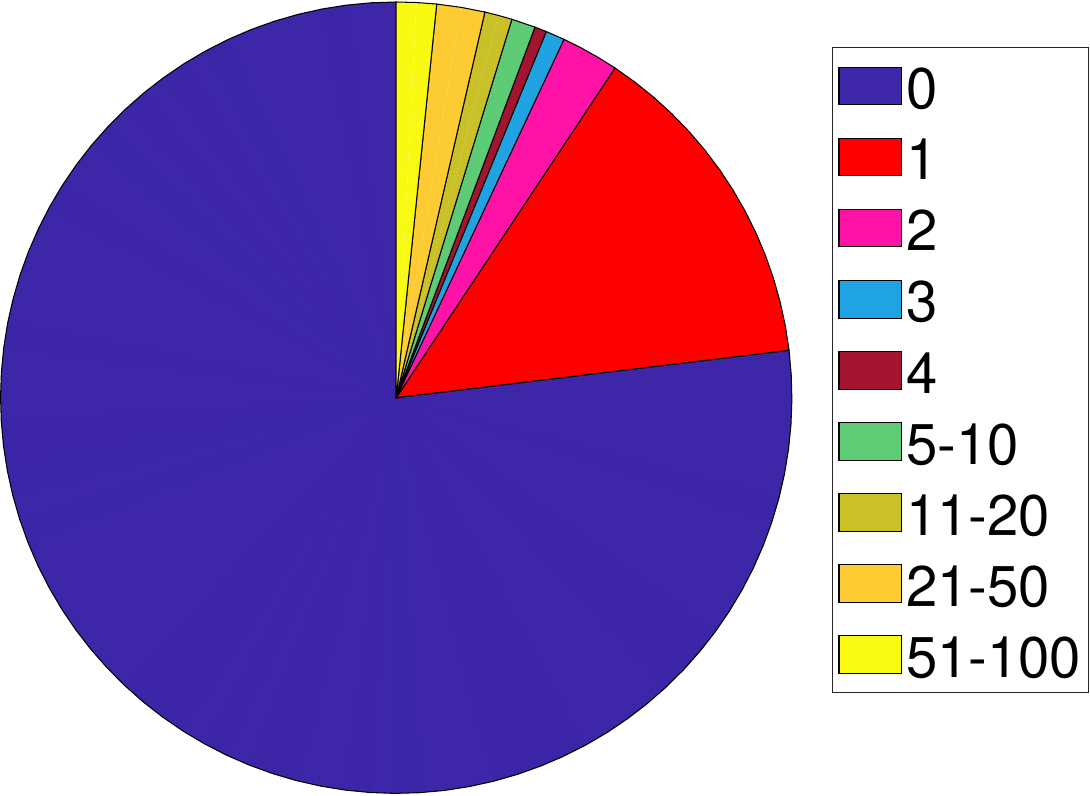}
    \caption{Distribution of the \emph{Gaia}~DR3 parameter $\tt{ipd\_frac\_multi\_peak}$ for the WBs in our sample prior to imposing the quality cut related to this parameter. For each WB, we take the star with the higher value. Although $\tt{ipd\_frac\_multi\_peak}$ can be as high as 100, it is typically 0, as expected for a single source.}
    \label{ipd_pie}
\end{figure}

If there is a clearly detected CB companion to a star in a WB and it is reasonably likely that the third star is bound to the WB, then the system is excluded from our analysis (Section~\ref{Gaia_DR3_sample}). However, CBs which are fainter and/or on a tighter orbit might not be clearly resolved. The contaminated star might then appear resolved into two sources in only some \emph{Gaia} focal plane transits. We exploit this by using the parameter $\tt{ipd\_frac\_multi\_peak}$ in the \emph{Gaia}~DR3 catalogue \citep*[see also section~5 of][]{Badry_2021}. This is the percentage of successful focal plane transits in which a source is detected as multiple peaks in the image. A genuinely isolated star would ideally give $\tt{ipd\_frac\_multi\_peak} = 0$, but since it might spuriously show up as multiple peaks in the \emph{Gaia} images due to issues like cosmic ray hits, we allow $\tt{ipd\_frac\_multi\_peak} \leq 2$ \citep*[a threshold of 2 was also used in][]{Pace_2022}. We reject any star that is detected as multiple peaks more frequently.

Figure~\ref{ipd_pie} shows the distribution of $\tt{ipd\_frac\_multi\_peak}$ for the WBs in our sample prior to imposing this quality cut, with the higher value shown for the two stars in each WB. It is apparent that our quality cut does not reduce the sample size very much. At the same time, it seems unlikely that a star really has a marginally detected companion if it appears to be isolated 98\% of the time. However, if a star frequently appears as multiple peaks in the \emph{Gaia} images, then we might reasonably suspect it to be part of a CB.

\subsubsection{Having a systemic RV}
\label{Systemic_RV_requirement}

Even though we are implementing a 2D version of the WBT using only motions within the sky plane, the systemic RV of each WB does still enter into our analysis, albeit scaled down substantially by the angular separation of the WB (Section~\ref{Systemic_RV}). For our technique to work, we are forced to reject WBs where the RV is unknown for both stars, which loses roughly 1/3 of our sample. We accept systems where the RV is known for only one star because the systemic RV is not needed very precisely and the RVs of the stars in a WB would differ by at most a few km/s, especially for the more widely separated systems where perspective effects are more important. Requiring both stars to have a measured RV reduces the sample size from 8611 to 5504, but the proportion of systems with $\widetilde{v} > 2.5$ (which is a rough proxy for the level of contamination) only drops from 10.2\% to 9.5\%.

\subsubsection{Similar RVs when both are available}
\label{Similar_RV_requirement}

While we do not require both stars in a WB to have a known RV, this is often the case. When available, we use this information to reject WBs whose two stars have RVs that differ from each other by more than triple the quadrature sum of the RV uncertainties or triple the Newtonian $v_c$, whichever is larger.\footnote{Even if we had perfect data, the RVs of the stars in a WB could still differ by order their Newtonian $v_c$.} This reduces the sample size by 3.5\% (we lose 308 WBs out of 8919 without this cut). Since almost 2/3 of the systems in our WB sample have a reported RV for both stars, the fact that only a small proportion of WBs consist of stars with discrepant RVs suggests that our selected WBs are mostly genuine. The use of systemic RVs constrained to km/s precision by \emph{Gaia} spectroscopy substantially reduces the scope for perspective effects to alter the results obtained in this contribution.

\subsubsection{The \texorpdfstring{$\widetilde{v}$}{v\_tilde} uncertainty}
\label{v_tilde_uncertainty}

The above quality cuts are designed to reduce systematic uncertainties in the WB parameters. These also have random uncertainties, which we need to quantify. As discussed in Section~\ref{Monte_Carlo_error_propagation}, we use Monte Carlo error propagation to obtain the uncertainty in the all-important $\widetilde{v}$ using $2^{12}$ trials. We find that the uncertainty is very small for a large proportion of the WBs in our sample \citep[as expected from a simpler analysis of \emph{Gaia}~DR2 results;][]{Banik_2019_line}. We therefore apply a quality cut such that
\begin{eqnarray}
    \text{Error in } \, \widetilde{v} ~\leq~ 0.1 \max \left( 1, \frac{\widetilde{v}}{2} \right) \, .
    \label{Max_vtilde_uncertainty}
\end{eqnarray}
We found that this achieves a good balance between the quality and quantity of data. The uncertainty is allowed to be somewhat larger at higher $\widetilde{v}$ because we use wider bins in $\widetilde{v}$ here when comparing to theory (Section~\ref{Binomial_statistics}). Moreover, such systems are less crucial for the WBT as they are not part of the main peak at $\widetilde{v} \la 2$, implying significant contamination from some source.

\begin{figure}
    \centering
    \includegraphics[width=0.47\textwidth]{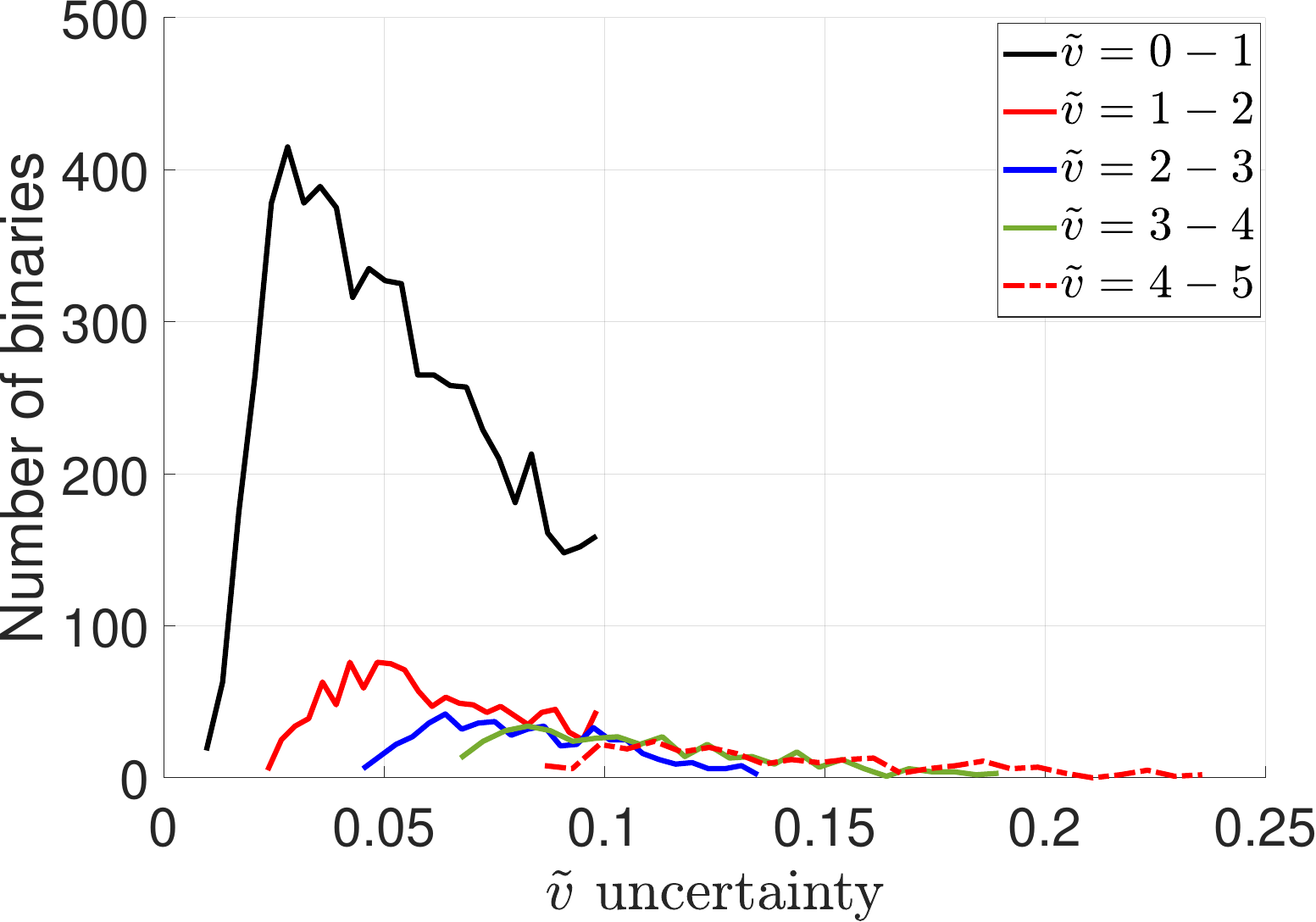}
    \caption{The frequency distribution of the $\widetilde{v}$ uncertainty as estimated from Monte Carlo error propagation of the $5 \times 5$ \emph{Gaia}~DR3 covariance matrix (Section~\ref{Monte_Carlo_error_propagation}). Different curves show results for different ranges of $\widetilde{v}$, as indicated in the legend. We only consider WBs whose $\widetilde{v}$ uncertainty is at most $0.1 \, \max \left( 1, \widetilde{v}/2 \right)$, truncating each error distribution at the right. The truncation at the left is caused by the 0.024~dex mass uncertainty at fixed $M_G$ imposing a minimum fractional uncertainty in $\widetilde{v}$.}
    \label{v_tilde_error_distribution}
\end{figure}

The distribution of $\widetilde{v}$ uncertainties is shown in Figure~\ref{v_tilde_error_distribution} for different $\widetilde{v}$ ranges of our WB sample. As the WBT is mostly concerned with the main peak to the $\widetilde{v}$ distribution at $\widetilde{v} \la 2$, our cut implies a maximum $\widetilde{v}$ uncertainty of only 0.1 in the most crucial parameter range. Since the mode of the main peak is at $\widetilde{v} \approx 0.5$, measurement uncertainties that add an extra 0.1 in quadrature would only boost the root mean square value by about 2\%, which is negligible in comparison to the predicted MOND enhancement of 20\%. Moreover, the typical $\widetilde{v}$ uncertainty for such systems is much smaller than the maximum permitted by our selection. Indeed, \citetalias{Pittordis_2019} found the $\widetilde{v}$ distribution in \emph{Gaia}~DR2 changed little when restricting to systems where the uncertainty is below 0.25 (compare their figures~12 and 13). We therefore neglect $\widetilde{v}$ uncertainties in the rest of our analysis, bearing in mind that our WB sample has already been chosen to have an accurate $\widetilde{v}$.

We need to integrate WB orbits for a range of masses covering the mass range of our sample. This is because larger mass systems have a larger MOND radius (Equation~\ref{MOND_radius}), reducing the impact of MOND at fixed separation \citepalias{Banik_2018_Centauri}. To limit the computational cost, we restrict our analysis to WBs with a total mass of $0.464 - 4.31 \, M_\odot$. This reduces our sample size very slightly. Including the remaining WBs would require a considerable broadening of the mass range covered by our WB orbit library or a substantial worsening of its mass resolution, but there would be very little gain in sample size. It would also leave our results more vulnerable to systematic issues with very low or very high mass stars.

\subsection{The WB sample}
\label{WB_sample}

\begin{figure}
    \centering
    \includegraphics[width=0.47\textwidth]{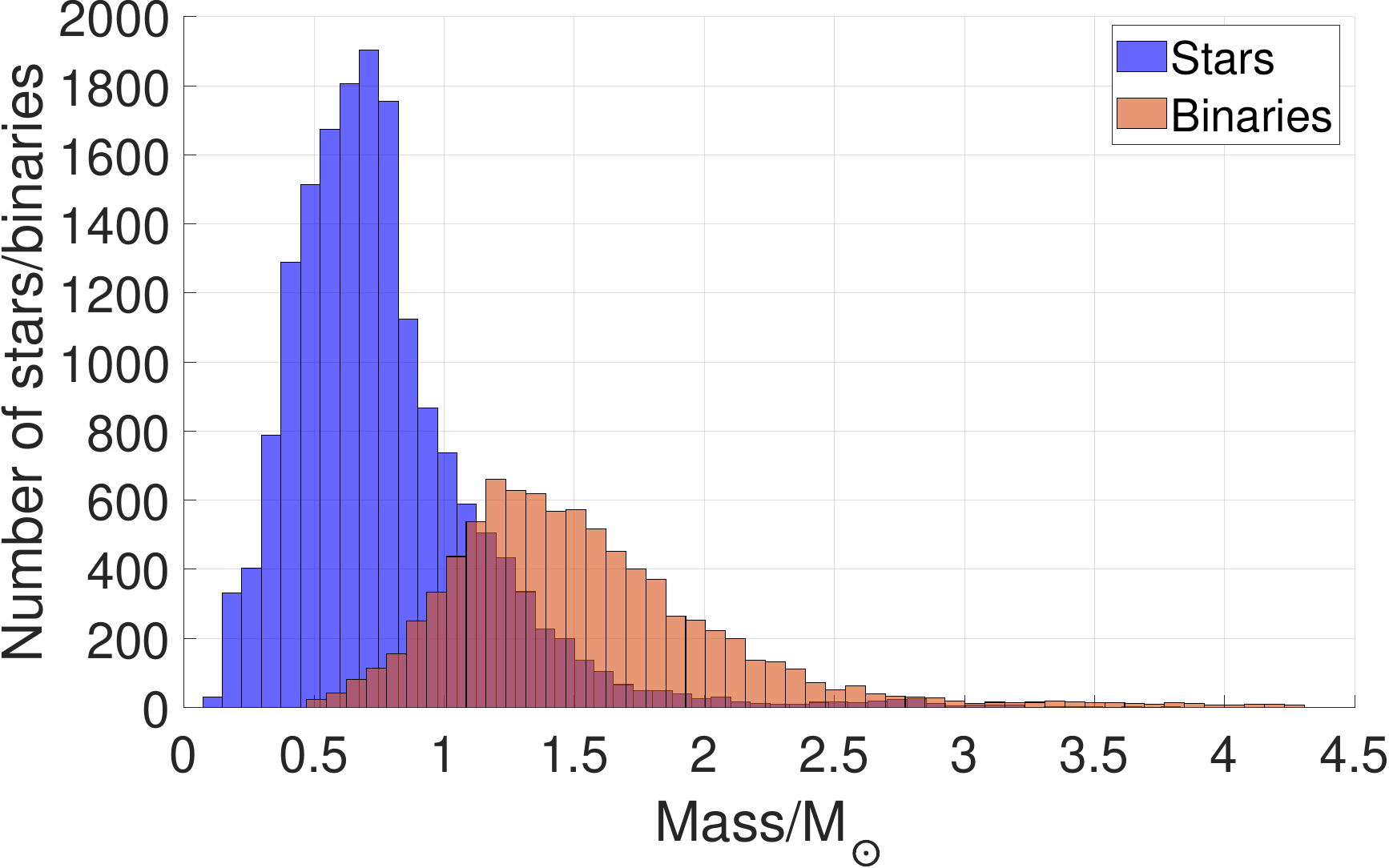}
    \caption{The mass distribution of all the stars in our WB sample (blue bars). The red bars show the distribution of the total mass of each binary. The outer edges of the outermost bins show the range of the data.}
    \label{Mass_distribution}
\end{figure}

After applying the quality cuts described so far, we are left with a final sample of 8611 WBs. Their mass distribution is shown in Figure~\ref{Mass_distribution}, with blue bars showing individual stars and red bars showing binary total masses. The latter are more relevant physically, so we will use $M$ to mean the binary total mass and $M_\star$ to denote the mass of a single star $-$ unless it is clear that the discussion refers to stars considered individually.

\begin{figure}
    \centering
    \includegraphics[width=0.47\textwidth]{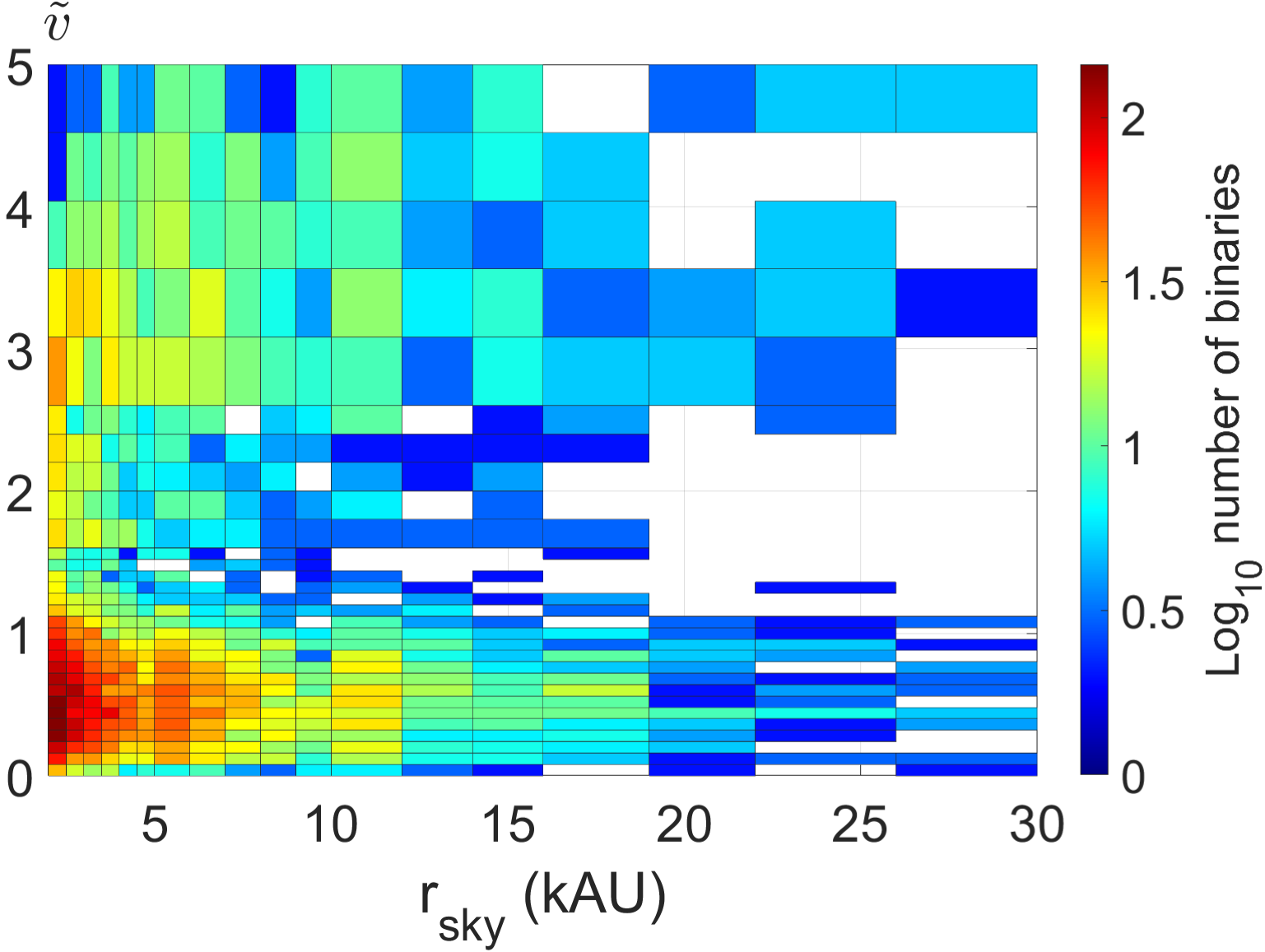}
    \caption{The number of WBs in different pixels of $\left( r_{\rm{sky}}, \widetilde{v} \right)$. Results are shown binned according to our pixellation scheme, which was fixed prior to any detailed analysis (the pixel widths increase gradually towards higher values). The comparison with theory is based on the likelihood of matching the number counts shown here, which we find by multiplying together the binomial likelihood for each pixel (Section~\ref{Binomial_statistics}). Notice the lack of WBs at $\widetilde{v} \approx 2$ for the highest $r_{\rm{sky}}$ (empty pixels are shown in white). We interpret this as due to dilution of the CB contamination over a wide $\widetilde{v}$ range due to the low Newtonian $v_c$ of the WB. There is also little LOS contamination here: it only becomes important when $r_{\rm{sky}}$ and $\widetilde{v}$ are both large (Section~\ref{Chance_alignments}).}
    \label{Observed_photo_tiled}
\end{figure}

As with previous detailed investigations into the WBT \citepalias{Pittordis_2019, Pittordis_2023}, we perform it by decomposing the data into pixels in the space of $r_{\rm{sky}}$ and $\widetilde{v}$ (Figure~\ref{Observed_photo_tiled}). It is clear that there are many more systems at low $r_{\rm{sky}}$ than at high $r_{\rm{sky}}$, in line with earlier results \citep[e.g.][]{Andrews_2017}. The main peak to the $\widetilde{v}$ distribution is quite prominent, with the mode located at $\widetilde{v} \approx 0.5$. There is also an extended tail going out to much larger $\widetilde{v}$, which was previously reported \citepalias{Pittordis_2019} and attributed to undetected CBs \citep{Belokurov_2020, Clarke_2020}. Interestingly, a gap appears in the distribution at $\widetilde{v} \approx 2$, but only at high $r_{\rm{sky}}$. We attribute this to the minimum in the $\widetilde{v}$ distribution at $\widetilde{v} \approx 1.5$, which is evident also at lower $r_{\rm{sky}}$. This minimum is caused by WBs having a sharp peak to their $\widetilde{v}$ distribution at $\widetilde{v} \approx 0.5$ followed by a rapid decline as a consequence of orbital mechanics and projection effects \citepalias[e.g.][]{Banik_2018_Centauri, Pittordis_2019}. There is an extended tail due to CB contamination, but this too would generally lead to a declining $\widetilde{v}$ distribution because a slower CB orbital velocity corresponds to a wider CB separation, which is less likely \citep[a more detailed explanation is given in][]{Banik_2021_plan}. However, we expect an increasing contribution from chance alignments at larger $\widetilde{v}$ because their distribution is expected to be linear in $\widetilde{v}$ (see Equation~\ref{LOS_contamination_pattern}). The minimum in the $\widetilde{v}$ distribution is not very apparent at $r_{\rm{sky}} \la 5$~kAU, where WBs are very common and chance alignments are relatively unimportant. The minimum becomes readily apparent at larger $r_{\rm{sky}}$ where WBs are quite rare, increasing the relative contribution of chance alignments because they should have a flat distribution with respect to $r_{\rm{sky}}$ (Equation~\ref{LOS_contamination_pattern}). The fact that WBs are much less common at high $r_{\rm{sky}}$ naturally leads to empty pixels around the minimum of the $\widetilde{v}$ distribution, with the $\widetilde{v}$ range covered by these empty pixels expected to gradually widen with increasing $r_{\rm{sky}}$.

It is clear that we need a detailed model including CB and LOS contamination if we are to infer the law of gravity at low accelerations and search for the predicted MOND signal. This is the subject of the next section.

\section{Modelling the WB dataset}
\label{Model}

The $\left( r_{\rm{sky}}, \widetilde{v} \right)$ distribution of our WB sample is a consequence of several simultaneously occurring physical processes and projection effects, which we try to model in this section to best reproduce Figure~\ref{Observed_photo_tiled}. In addition to WB orbital motion (Section~\ref{WB_population}), we also consider how an undetected CB population affects the observables (Section~\ref{Undetected_companions}) and take into account the possibility of some LOS contamination (Section~\ref{Chance_alignments}). This might be field stars, but it could also be from stars that formed in the same star cluster. Given the results of \citetalias{Pittordis_2019} and the excellent quality of \emph{Gaia}~DR3, we expect LOS contamination to only slightly affect our WB sample and to become important only at large separation and $\widetilde{v}$.

LOS contamination can be included in our model fairly easily. Moreover, detailed WB orbit modelling has previously been conducted in Newtonian and Milgromian gravity in a manner that can readily be repurposed for the WBT \citepalias{Banik_2018_Centauri}. As a result, the main complication that we need to handle is the possibility of an undetected CB companion around 0, 1, or 2 of the stars in a WB. Since recoil motion and additional mass created by an undetected CB physically affect the WB rather than merely contribute an extra population to the statistics, including CB contamination greatly increases the complexity and computational cost of the WBT. We were eventually able to devise a detailed plan that captures the essential physics in a computationally feasible way \citep{Banik_2021_plan}. We refer the reader to that work for a more detailed look at the computational techniques needed to implement the WBT in a reasonable timeframe. This allows us to prepare the first detailed model of the WB population whose parameter space is thoroughly explored using standard statistical techniques.

\subsection{The WB population}
\label{WB_population}

We set up a library of WB orbits using the prior work of \citetalias{Banik_2018_Centauri} with minimal adjustments, which we briefly describe. We need to run calculations over a wider mass range in order to cover the range of WB masses in our sample (red bars in Figure~\ref{Mass_distribution}). We also found that the computational cost of including CB contamination can be greatly reduced if instead of recording results with respect to the projected separation $r_{\rm{sky}}$, we use
\begin{eqnarray}
    \widetilde{r}_{\rm{sky}} ~\equiv~ \frac{r_{\rm{sky}}}{a} \, ,
    \label{rt_sky_definition}
\end{eqnarray}
where $a$ is the semi-major axis of the WB (this is motivated in Section~\ref{aint_distribution}). An incidental benefit of this approach is that at low $a$, recording probabilities into different bins in $\widetilde{r}_{\rm{sky}}$ rather than $r_{\rm{sky}}$ improves the resolution in $r_{\rm{sky}}$. Since we record results separately for different $a$ and only marginalize over $a$ at a much later stage, we can recover $r_{\rm{sky}} \equiv a\widetilde{r}_{\rm{sky}}$ as and when needed, which is necessary for a comparison with observations because $a$ is generally not known.

In MOND, the Galactic EFE plays a crucial role \citepalias{Banik_2018_Centauri}. This can be estimated as $g_e = v_{c, \odot}^2/R_0$, where $R_0$ is the Galactocentric distance of the Sun and $v_{c, \odot}$ is the Galactic rotation curve amplitude at the Solar circle (sometimes called the Local Standard of Rest or LSR). As in section~3.6 of \citetalias{Banik_2018_Centauri}, we assume that $v_{c, \odot} = 232.8$~km/s and $R_0 = 8.2$~kpc \citep{McMillan_2017}, which implies that $g_e = 1.785 \, a_{_0}$. Subsequent studies show that $R_0$ must be very close to 8.2~kpc based on combining astrometry and spectroscopy of the star S2 near the Galactic centre black hole \citep{Gravity_2019}. Since its proper motion is precisely known \citep*{Reid_2004, Gordon_2023}, the Galactocentric tangential velocity of the Sun now has very little uncertainty, constraining the LSR velocity to very close to 233~km/s \citep{McGaugh_2018, Zhou_2023} if the non-circular motion of the Sun is taken to be precisely known from \citet*{Schonrich_2010}. The latter is somewhat more uncertain than the other parameters entering into the LSR velocity \citep{Schonrich_2012, Francis_2014, Bovy_2015}, but it is still clear that $v_{c, \odot}$ is very unlikely to differ from our assumed value by more than 10~km/s, which implies that uncertainty in $v_{c, \odot}$ has little effect on the predicted MOND signal in local WBs (Section~\ref{MOND_interpolating_function}). Moreover, $g_e$ has recently been directly determined by the measurement of the Solar System's acceleration relative to distant quasars \citep{Klioner_2021}. Their work discovered the changing aberration angle of distant quasars due to the changing velocity of the Solar System barycentre. Its directly measured acceleration points directly towards the Galactic centre within a small uncertainty (see their figure~10) and has a magnitude of $g_e = 1.94 \pm 0.13 \, a_{_0}$. Since gravitational fields from objects beyond the Galaxy are thought to be negligible in comparison to the Galactic gravitational field, this provides a completely independent and very direct confirmation of the kinematic estimate of $\bm{g}_e$ used in \citetalias{Banik_2018_Centauri} and in this contribution.

\subsubsection{Orbit modelling}
\label{WB_orbit_modelling}

WBs are bound together by their mutual gravity. This is easy to calculate in the Newtonian case but is somewhat more complicated in MOND, as discussed next. Our calculation of the Milgromian gravitational field and integration of WB orbits in this field use the same approach as \citetalias{Banik_2018_Centauri}, to which we refer the reader for further details.

In an isolated spherically symmetric mass distribution, the asymptotic MOND behaviour is given by Equation~\ref{g_cases}. To interpolate between the Newtonian and deep-MOND regimes, we have to use an interpolating function $\nu$ with argument $g_{_N}$ such that $\bm{g} = \nu \bm{g}_{_N}$. In the quasilinear formulation of MOND \citep[QUMOND;][]{QUMOND}, this can be generalized to a more complicated geometry by taking the divergence of both sides.
\begin{eqnarray}
    \nabla \cdot \bm{g} ~=~ \nabla \cdot \left( \nu \bm{g}_{_N} \right).
    \label{QUMOND_governing_equation}
\end{eqnarray}
In this contribution, we adopt the simple interpolating function \citep{Famaey_2005} for reasons discussed in section~7.1 of \citetalias{Banik_2018_Centauri}.
\begin{eqnarray}
    \nu ~=~ \frac{1}{2} + \sqrt{\frac{1}{4} + \frac{a_{_0}}{g_{_N}}} \, .
    \label{Simple_interpolating_function}
\end{eqnarray}
The WBT depends significantly on the adopted interpolating function because local WBs have an intermediate total acceleration. Using a very sharp transition can cause WBs to be completely Newtonian even if MOND is correct. However, this would not be consistent with the RAR (we discuss this further in Section~\ref{MOND_interpolating_function}).

Since local WBs are significantly affected by the Galactic EFE, we need to include the Galactic $\bm{g}_e$. This is assumed to point directly towards the Galactic centre and to have a magnitude of $1.785 \, a_{_0}$ based on the Galactic rotation curve, as discussed above. In QUMOND, we need the Newtonian-equivalent external field $\bm{g}_{_{N,e}}$, which is what the Galactic gravity on the Solar neighbourhood would have been in Newtonian gravity without DM. As discussed in section~9.3.1 of \citetalias{Banik_2018_Centauri}, we can use the spherically symmetric relation between $\bm{g}_e$ and $\bm{g}_{_{N,e}}$ because the Sun is many disc scale lengths from the Galactic centre, reducing the importance of the Galaxy's disc geometry.
\begin{eqnarray}
    \bm{g}_e \, \nu \left( g_e \right) ~=~ \bm{g}_{_{N,e}} \, .
    \label{gNe}
\end{eqnarray}
Solving this implicitly gives $g_{_{N, e}} = 1.144 \, a_{_0}$ in the Solar neighbourhood. The Newtonian gravity in a WB is then found by adding $\bm{g}_{_{N,e}}$ to the contribution from the WB itself, which is called $\bm{g}_{_{N,i}}$ in some works as it is internal to the WB.

To limit the computational cost, we treat each WB as a test particle orbiting a point mass containing the binary total mass $M$. This approach is valid in Newtonian gravity, but it was shown in section~7.3 of \citetalias{Banik_2018_Centauri} that it remains a very good approximation in MOND. This is because the fairly strong Galactic EFE causes local WBs to lack an extended regime in which they are isolated and at low accelerations. For all intents and purposes, WBs are either in the Newtonian regime or they are dominated by the EFE. In either case, MOND becomes linear in the mass distribution, allowing the potentials of the two stars to be superposed.

With this approximation, the problem becomes axisymmetric about $\bm{g}_e$. We then solve Equation~\ref{QUMOND_governing_equation} using the `ring library' protocol discussed in \citetalias{Banik_2018_Centauri}. This involves finding the source term $\nabla \cdot \left( \nu \bm{g}_{_N} \right)$ at every point in a 2D grid and then using direct summation to find $\bm{g}$. The calculations are accelerated at off-axis locations by scaling the previously calculated Newtonian gravity of a unit mass and unit radius ring, avoiding the need to sum contributions to $\bm{g}$ from different positions around each ring. An analytic allowance is made for $\nabla \cdot \left( \nu \bm{g}_{_N} \right)$ being non-zero outside the rather large region used in the integration (the extra contribution to the potential is given in equation~9 of \citetalias{Banik_2018_Centauri}; see their appendix~A for a derivation).

\begin{figure}
    \centering
    \includegraphics[width=0.47\textwidth]{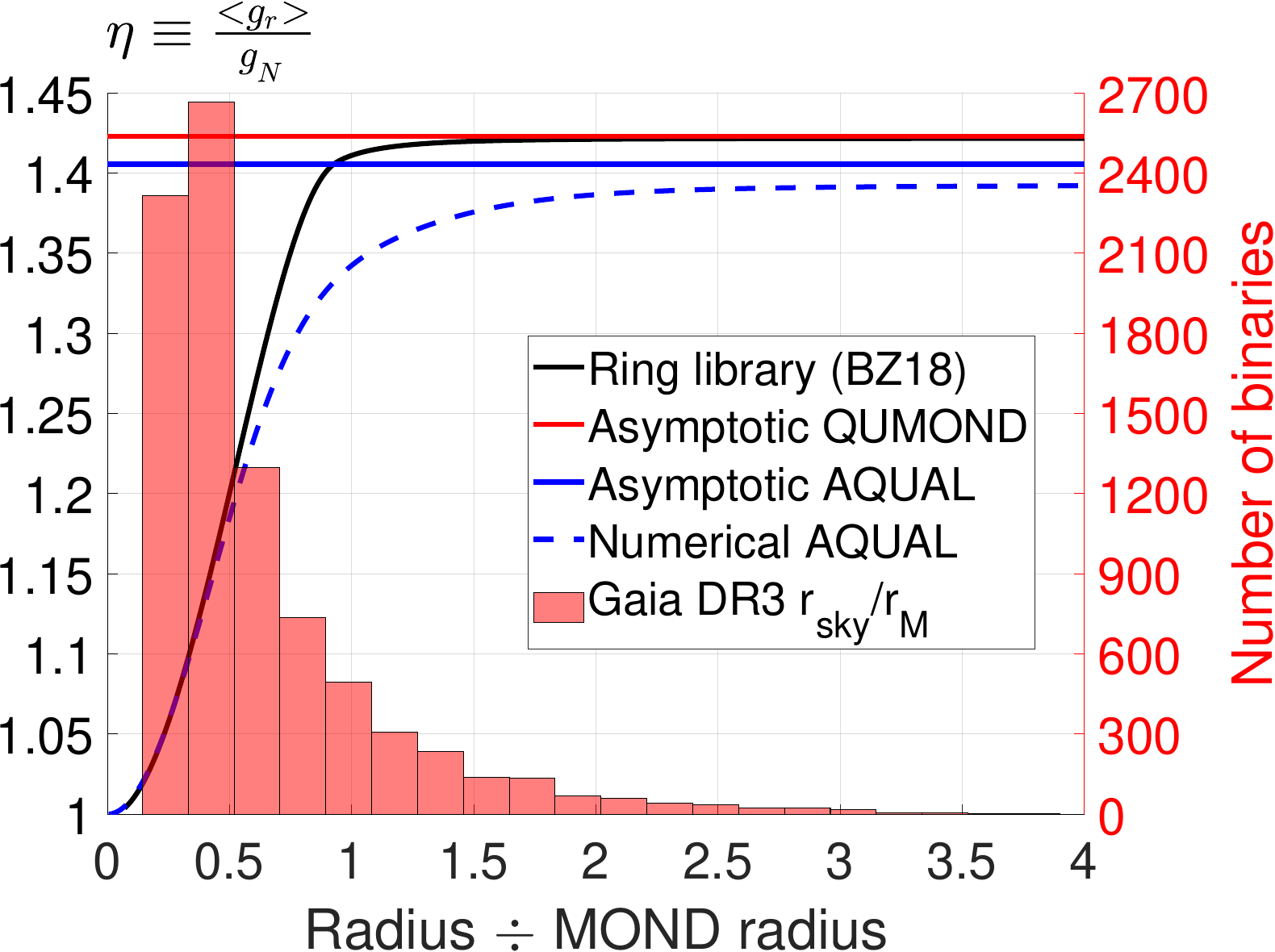}
    \caption{The azimuthally averaged boost factor $\eta$ to the radially inward gravity compared to the Newtonian expectation for WBs with different separation relative to their MOND radius, assuming the simple interpolating function (Equation~\ref{Simple_interpolating_function}). The solid red (blue) horizontal line shows the asymptotic analytic expectation in QUMOND (AQUAL), which we give in Equation~\ref{eta_QUMOND} (\ref{eta_AQUAL}). The numerical QUMOND result central to our study \citepalias[solid black line;][]{Banik_2018_Centauri} has the expected asymptotic behaviour. The dashed blue line shows equation~15 of \citet{Chae_Milgrom_2022}, which has the wrong asymptotic limit (mismatch between blue lines towards the right) and is based on numerical calculations that only deal with a weak external field rather than a dominant one (see the text). The red histogram shows the distribution of $r_{\rm{sky}}/r_{_M}$ for the WBs in our sample. The outer edges of the most extreme bins show the full range of the dataset.}
    \label{Radial_gravity_boost}
\end{figure}

The Milgromian potential of a point mass is not spherically symmetric in the presence of the EFE \citep{Milgrom_1986, QUMOND, Banik_2018_EFE}. This has some interesting consequences \citep{Candlish_2018, Thomas_2018, Banik_2019_spacecraft, Banik_2020_M33, Kroupa_2022}. Since the WB separation $\bm{r}$ is expected to sample a range of directions relative to the EFE, the WBT is mostly sensitive to the angle-averaged inward radial gravity, which we denote $\langle g_r \rangle$. Since the use of $\widetilde{v}$ rather than actual velocities effectively divides out the Newtonian prediction $g_{_N}$, the WBT is really about the parameter
\begin{eqnarray}
    \eta ~\equiv~ \frac{\langle g_r \rangle}{g_{_N}} \, ,
    \label{eta_definition}
\end{eqnarray}
whose square root is probed by velocity measurements. We use Figure~\ref{Radial_gravity_boost} to show $\eta$ as a function of $r/r_{_M}$ (solid black line).\footnote{\citet{Zonoozi_2021} provide an analytic fit to these results for arbitrary $g_e$ with the correct asymptotic behaviour (see their equations~23 and 24).} There is no boost to the Newtonian expectation when $r \ll r_{_M}$, but this changes rapidly until $r \approx r_{_M}$, beyond which the Milgromian enhancement to $\langle g_r \rangle$ saturates at just over 40\% due to the EFE. Without it, the boost factor would continue growing linearly with $r$. In the asymptotic or quasi-Newtonian regime \citep[discussed further in section~3.3 of][]{Asencio_2022} where the enhancement factor has reached its maximum, the WB gravity can be treated as a small perturbation about the Galactic gravity, making the potential analytic \citep[equation 30 of][]{Banik_2020_M33}. The analytic expectation is shown using a horizontal red line, revealing an excellent match to the numerical result when $r > r_{_M}$.

In this contribution, we focus on the QUMOND formulation because it is more computer-friendly than the earlier aquadratic Lagrangian formulation \citep[AQUAL;][]{Bekenstein_Milgrom_1984}. The asymptotic AQUAL result is shown in Figure~\ref{Radial_gravity_boost} using a horizontal blue line \citepalias[equation~35 of][]{Banik_2018_Centauri}. As discussed there, this result is very similar to the QUMOND result for the same interpolating function, i.e. the same relation between $\bm{g}$ and $\bm{g}_{_N}$ in spherical symmetry. Numerical AQUAL results are not available in the transition zone where the EFE is neither negligible nor dominant, but we can get a general idea using equation~15 of \citet{Chae_Milgrom_2022}, which is a fit to numerical results for a weak EFE as relevant to galaxy rotation curves. We show this using a dashed blue line. It is clear that this equation has the wrong asymptotic limit, as can also be seen from the equation: when the EFE dominates, it takes the argument $1.1 \, g_{_{N,e}}$ rather than $g_{_{N,e}}$. Even if this issue is fixed, the functional form is not compatible with the EFE-dominated point mass potential in AQUAL \citep[equation~66 of][]{QUMOND} or the resulting radial gravity \citepalias[equation~32 of][]{Banik_2018_Centauri}. Despite these issues when the EFE is strong, equation~15 of \citet{Chae_Milgrom_2022} should still give a reasonable idea of the AQUAL enhancement to $\langle g_r \rangle$ when $r \ll r_{_M}$ because it is a fit to numerical results in this regime. We see that when $r \la 0.5 \, r_{_M}$, the AQUAL and QUMOND results are almost identical. Therefore, results for the WBT are not sensitive to which formulation of MOND we use. We discuss this further in Section~\ref{MOND_interpolating_function}.

Another aspect of Figure~\ref{Radial_gravity_boost} is the distribution of $r_{\rm{sky}}/r_{_M}$, which we show using red bars (the left edge of the first bar shows the minimum value). Our sample covers down to $r_{\rm{sky}}/r_{_M} = 0.14$, so we consider it to be reasonably sensitive down to $r/r_{_M} = 0.2$, leaving some allowance for 3D separations exceeding $r_{\rm{sky}}$. A WB separated by only $r_{_M}/5$ has an acceleration of $g_{_N} = 25 \, a_{_0}$ and can thus be considered isolated. In this case, $\nu = 1/2 + \sqrt{1/4 + 1/25} = 1.04$, which is very small compared to the predicted enhancement to $g_{_N}$ for more widely separated WBs. This is even truer for interpolating functions that have a somewhat faster return to Newtonian behaviour at high accelerations in order to better match Solar System ephemerides \citep{Hees_2014, Hees_2016}. At the other end, our sample probes well into the asymptotic regime where we would expect almost the maximum possible enhancement to $g_{_N}$ given the Galactic EFE. Our sample thus covers a wide enough range of internal WB accelerations to allow a sensitive search for the predicted MOND effect. While our WBs experience a narrower range of total gravitational field strengths because the Galactic EFE essentially imposes a lower limit, we argue later that this is not a major limitation given rotation curve constraints on the MOND interpolating function in the transition zone (Section~\ref{MOND_interpolating_function}).

We use the assumed gravity law to integrate WB orbits with a range of mass $M$, semi-major axis $a$, and orbital eccentricity $e$, defining $a$ and $e$ for a generalized gravity theory as in section~2.3.1 of \citetalias{Banik_2018_Centauri}. Following their approach, each orbit is started at a separation of $a$ and integrated for 20 revolutions. This should be more than sufficient in the Newtonian case, but it is also adequate in MOND because the resulting statistical distribution is similar to that yielded by a much longer 5~Gyr integration (see their figure~15). We also consider a range of inclinations $i$ between the orbital pole and the Galactic centre direction defining the EFE. Results for different inclinations are marginalized assuming a prior of $\sin i$, which is appropriate for an isotropic distribution of orbital poles. Orbits are truncated if the stars get to within 50~AU or become more distant than 100~kAU as we consider such WBs to have crashed together or to have likely become unbound at some point by perturbations from passing stars, respectively \citepalias[section~2.3.2 of][]{Banik_2018_Centauri}. At each timestep, we consider a dense 2D grid of viewing directions and compute the $\left( \widetilde{r}_{\rm{sky}}, \widetilde{v} \right)$ inferred by a distant observer along the chosen LOS, with results statistically weighted by the solid angle covered by each LOS. We also weight the results at different timesteps by the duration of the timestep, which varies because we use an adaptive timestep to reduce the computational cost. In this way, we build up a simulated $\left( \widetilde{r}_{\rm{sky}}, \widetilde{v} \right)$ distribution for different $\left( M, a, e \right)$.

In the Newtonian case, it is not necessary to consider different $M$ or $i$, but we use a $5\times$ higher resolution in $e$ because the $\widetilde{v}$ distribution at fixed $e$ has sharp peaks corresponding to pericentre and apocentre and $e$ does not change in a Newtonian orbit integration \citep[unlike in MOND; see][]{Pauco_2016}. To get a smooth distribution when marginalizing over different $e$, \citetalias{Banik_2018_Centauri} found that a rather high resolution is required in $e$ for the Newtonian case. This is not needed in MOND because the peaks are blurred out by changes in the shape of the orbit over time: typically about 8 orbits are needed to go from nearly circular to very eccentric back to nearly circular (see their figure~20).

\subsubsection{Orbital parameter distribution}
\label{WB_orbital_parameter_distribution}

The mass distribution of WBs is obtained directly from that of our WB sample (Figure~\ref{Mass_distribution}). We count up how many WBs are in the mass range covered by each value of $M$ at which we perform an orbit integration. This directly determines the relative statistical weights of different $M$ values in our WB orbit library.

Following \citet{Andrews_2017}, we assume that the semi-major axes of the WBs follow a broken power law in $a$. The logarithmic slope at low $a$ is fixed to $-1.6$, which is known to be valid down to at least 0.5~kAU \citep{Badry_2018}. While this can in principle extend up to arbitrarily large $a$, we assume that it breaks down at $a = a_{\rm{break}}$, beyond which the logarithmic slope becomes $\beta < -1$ to ensure a convergent integral. Our assumed distribution of $a$ is therefore given by
\begin{eqnarray}
    P \left( a \right) ~\propto~ \begin{cases}
    a^{-1.6} \, , & \textrm{if} ~a < a_{\rm{break}} \, , \\
    a^{\beta} \, , & \textrm{if} ~a > a_{\rm{break}} \, ,
    \end{cases}
    \label{a_cases}
\end{eqnarray}
where $a_{\rm{break}}$ and $\beta$ are free parameters that our analysis infers from the data. Note that since we are modelling the distribution of our sample of WBs rather than the full WB population, the inferred values of these parameters may be somewhat biased. In particular, the requirement to have accurate $\widetilde{v}$ measurements is quite difficult to satisfy at large separations because the Newtonian $v_c$ is low. This is likely to reduce $\beta$.

Following \citet{Hwang_2022}, we parametrize the distribution of orbital eccentricities using a power law with index $\gamma$.
\begin{eqnarray}
    P \left( e \right) ~=~ \left( \gamma + 1 \right) e^\gamma \, .
    \label{gamma_definition}
\end{eqnarray}
$\gamma$ is left as a free parameter that is allowed to vary over the range $0-4$ in all our analyses. We expect a nearly thermal distribution of eccentricities \citep[$\gamma = 1$;][]{Jeans_1919, Ambartsumian_1937} $-$ this is discussed in more detail in section~4.2 of \citet{Kroupa_2008}. We use a flat prior on $\gamma$ for our nominal analysis. However, \citet{Hwang_2022} found that a slightly superthermal distribution ($\gamma > 1$) better fits the observed distribution of the angle $\psi$ between $\bm{r}_{\rm{sky}}$ and $\bm{v}_{\rm{sky}}$ \citepalias[this is related to the WB orthogonality test outlined in section~8.2.1 of][]{Banik_2018_Centauri}. We therefore run a variation of our nominal analysis in which we assume a Gaussian prior of $\gamma = 1.32 \pm 0.09$, which should shrink the uncertainties somewhat. We note that since CB contamination would tend to flatten out the distribution of $\psi$ and thereby drive the inferred eccentricity distribution towards thermal, it is quite likely that $\gamma$ is slightly higher than inferred by \citet{Hwang_2022} given that those authors do not consider undetected CBs. On the other hand, their figure~10 shows that $\gamma$ can be overestimated slightly at large separations for an underlying slightly superthermal distribution.

\subsubsection{Interpolating the gravity law}
\label{Gravity_law_interpolation}

An important aspect of our analysis is that we consider an arbitrary gravity law on a sliding scale between Newtonian and Milgromian. For this purpose, we introduce the gravity law parameter $\alpha_{\rm{grav}}$, which is the most important parameter for the WBT. We define this such that $\alpha_{\rm{grav}} = 0$ represents Newtonian gravity and $\alpha_{\rm{grav}} = 1$ represents QUMOND with the simple interpolating function. We allow $\alpha_{\rm{grav}}$ values somewhat outside this range to capture the possibility of behaviour different to either theory. The most plausible outcomes to the WBT are a strong preference for $\alpha_{\rm{grav}}$ values very close to 0 or 1, with the large sample size hopefully tightening the error budget enough that both possibilities are not simultaneously consistent with the observations \citepalias[a few thousand WBs should be sufficient; see][]{Banik_2018_Centauri}. Since the WBT will have some uncertainties, we allow $\alpha_{\rm{grav}}$ to lie in the range $\left( -2, 3.6 \right)$. Negative values imply that the gravity in the asymptotic regime has a lower than Newtonian normalization, while $\alpha_{\rm{grav}} > 1$ implies a normalization higher than expected in MOND.

To set up a parametrized gravity law, we obtain the simulated $\left( \widetilde{r}_{\rm{sky}}, \widetilde{v} \right)$ distribution in both gravity theories for each $\left( M, a, \gamma \right)$, with different orbital eccentricities marginalized over for the adopted $\gamma$ using the eccentricity distribution it defines (Equation~\ref{gamma_definition}). The main idea of our parametrization is to transform the Newtonian $\left( \widetilde{r}_{\rm{sky}}, \widetilde{v} \right)$ distribution into the Milgromian one and then interpolate the coefficients involved in the transformation so that arbitrary gravity laws can be considered.\footnote{We marginalize over $M$ and $a$ at a later stage.} To simplify our analysis, we consider a simple 2D stretch in which the probability distribution is stretched along the $\widetilde{r}_{\rm{sky}}$ axis by a factor $S_r$ and then along the $\widetilde{v}$ axis by a factor $S_v$, with $S_r \neq S_v$ in general. These are found by running a 2D gradient descent in $\left( S_r, S_v \right)$ to minimize the sum of squared residuals between the stretched Newtonian array and the MOND array \citep{Fletcher_1963}. While a simple 2D stretch applied to the former does provide a good approximation to the latter, there are still some residuals, which we calculate for each pixel. This array of residuals is itself scaled by a factor of $\alpha_{\rm{grav}}$ and stretched in 2D by some factors depending on $\alpha_{\rm{grav}}$ and $\left( S_r, S_v \right)$. The idea behind applying this `corrections array' is to recover the Newtonian (Milgromian) $\left( \widetilde{r}_{\rm{sky}}, \widetilde{v} \right)$ distribution exactly if $\alpha_{\rm{grav}} = 0$ (1). If we consider an intermediate case where $\alpha_{\rm{grav}} = 0.5$, we would expect the $\widetilde{v}$ distribution to stop at a value about halfway between the maximum which arises in the Newtonian and MOND distributions. To prevent the corrections array from creating non-zero contributions all the way out to the maximum $\widetilde{v}$ which arises in MOND, the residuals array itself needs to be stretched somewhat, though in this case we expect $S_v <1$. Since no corrections are required if $\alpha_{\rm{grav}} = 0$, we need to avoid applying any corrections in this case, which is achieved by our scaling factor of $\alpha_{\rm{grav}}$. We use linear interpolation in all the stretch and scaling factors discussed above with respect to $\alpha_{\rm{grav}}$. To avoid negative probabilities due to negative values in the corrections array, any negative entries in the resulting $\left( \widetilde{r}_{\rm{sky}}, \widetilde{v} \right)$ distribution are converted to zero and the array renormalized.

\subsection{Undetected companions}
\label{Undetected_companions}

As discussed in Section~\ref{WB_importance}, undetected companions are the main source of contamination to the WBT given the low measurement uncertainties (Figure~\ref{v_tilde_error_distribution}) and the low LOS contamination fraction evident from the low number counts in the rather wide pixels at high $r_{\rm{sky}}$ and $\widetilde{v}$ (Figure~\ref{Observed_photo_tiled}). In particular, the declining $\widetilde{v}$ distribution it reveals is a clear sign of some process that broadens the WB distribution, which we otherwise expect to be concentrated at $\widetilde{v} \la 2$ for any plausible gravity theory \citepalias{Pittordis_2019}.

We therefore need to include the possibility that a star in a WB is actually itself part of a CB, even though the CB companion is undetected. Including CBs proved to be by far the most challenging aspect of the WBT, so we had to make some simplifying assumptions to keep the complexity and computational cost manageable. Due to their much smaller separation, we assume that the CB is completely Newtonian internally, even if MOND affects the motion of its barycentre around the other star in the WB.\footnote{\citet{Bekenstein_Milgrom_1984} showed that a body with high internal accelerations can still exhibit MONDian behaviour for its barycentre.} It will become clear later that the undetected CBs relevant to the WBT indeed have only a small separation compared to the WB, justifying this approximation.

\subsubsection{Impact on WB observables}
\label{CB_impact}

Undetected companions have two main effects on a WB: increased mass and induced recoil velocity. Since the CB companion is expected to tightly orbit the star it contaminates, we neglect the impact of the CB on the WB's $r_{\rm{sky}}$. For this reason, we also assume that the light emitted by the `undetected' star is blended with that of the contaminated star, i.e. they are unresolved by \emph{Gaia}. This is reasonable given that resolved third star companions would cause the WB to be removed from our sample at an early stage (Section~\ref{Gaia_DR3_sample}). Indeed, we will see later that our analysis prefers to avoid CBs whose separation is significant compared to that of the WB.

\begin{figure}
    \centering
    \includegraphics[width=0.47\textwidth]{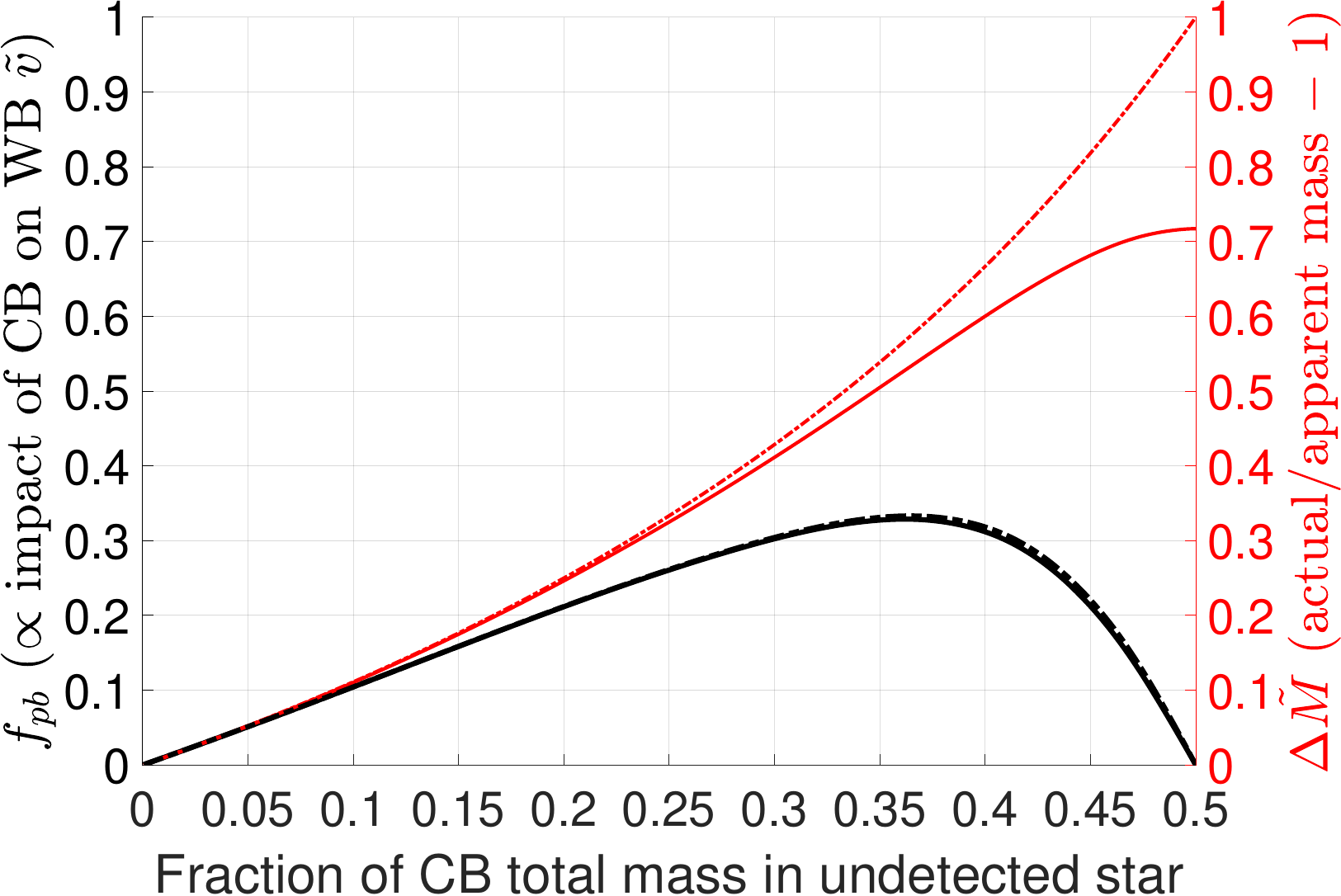}
    \caption{How the fraction of a CB's total mass in its undetected component affects the offset between its photocentre and barycentre (black lines) and the fraction by which the total mass exceeds the mass estimated by observers based on the luminosity (red lines). The scale for the left $y$-axis is arbitrary: the important aspect is the functional form. The dot-dashed lines neglect the increase to the estimated mass caused by the luminosity of the undetected component. The solid lines include this effect, which helps to somewhat mitigate the impact of CB contamination. The solid black line is proportional to the CB's impact on the WB $\widetilde{v}$ under the assumption that an undetected CB increases the estimated mass of both WB components by a similar fraction. Only the solid lines shown here enter into our model as $\Delta \widetilde{M}$ (red; Equation~\ref{dM_tilde_definition}) and $f_{\rm{pb}}$ (black; Equation~\ref{f_pb_definition}).}
    \label{v_factors}
\end{figure}

The actual mass of the CB exceeds its estimated mass by a factor
\begin{eqnarray}
    \Delta \widetilde{M} ~\equiv~ \frac{\text{Total CB mass}}{\text{Estimated CB mass}} \, - \, 1\, .
    \label{dM_tilde_definition}
\end{eqnarray}
This is positive because the mass-luminosity relation is very steep (Figure~\ref{Mass_luminosity_relation}), so blended light from the `undetected' companion raises the estimated CB mass by less than the companion's mass. As a result, two unresolved Sun-like stars would appear to have a total mass of only $1.19 \, M_\odot$ even though their total mass is actually $2 \, M_\odot$, yielding $\Delta \widetilde{M} = 0.68$. Since the mass-luminosity relation is not an exact power law, $\Delta \widetilde{M}$ depends on the masses of the stars involved. We handle this complication by considering as a prior the mass distribution of all the stars in our WB sample (blue bars in Figure~\ref{Mass_distribution}). For a fixed fraction $\widetilde{q} \leq 1/2$ of the CB total mass in the undetected companion, we consider a dense grid of primary star masses covering the range shown there.\footnote{`Primary' here refers to the more massive star in the CB, which may itself be a sub-dominant component of the total WB mass.} In each case, we use the mass-luminosity relation from \citet{Pecaut_2013} to obtain the fraction $\widetilde{L}$ of the CB's total luminosity contributed by the undetected companion $-$ this will be important later.\footnote{For simplicity, we do not implement the FLAME calibration discussed in Section~\ref{FLAME_calibration} when converting masses to luminosities or vice versa in this section.} We then numerically invert Equation~\ref{Cubic_mass_luminosity_relation} using the Newton-Raphson algorithm to find the estimated mass of the CB given the total \emph{Gaia}-band luminosity of the two stars in it, thus following the same approach as observers would use without knowing about the CB. In this way, we obtain the relation between $\Delta \widetilde{M}$ and $\widetilde{q}$, which we show as the solid red line in Figure~\ref{v_factors}. We also use a dotted red line to show the results if we neglect the blended light contributed by the undetected companion. The results are similar either way unless the CB consists of two stars with an almost equal mass: in this limit, blended light reduces $\Delta \widetilde{M}$ from 1 to 0.72.

The impact of blended light is more significant when it comes to the recoil velocity induced by an undetected companion. In the limit that the CB has two equal mass stars such that $\widetilde{q} = 1/2$, the photocentre and barycentre would coincide and move only due to the WB orbit. This is true regardless of the relative velocity $\bm{v}$ of the CB orbit. In general, the velocity of the primary star relative to the CB barycentre would be $\widetilde{q} \bm{v}$, while the undetected star would move at $-\left( 1 - \widetilde{q} \right) \bm{v}$, which is in the opposite direction. Once these velocities are averaged weighted by the mass fractions, the CB barycentre is of course static. But if we instead weight the stars using their luminosities, we get that the photocentre velocity is
\begin{eqnarray}
    \bm{v}_{\rm{phot}} ~=~ \left[\widetilde{q} \left( 1 - \widetilde{L} \right) - \widetilde{L} \left( 1 - \widetilde{q} \right) \right] \bm{v} ~=~ \left( \widetilde{q} - \widetilde{L} \right) \bm{v} \, .
\end{eqnarray}
Since we are only interested in the magnitude of $\bm{v}_{\rm{phot}}$, the impact on the observed WB relative velocity (based on photocentres) is $\propto \left| \widetilde{q} - \widetilde{L} \right|$ \citep*[as also found by][]{Penoyre_2022}. Moreover, $\bm{v}$ is itself proportional to the Keplerian velocity of the CB, which depends on its total mass and separation. Finally, the impact on $\widetilde{\bm{v}}$ also needs to take into account that blended light increases the apparent mass of the WB and thus its Newtonian $v_c$ (Equation~\ref{v_tilde_definition}), but the extra mass is underestimated thanks to the steep mass-luminosity relation. We assume for simplicity that both stars in a WB are affected by CBs with a similar mass ratio to their primary star.

Putting all this together, we get that the impact on $\widetilde{\bm{v}}$ scales with the parameter
\begin{eqnarray}
    f_{\rm{pb}} ~=~ \left| \widetilde{q} - \widetilde{L} \right| \overbrace{\sqrt{1 + \Delta \widetilde{M}}}^{f_M} \, .
    \label{f_pb_definition}
\end{eqnarray}
This is shown as the solid black line in Figure~\ref{v_factors}. The dotted black line shows the result if $\Delta \widetilde{M}$ is calculated neglecting blended light. It is clear that $f_{\rm{pb}}$ is not sensitive to details of how $\Delta \widetilde{M}$ is calculated for the purposes of the above equation. This is because if $\widetilde{q} \ll 1/2$ and the CB has an extreme mass ratio, then the undetected star contributes almost no blended light, making $\Delta \widetilde{M}$ insensitive to whether we consider this light. In the opposite limit that $\widetilde{q} \approx 1/2$, the stars in the CB are very similar such that $\widetilde{q} \approx \widetilde{L}$, making $f_{\rm{pb}}$ very small and thus rendering irrelevant small differences in $f_M$.

\subsubsection{Mass ratio distribution}
\label{qint_distribution}

\begin{figure}
    \centering
    \includegraphics[width=0.47\textwidth]{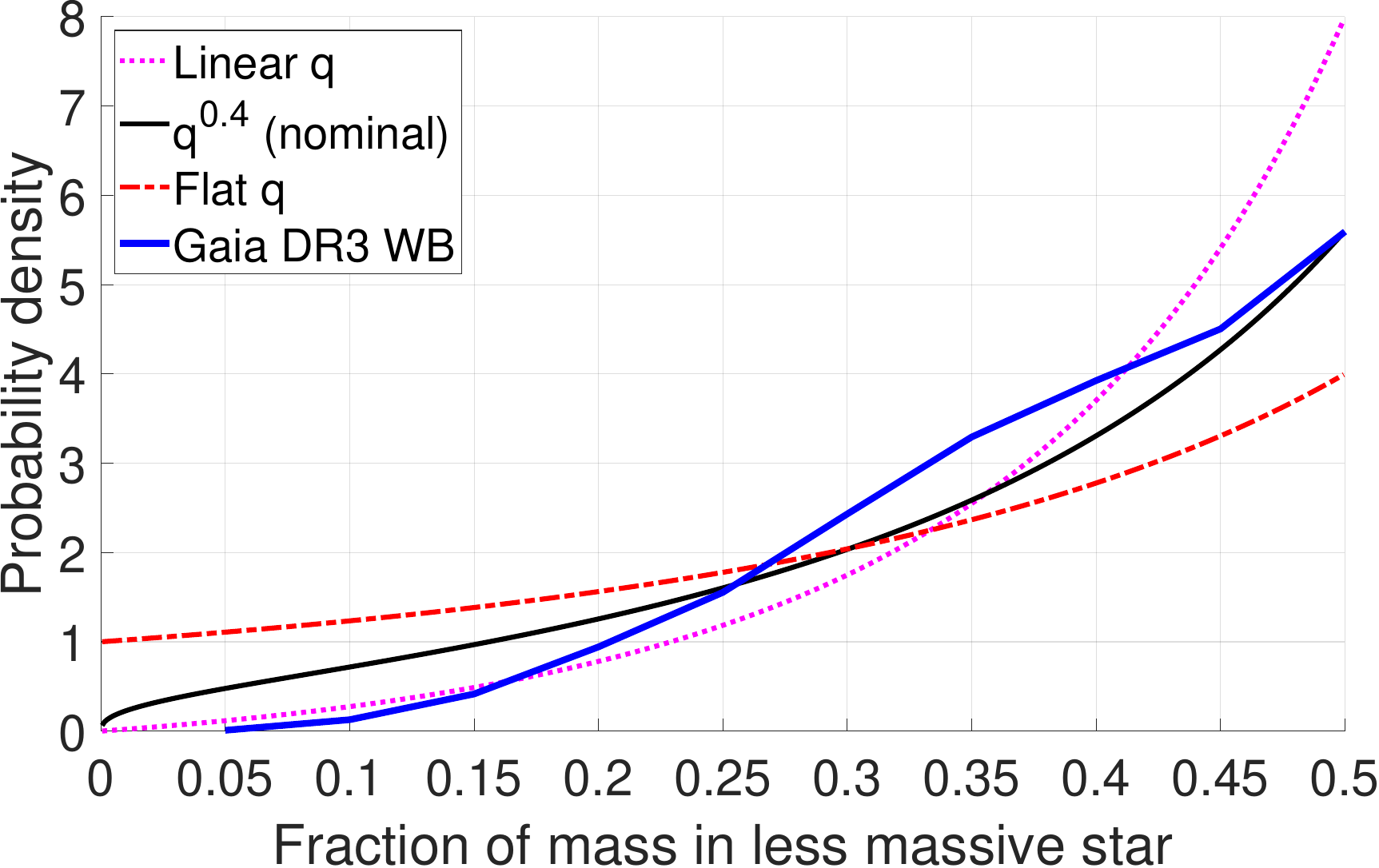}
    \caption{The mass ratio distribution of the WBs in our \emph{Gaia}~DR3 sample (thick solid blue line). We consider three different forms for the undetected CB population. The nominal assumption of a $q^{0.4}$ dependency (solid black line) is based on \citet{Korntreff_2012}, with $q \leq 1$ being the mass ratio between the stars in each binary. This model fits the WB distribution fairly well, especially given that selection effects would make it hard to find binaries with $q \ll 1$. We also consider the case of a flat distribution in $q$ (dot-dashed red line) and a linear distribution (dotted magenta line). These should bracket the range of possible $q$ distributions.}
    \label{q_distribution_comparison}
\end{figure}

The impact of an undetected CB on the WB observables depends on the fraction of the CB mass in the undetected star (Figure~\ref{v_factors}). This requires us to assume some distribution for $\widetilde{q} \leq 1/2$, or equivalently for the mass ratio $q \equiv \widetilde{q}/\left( 1 - \widetilde{q} \right) \leq 1$. Following \citet*{Korntreff_2012}, we assume that
\begin{eqnarray}
    P \left( q \right) \, \propto \, q^{0.4} \, , \quad q \leq 1 \, .
    \label{q_distribution_nominal}
\end{eqnarray}
We expect this to broadly match the distribution of WB mass ratios. We therefore use Figure~\ref{q_distribution_comparison} to show the mass fraction in the less massive star, which is the quantity more relevant for our analysis (Equation~\ref{dM_tilde_total}). The $q^{0.4}$ assumption is shown as a solid black line, while the distribution for our WB sample is shown as a solid blue line. It is apparent that our adopted $q^{0.4}$ distribution provides a good match to the WB distribution. The WB sample contains a deficit of systems with a very extreme mass ratio, but this is almost certainly due to selection effects: low mass stars are very faint (Figure~\ref{Mass_luminosity_relation}). Even if they are not present in our WB sample, we still need to consider extreme mass ratio CBs in our analysis because they can affect the WB observables if the CB has a small separation.

To bracket the possible uncertainties in the CB mass ratio distribution, we also consider a flat distribution of $q$ and a linear distribution, which we show on Figure~\ref{q_distribution_comparison} as the dot-dashed red and dotted magenta line, respectively. A flat distribution is a rough approximation to the three-part power law recommended in figure~7 of \citet{Badry_2019_twin}. We will see later that using instead a flat or linear prior on $q$ gives similar results to our our nominal $q^{0.4}$ assumption, so our conclusions are similar for any power-law distribution of $q$ with exponent in the range $0-1$ (Section~\ref{CB_mass_ratio_distribution}).

\subsubsection{Semi-major axis distribution}
\label{aint_distribution}

The Keplerian orbital velocity of the CB is affected by its semi-major axis $a_{\rm{int}}$. Following \citetalias{Pittordis_2023}, we impose an upper limit of $ak_{\rm{CB}}$, where $a$ is the semi-major axis of the WB. Thus, we require that
\begin{eqnarray}
    \widetilde{a}_{\rm{int}} ~\equiv~ \frac{a_{\rm{int}}}{a} < k_{\rm{CB}} \, .
    \label{k_CB_defintion}
\end{eqnarray}
We work with $\widetilde{a}_{\rm{int}}$ rather than $a_{\rm{int}}$ because the ratio between the CB and WB orbital velocities depends on the ratio of their semi-major axes. The maximum allowed value of this ratio is $k_{\rm{CB}}$, which is a free parameter in our analysis.

Since we remove systems with an obvious third star, $k_{\rm{CB}}$ refers to the maximum $\widetilde{a}_{\rm{int}}$ for CB companions to the WBs in our sample, bearing in mind that we do not consider stars which have a discernible companion within $0.5\arcsec$ with a reasonably consistent parallax and proper motion (Section~\ref{Gaia_DR3_sample}). To convert this to a separation in AU, we note that as our sample goes out to 250~pc, we can expect a reasonable number of systems out to 200~pc. Thus, we expect WBs in our sample to typically not have CB companions with projected separation beyond 100~AU. The upper limit to the actual separation would be larger thanks to projection effects, but their impact on $a_{\rm{int}}$ would be counteracted by the high likelihood of the orbit being quite eccentric \citep{Hwang_2022} and the fact that orbital phases near apocentre are more frequent. It is possible for a sufficiently faint star to go undetected at an even larger separation, but it would have to evade our faint star catalogue used to reduce CB contamination (Section~\ref{Gaia_DR3_sample}). This catalogue goes down to an apparent magnitude of $m_G = 20$, which at a distance of 200~pc corresponds to $M_G = 20 - 5 \log_{10} 20 = 13.5$ and thus a mass of about $0.1 \, M_\odot$ (Figure~\ref{Mass_luminosity_relation}). As a result, CB companions beyond $0.5\arcsec$ from the contaminated star can remain undetected only in a narrow range of mass at the very bottom of the main sequence. Such a low mass would also reduce the perturbation to the velocity of the contaminated star (low $f_{\rm{pb}}$ in Equation~\ref{f_pb_definition}). Combined with the large separation, this makes it unlikely that a distant low-mass companion which evades our faint star catalogue would actually be relevant to our analysis. We thus assume that the maximum relevant $a_{\rm{int}} \approx 100$~AU. To convert this to $k_{\rm{CB}}$, we note that a reasonable $a$ for the WBs might be 40~kAU because our WB sample goes up to $r_{\rm{sky}} = 30$~kAU and we need to consider all viewing angles and orbital phases. For this reason, our WB orbit library goes up to $a = 57$~kAU, though we expect our analysis to assign only very small statistical weights at such high $a$ due to its declining distribution (Equation~\ref{a_cases}) and the fact that we truncate orbits when they go beyond 100~kAU to mimic disruption caused by Galactic tides and encounters with other stars. Taking the ratio of the above estimates for $a_{\rm{int}}$ and $a$ suggests that $k_{\rm{CB}} \approx 2.5$\textperthousand. We do not impose this in our analysis, but we do set a prior that it should be at least 1\textperthousand\ to avoid numerical difficulties. This is very conservative because it implies a maximum $a_{\rm{int}}$ of about 40~AU, which corresponds to just $0.2\arcsec$ for a WB 200~pc away. One can easily envisage that a CB with a larger separation would not be detected at this distance.

At the opposite end, the lowest $a_{\rm{int}}$ that we need to consider is set by the fact that CBs on very small orbits have a short orbital period, leading to rapid but low amplitude astrometric oscillations that have little impact on the inferred space velocity of the contaminated star. This phenomenon is not directly included in our analysis, which for simplicity assumes that the stars in a CB move at constant velocity over the whole \emph{Gaia} observing baseline. In Appendix~\ref{Non_linear_orbital_motion}, we estimate that this approximation breaks down when $a_{\rm{int}} \la 3.2$~AU, below which the impact of the CB rapidly becomes much smaller than our estimate based on uniform linear motion.

Based on the above considerations, we allow CBs to have a 1.5~dex range of $\widetilde{a}_{\rm{int}}$ with a maximum at $k_{\rm{CB}} > 1$\textperthousand. We assume that $\widetilde{a}_{\rm{int}}$ has a flat distribution in log-space, as in the classical \"Opik law \citep{Opik_1924}. The logarithmic distribution of $a_{\rm{int}}$ does indeed seem to be rather flat between 10 and 100~AU \citep[figure 2 of][]{Offner_2023}, so this ought to be a reasonable assumption. We will see later that our results are not much affected by using a somewhat different shape or width to the $a_{\rm{int}}$ distribution (Section~\ref{a_int_distribution}).

In addition to properties of the CB, how it affects $\widetilde{\bm{v}}$ of the WB also depends on the latter's projected separation. At fixed $a$, the impact on $\widetilde{\bm{v}}$ is reduced at low $r_{\rm{sky}}$ because this raises the predicted Newtonian $v_c$ of the WB (Equation~\ref{v_tilde_definition}). Considering also that the CB orbital velocity $\propto 1/\sqrt{\widetilde{a}_{\rm{int}}}$, the perturbation to the WB $\widetilde{\bm{v}}$ scales as
\begin{eqnarray}
    \Delta \widetilde{\bm{v}} ~\propto \sqrt{\frac{\widetilde{r}_{\rm{sky}}}{\widetilde{a}_{\rm{int}}}} \, .
\end{eqnarray}
The lack of any explicit dependence on $a$ greatly reduces the computational cost of the WBT, but this is only possible because the WB libraries are stored with respect to $\widetilde{r}_{\rm{sky}}$ rather than $r_{\rm{sky}}$. The only slight downside is that we need the latter for a comparison with observations, so we subsequently need to find $r_{\rm{sky}} \equiv a \widetilde{r}_{\rm{sky}}$.

\subsubsection{Convolving the CB and WB libraries}
\label{CB_WB_convolution}

The fractional increase to the total mass of a WB is found by combining $\Delta \widetilde{M}$ for each of the stars in it, which we denote using A and B subscripts.
\begin{eqnarray}
    \Delta \widetilde{M} ~=~ \widetilde{q}_A \Delta \widetilde{M}_A \, + \, \widetilde{q}_B \Delta \widetilde{M}_B \, ,
    \label{dM_tilde_total}
\end{eqnarray}
where $\widetilde{q}_A \equiv 1 - \widetilde{q}_B$ is the fraction of the WB total mass in star~A and $\Delta \widetilde{M}_A$ is the fraction by which its actual mass exceeds its apparent mass (Equation~\ref{dM_tilde_definition}). If one of the WB stars is not in a CB, then $\Delta \widetilde{M} = 0$ for that star. We assume that $\widetilde{q}_A$ and $\widetilde{q}_B$ have an identical distribution, which we obtain from our WB sample (solid blue line in Figure~\ref{q_distribution_comparison}). This is not exactly correct because undetected CBs can alter the mass ratio. However, we expect this to be only a very minor issue because even in the extreme case of having two equal Solar mass stars in a WB where one of the stars has an undetected equal mass companion, observers would infer a mass fraction in the less luminous component of 0.46 rather than the correct value of 1/3. The error would be smaller in more typical cases, especially as both stars in the WB could actually be CBs. We therefore expect the WB mass ratio inferred from luminosities to be quite accurate.

To obtain the relative velocity between the photocentres and barycentres of the CBs making up a WB, we again need to combine the contributions from both CBs. In this case, the scaling factors are $\sqrt{\widetilde{q}_{A}}$ and $\sqrt{\widetilde{q}_{B}}$ because the Keplerian orbital velocity of a CB scales with the square root of its mass. Allowing also for the increased WB orbital velocity due to its mass being higher than estimated, we get that the WB $\widetilde{\bm{v}}$ should be revised as follows:
\begin{eqnarray}
    \widetilde{\bm{v}} ~\to~ \widetilde{\bm{v}} f_M \, + \, \Delta \widetilde{\bm{v}}_A \sqrt{\widetilde{q}_A} \, + \, \Delta \widetilde{\bm{v}}_B \sqrt{\widetilde{q}_B} \, ,
    \label{dv_tilde_total}
\end{eqnarray}
where $f_M$ was defined in Equation~\ref{f_pb_definition} and $\Delta \widetilde{\bm{v}}_A$ is the change to $\widetilde{\bm{v}}$ of the WB if all its mass was in star~A such that $\widetilde{q}_A = 1$ ($\Delta \widetilde{\bm{v}}_B$ is defined analogously). Since $\Delta \widetilde{\bm{v}}$ induced by each CB is a 2D vector, we also need to allow for partial cancellation between the photocentre-barycentre offsets arising from each CB. This leads to 4 possible outcomes for whether the contributions along each direction are parallel or opposite.

Following \citet{Badry_2019_twin}, we implement a $\delta$-function in the probability distribution of the CB mass ratio. Such an equal mass ratio population is also evident in our colour-magnitude diagram (dashed grey lines in Figure~\ref{Colour_magnitude_diagram}), with our analysis suggesting that the equal mass likelihood $P_{\rm{eqm}} = 0.04$ (see Appendix~\ref{Equal_mass_binaries}). The origin of such a twin population is unclear, but it could be related to gas accretion onto a forming binary star tending to equalize the masses \citep{Tokovinin_2017, Tokovinin_2020, Tokovinin_2023}. To include this, we combine the CB perturbations calculated assuming a smooth distribution of the CB mass ratio with similar calculations assuming an exactly equal mass ratio for the CB, weighting the former at $\left( 1 - P_{\rm{eqm}} \right)$ and the latter at $P_{\rm{eqm}}$ \citep[further details are provided in][]{Banik_2021_plan}. We demonstrate later that changing the assumed CB mass ratio distribution has very little impact on our results (Section~\ref{CB_mass_ratio_distribution}).

The overall impact of CBs on the WB population is modulated by the likelihood $f_{\rm{CB}}$ that a star in our WB sample has an undetected CB companion. We assume that CB contamination of the two stars in each WB occurs independently, so the likelihood of both stars being contaminated is $f_{\rm{CB}}^2$, which is not negligible. We found that one of the most computationally expensive parts of the WBT is calculating the CB perturbations to the WB parameters in this double contaminated case. This is due to the need to consider a range of CB semi-major axes, velocity perturbations, and mass contributions from both CBs, which almost doubles the already large number of nested loops.

\begin{table}
    \centering
    \caption{The $r_{\rm{sky}}$ and $\widetilde{v}$ range used for the WBT and the binning scheme, which we fixed in advance to limit the possibility of bias \citep{Banik_2021_plan}.}
    \begin{tabular}{cc}
    \hline
    $r_{\rm{sky}}$ (kAU) & $\widetilde{v}$ \\ \hline
    $2-5$ (6 bins) & $0-1.6$ (20 bins) \\
    $5-10$ (5 bins) & $1.6 - 2.6$ (5 bins) \\
    $10-16$ (3 bins) & $2.6 - 5$ (5 bins) \\
    $16-22$ (2 bins) & \\
    $22-30$ (2 bins) & \\ \hline
    \end{tabular}
    \label{Pixellation_decision}
\end{table}

Combining the CB perturbations with the WB library is non-trivial because CBs cannot be considered as adding an extra population to the statistics $-$ they physically affect the WB. As summarized in Equation~\ref{dv_tilde_total}, we need to scale up the WB $\widetilde{\bm{v}}$ to account for the fractional mass increase $\Delta \widetilde{M}$ induced by the CB(s) and then add the velocity perturbations from each CB in the directions parallel and orthogonal to $\widetilde{\bm{v}}$ of the WB, which is a 2D vector in the sky plane. We then use the Pythagoras Rule to calculate the resultant $\widetilde{v}$, or more precisely the appropriate $\widetilde{v}$ pixel (Table~\ref{Pixellation_decision}). Since there is a reasonable chance that the resultant $\widetilde{v} > 5$, CBs can reduce the overall normalization of the $\left( r_{\rm{sky}}, \widetilde{v} \right)$ distribution in our WB library. This may well have happened in reality, but since our \emph{Gaia}~DR3 sample only contains WBs with $\widetilde{v} < 5$, we renormalize the $\left( r_{\rm{sky}}, \widetilde{v} \right)$ distribution that we obtain after the CB-WB convolution step. The computational cost of convolving the CB and WB libraries is kept low using the techniques detailed in \citet{Banik_2021_plan}.

\subsection{Chance alignments}
\label{Chance_alignments}

So far we have only considered bound systems. It is inevitable that some of the WBs in our sample will be unbound. Observations are unlikely to catch a WB in the process of disruption due to a passing third star or a molecular cloud \citepalias[see section~8.1 of][]{Banik_2018_Centauri}. Even so, chance alignments of stars can still arise, especially when we get to large $r_{\rm{sky}}$ and $\widetilde{v}$ \citepalias[see figure~7 of][]{Pittordis_2019}. The likelihood is enhanced due to the possibility of the stars being born in the same star cluster which later dissolved \citep{Oh_2017, Dinnbier_2022}, giving them a much lower relative velocity than the Galactic velocity dispersion. This raises the chance of observing the stars close to each other on the sky, which could cause them to be misidentified as a WB given the \emph{Gaia} parallax uncertainty. LOS contamination should be reduced somewhat by excluding regions of the sky with an enhanced density of stars, e.g. known star clusters \citepalias[table~1 of][]{Pittordis_2023}.

To limit the complexity of our analysis, we set up a very simple model for LOS contamination under the assumption that the separation and relative velocity of stars born in the same cluster can greatly exceed that of the WBs in our sample. Since $\bm{r}_{\rm{sky}}$ and $\bm{v}_{\rm{sky}}$ are both 2D vectors within the sky plane, we assume that the number density of chance-aligned WBs scales as
\begin{eqnarray}
    \frac{dN_{\rm{LOS}}}{dr_{\rm{sky}} \, dv_{\rm{sky}}} ~\propto~ r_{\rm{sky}} v_{\rm{sky}} \, .
\end{eqnarray}
Bearing in mind that $v_{\rm{sky}} \propto \widetilde{v}/\sqrt{r_{\rm{sky}}}$ (Equation~\ref{v_tilde_definition}), the population distribution of LOS contamination is
\begin{eqnarray}
    \frac{dN_{\rm{LOS}}}{dr_{\rm{sky}} \, d\widetilde{v}} ~\propto~ \widetilde{v} \, .
    \label{LOS_contamination_pattern}
\end{eqnarray}
The dependence on $r_{\rm{sky}}$ cancels out because raising $r_{\rm{sky}}$ reduces the range of $\bm{v}_{\rm{sky}}$ for the same range in $\widetilde{v}$.

We integrate this distribution across each of the $r_{\rm{sky}}$ and $\widetilde{v}$ pixels shown in Table~\ref{Pixellation_decision} to obtain the relative number of chance-aligned systems that we expect to find in each pixel. We then normalize the results to a sum of $f_{\rm{LOS}}$ and combine with the result of our CB-WB convolution (Section~\ref{CB_WB_convolution}) weighted by $\left( 1 - f_{\rm{LOS}} \right)$, where $f_{\rm{LOS}}$ is the LOS contamination fraction. We leave $f_{\rm{LOS}}$ as a free parameter in our analysis. Since there are only a few WBs in our catalogue at high $r_{\rm{sky}}$ and $\widetilde{v}$ where LOS contamination would be most significant (Figure~\ref{Observed_photo_tiled}), we expect $f_{\rm{LOS}}$ to be at most a few percent.

\subsection{The simulated distribution of WBs}
\label{Simulated_WB_distribution}

\begin{figure*}
    \centering
    \includegraphics[width=0.48\textwidth]{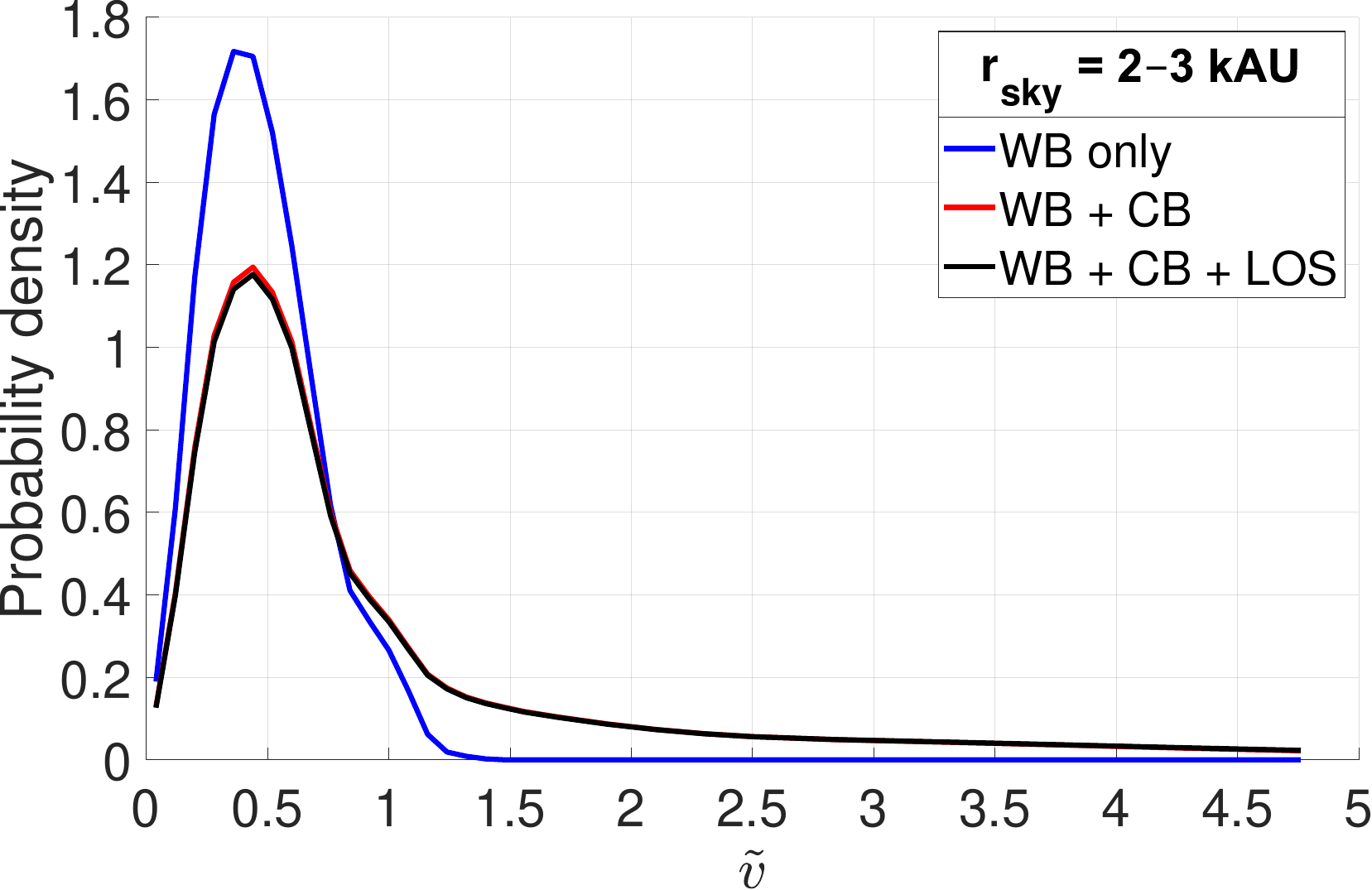}
    \hfill
    \includegraphics[width=0.48\textwidth]{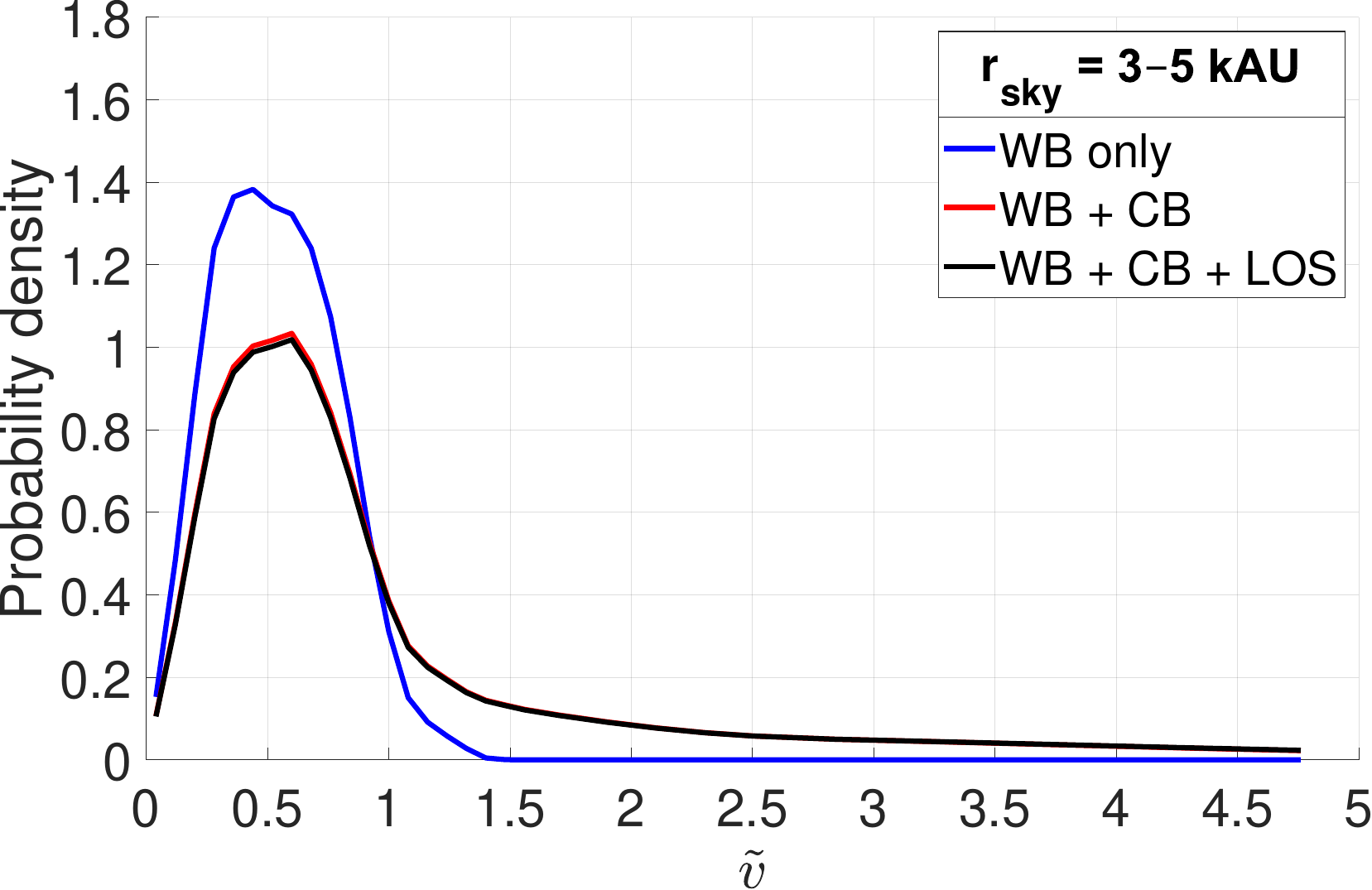}
    \includegraphics[width=0.48\textwidth]{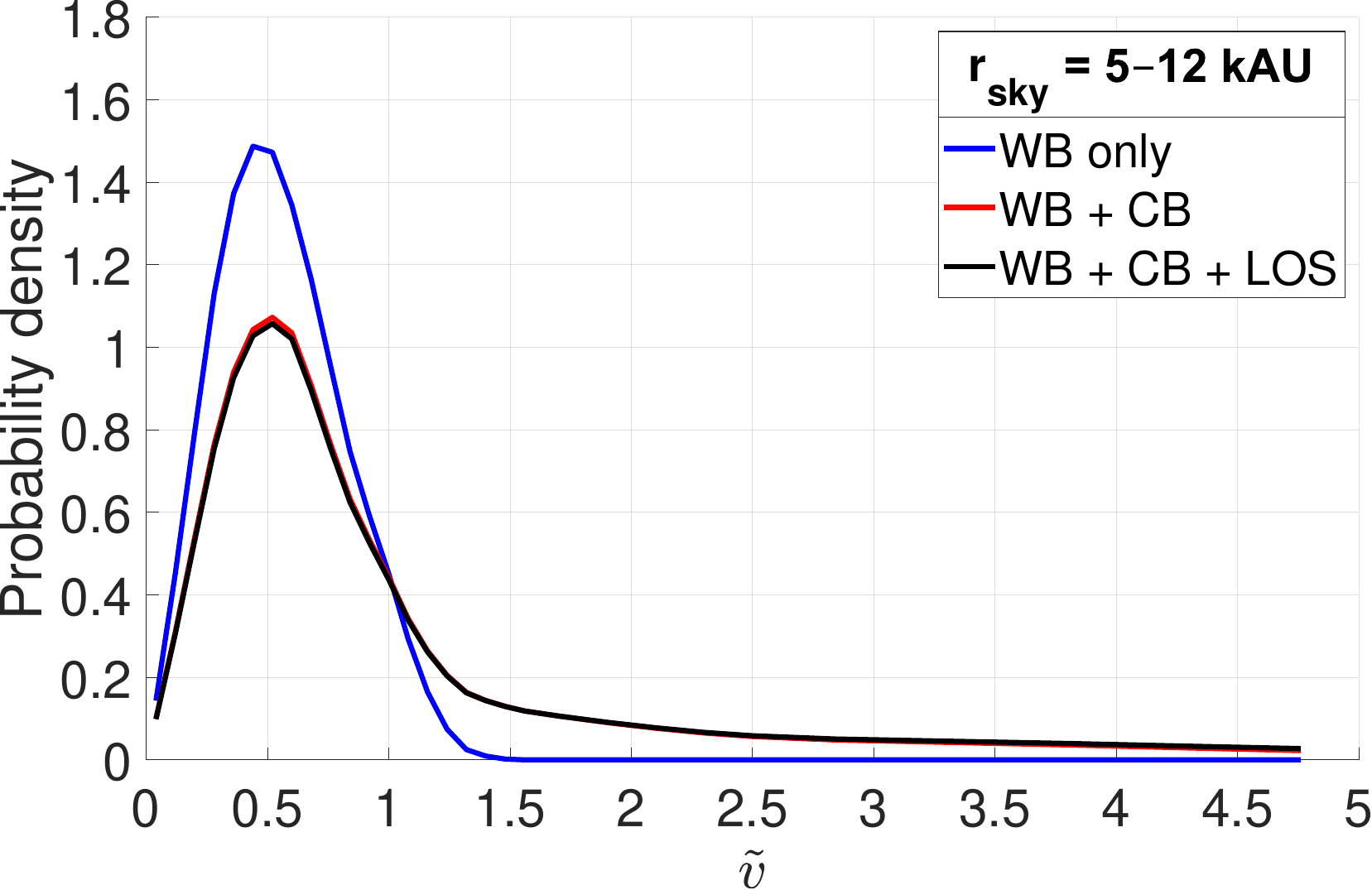}
    \hfill
    \includegraphics[width=0.48\textwidth]{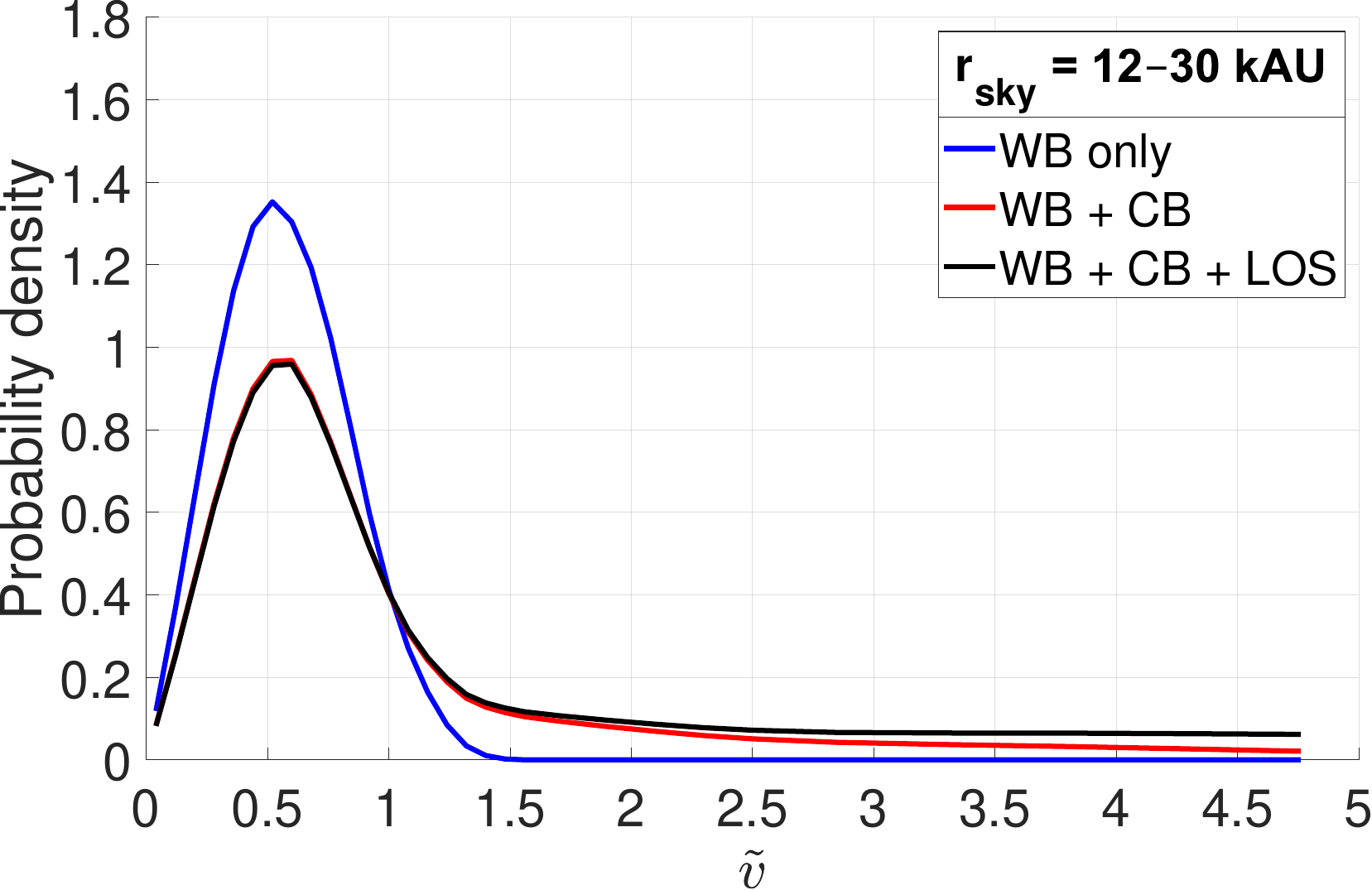}
    \caption{How we build up the simulated $\widetilde{v}$ distribution in our best MOND model (Section~\ref{Gradient_ascent_fixed_gravity}), with results shown in four different $r_{\rm{sky}}$ ranges (different panels). In each case, the distribution due to WBs alone (blue) is significantly altered by including CBs (red). Further adding LOS contamination yields our final simulated distribution (black), which is used to normalize the results shown here. This helps to highlight that including LOS contamination slightly raises the predicted number of WBs at high $r_{\rm{sky}}$.}
    \label{Photo_buildup_MOND}
\end{figure*}

To simulate the distribution of WBs, we tie together a model for the WB population (Section~\ref{WB_population}) with a model for an undetected population of CB companions (Section~\ref{Undetected_companions}) and then allow for chance alignments (Section~\ref{Chance_alignments}). We use Figure~\ref{Photo_buildup_MOND} to illustrate the separate stages for the particular example of our best-fitting MOND model (Section~\ref{Gradient_ascent_fixed_gravity}). The simulated $\widetilde{v}$ distribution is shown in four different $r_{\rm{sky}}$ ranges considering only WBs (blue curves), including CBs (red curves), and finally including also the LOS contamination (black curves). Since only a small proportion of the WB sample is expected to lie in any single $r_{\rm{sky}}$ interval used here, we normalize the $\widetilde{v}$ distributions using the final simulated distribution, so only the black curve in each panel is guaranteed to have an integral of 1. WBs alone almost never yield $\widetilde{v} \ga 1.5$, even in MOND $-$ the theoretical limit of 1.7 only refers to systems on highly eccentric orbits seen close to pericentre from a direction nearly orthogonal to the relative velocity \citepalias[this is also apparent in figure~3 of][]{Banik_2018_Centauri}. CB companions create an extended declining tail that is easily able to reach much higher $\widetilde{v}$. We capture much of this extended tail by going up to $\widetilde{v} = 5$, a choice we discussed in Section~\ref{Gaia_DR3_sample} and fixed in advance of the WBT to mitigate possible biases \citep{Banik_2021_plan}. This extra leverage allows our analysis to constrain the CB population, which dominates the behaviour once $\widetilde{v} \ga 2$. A flat log-space distribution of $a_{\rm{int}}$ \citep{Opik_1924} yields a $\widetilde{v}$ distribution that is also flat in log-space in the CB-dominated region, helping to explain why the distribution here is $\appropto 1/\widetilde{v}$. We expect this to flatten out and ultimately start rising at very high $\widetilde{v}$ due to LOS contamination. This is most evident in the highest $r_{\rm{sky}}$ interval, where there is less contribution from only the gravitationally bound systems because WBs have a declining distribution of $r_{\rm{sky}}$ (Equation~\ref{a_cases}). Our WB sample reaches high enough $\widetilde{v}$ for its distribution to become flat, but a clearly rising trend is not apparent because we have limited the considered $\widetilde{v}$ range to avoid being swamped by chance alignments. Other studies which consider a wider parameter range do show a rising trend \citep[e.g. see figure~12 of][]{Badry_2021}.

\subsection{Comparison with observations}
\label{Binomial_statistics}

We compare the observed number of WBs in each $\left( r_{\rm{sky}}, \widetilde{v} \right)$ pixel with the expectation of our model for that pixel. We use binomial statistics because we are dealing with integer statistics and the total sample size $N$ is finite. Thus, the binomial likelihood of observing $k$ systems in a pixel is
\begin{eqnarray}
    P_{\rm{pixel}} ~=~ \frac{N!}{\left( N - k \right)! k!} p^k \left( 1 - p \right)^{N - k} \, ,
    \label{P_pixel}
\end{eqnarray}
where $p$ is the fraction of the total number of WBs in the comparison region ($r_{\rm{sky}} = 2-30$~kAU, $\widetilde{v} < 5$) which should be located within the pixel under consideration. For computational reasons, we use the logarithmic version of this equation. The log-likelihood $\ln P$ of the model is then given by considering all 540 pixels summarized in Table~\ref{Pixellation_decision}.
\begin{eqnarray}
    \ln P ~=~ \sum_{\rm{Pixels}} \ln P_{\rm{pixel}} \, .
    \label{P_total}
\end{eqnarray}
This is used to infer the optimal values and confidence intervals for our model parameters associated with the CBs ($f_{\rm{CB}}$, $k_{\rm{CB}}$, and $\gamma$), LOS contamination ($f_{\rm{LOS}}$), and the WBs ($a_{\rm{break}}$, $\beta$, and $\alpha_{\rm{grav}}$). Note that $\gamma$ affects both the CB and WB populations because they are assumed to follow the same eccentricity distribution for simplicity. The CBs are assumed to follow an \"Opik law distribution of semi-major axes \citep{Opik_1924} over a 1.5~dex range (Section~\ref{aint_distribution}), while the WBs follow a broken power law that reduces to $a^{-1.6}$ at small separations (Equation~\ref{a_cases}).

Notice that regardless of how many WBs a pixel contains, $\ln P_{\rm{pixel}} < 0$ in every pixel as its predicted $p$ is always non-zero once we include CB and LOS contamination. Since we have quite a large number of WBs spread across a large number of pixels, we expect $\ln P \ll 0$ \citep[a similar situation is evident in table~2 of][]{Asencio_2022}. As a crude estimate, we can suppose that our model predicts $8611/540 = 16$ WBs in each of our 540 pixels, so the actual number of WBs in any pixel is expected to nearly follow a Poisson distribution with a standard deviation of $\sqrt{16} = 4$. If the model works well, then the actual number of WBs would be approximately uniformly distributed over the range $12-20$, with all of these outcomes having a likelihood of $P_{\rm{pixel}} = 1/9$. The combined likelihood would then be $\ln P = -540 \, \ln 9 = -1200$. We therefore expect to get a combined log-likelihood of about this much, but this is of course only a very crude estimate. In reality, not all pixels would be expected to have the same number of WBs, and even if they were, outcomes in the tail of the Poisson distribution are less likely than the inverse of the Poisson noise. Moreover, it is inevitable that our model does not capture all the subtleties of the actual WB population because of the simplifying assumptions made to keep the complexity and computational cost manageable. Nonetheless, we will see later that the above estimate is fairly accurate.

\section{Results}
\label{Results}

\begin{figure}
    \centering
    \includegraphics[width=0.47\textwidth]{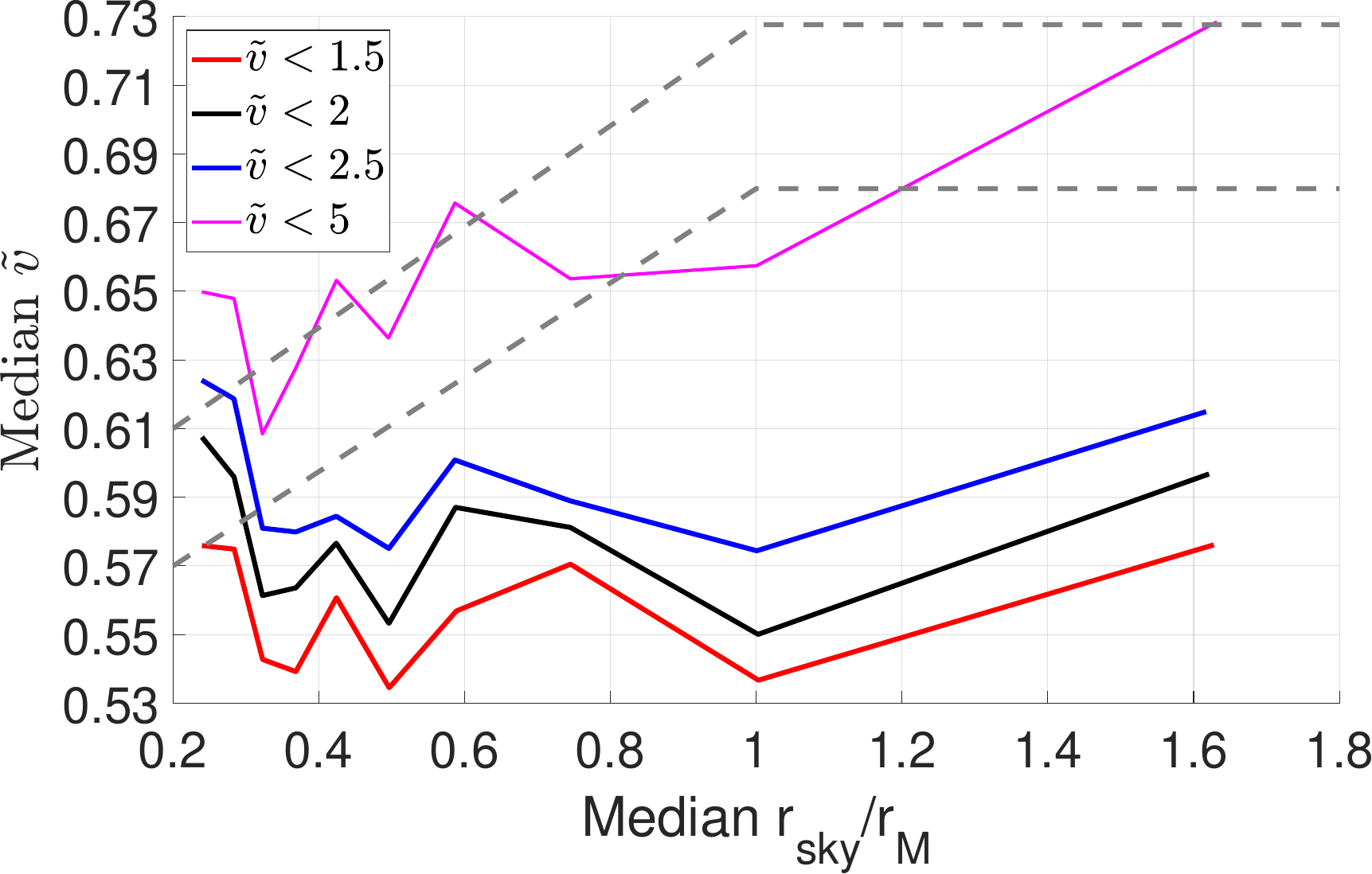}
    \caption{The median $\widetilde{v}$ of our WB sample as a function of the median $r_{\rm{sky}}/r_{_M}$, shown for ten bins in the latter which each contain 861 WBs except the last bin which has 862. The solid lines show results with four different upper limits to $\widetilde{v}$ of 1.5 (red), 2 (black), 2.5 (blue), and 5 (thin magenta). Qualitatively, the MOND expectation is that the median $\widetilde{v}$ rises for larger $r_{\rm{sky}}/r_{_M}$ and then becomes flat once the Galactic gravity dominates. We indicate this by two dashed grey lines with different normalizations. Their shape is based on Figure~\ref{Radial_gravity_boost}, which implies that the main peak in the $\widetilde{v}$ distribution for $\widetilde{v} \la 2$ should broaden at low accelerations. Neglecting projection effects and assuming the current separation is representative of the whole orbit, this would lead to a rise in the median $\widetilde{v}$ by a factor of $\sqrt{\eta} = 1.2$ that occurs roughly linearly over the range $r_{\rm{sky}}/r_{_M} = 0.2-1$ before rapidly flattening out. The Newtonian prediction is a flat relation, which is consistent with the data up to $\widetilde{v} < 2.5$, within which lies 89.8\% of the sample. The full sample of WBs (with $\widetilde{v} < 5$) does show a rising median that looks qualitatively similar to the MOND expectation, but since this rise is not seen for the subsamples restricted to the main peak region of the $\widetilde{v}$ distribution where $\widetilde{v}$ is low enough to be physically plausible for an uncontaminated system, we attribute this behaviour to LOS contamination becoming more important at larger separations (Section~\ref{Chance_alignments}).}
    \label{Median_vtilde_values}
\end{figure}

Before presenting the results of our detailed statistical analysis discussed in the previous section, we first look for trends in the median $\widetilde{v}$ of our WB population with $r_{\rm{sky}}/r_{_M}$, which is a proxy for the internal acceleration of each WB. For this purpose, we sort our WBs in order of $r_{\rm{sky}}/r_{_M}$ and use this to create ten equally sized subsamples with no overlap. The thin magenta line in Figure~\ref{Median_vtilde_values} shows the median $\widetilde{v}$ as a function of the median $r_{\rm{sky}}/r_{_M}$ of each subsample. A clear rising trend is evident, with the increase being close to the 20\% expected in MOND. However, LOS contamination would be more important at high $r_{\rm{sky}}$ and high $\widetilde{v}$, potentially inflating the median $\widetilde{v}$ in a manner which appears acceleration-dependent. To mitigate against this possibility, we note that MOND would cause an acceleration-dependent broadening to the main peak of the $\widetilde{v}$ distribution at $\widetilde{v} \la 2$ \citepalias[Figure~\ref{Observed_photo_tiled}; see also figure~7 of][]{Pittordis_2019}. In MOND, we expect almost no WBs with $\widetilde{v} > 1.5$ in the absence of contaminating effects \citepalias[see figure~3 of][]{Banik_2018_Centauri}. We therefore consider the median $\widetilde{v}$ of only those WBs in each $r_{\rm{sky}}/r_{_M}$ bin with $\widetilde{v} < 1.5$, 2, or 2.5, which we show in Figure~\ref{Median_vtilde_values} using the red, black, and blue line, respectively. In all three cases, the rising trend disappears: the median $\widetilde{v}$ becomes flat with respect to our proxy for the internal acceleration. There is of course some scatter, but we would expect a random scatter of about $0.6/\sqrt{861} = 0.02$ due to the finite number of WBs, which is roughly in line with the actual scatter.

Since the upper limit to $\widetilde{v}$ should substantially alleviate the impact of CB and LOS contamination without much affecting a genuine MOND signal, it is clear that the apparent MOND signal discussed above is almost certainly a consequence of contaminating effects, highlighting how easily one can reach erroneous conclusions from the WBT using simple population statistics without a good understanding of the astrophysical systematics \citep{McCulloch_2019, Hernandez_2019_WB, Hernandez_2022}. Focusing on WBs with $\widetilde{v} \la 2$, the very weak dependence of the median $\widetilde{v}$ on $r_{\rm{sky}}/r_{_M}$ despite the significant range covered by our sample strongly suggests that the WBT will return a Newtonian result. We highlight this using the dashed grey lines in Figure~\ref{Median_vtilde_values}, which approximately show the MOND prediction that the median $\widetilde{v}$ should rise by a factor of $\sqrt{\eta} = 1.2$ over the range $r_{\rm{sky}}/r_{_M} = 0.2-1$. This is motivated by Figure~\ref{Radial_gravity_boost}, which shows that the enhancement factor to the radial Newtonian gravity rises from 1 up to its asymptotic value roughly linearly over this range of $r/r_{_M}$ and is almost flat afterwards. While this is a rather crude way of considering the situation because it neglects projection and orbital phase effects, it should approximately capture the expected broadening of the $\widetilde{v}$ distribution in the main peak region as one considers WB subsamples with a lower internal acceleration. The expected Newtonian behaviour is not shown as there is already a gridline at 0.57. This provides a much better fit to the median $\widetilde{v}$ for the $\widetilde{v} < 2$ subsample than the approximate MOND expectation mentioned above.

This strong hint that the WBT will disfavour MOND is in line with the detailed analysis of \citetalias{Pittordis_2023}, who prepared a model for the WB population similar to ours and found a strong preference for Newtonian gravity over MOND. Their Newtonian model fits the observations surprisingly well given their limited exploration of the parameter space (see their figure~12). Moreover, the right panel of their figure~17 shows that WBs have a median $v_{\rm{sky}} \propto 1/\sqrt{r_{\rm{sky}}}$, especially when restricting to only those WBs with $\widetilde{v} \la 2$. This is the expected behaviour if Kepler's Third Law works even in the low-acceleration regime, as illustrated by the dashed lines on this figure.

The lack of a trend in the median $\widetilde{v}$ is unlikely to be caused by observational uncertainties, which would generally lead to a larger $\widetilde{v}$ uncertainty at high $r_{\rm{sky}}$ due to the lower Newtonian $v_c$ (Equation~\ref{v_tilde_definition}). This would if anything create the appearance of a MOND-like signal in actually Newtonian data rather than precisely cancel out a genuine MOND signal (c.f. Section~\ref{Other_WBT_results}). Moreover, the $\widetilde{v}$ uncertainties in our WB sample are very small (Figure~\ref{v_tilde_error_distribution}).

Although our results favour the Newtonian model, the distribution of $\widetilde{v}$ contains much more information than just its median value. Before drawing strong conclusions about the gravity law, we try to exploit this information using a forward model that considers the most relevant factors (Section~\ref{Model}). We now turn to the results of a comparison between this model and the WB dataset.

\subsection{Gradient ascent with a fixed gravity law}
\label{Gradient_ascent_fixed_gravity}

To check if Newtonian gravity indeed provides a better fit to the data than MOND, we fix the gravity law to Newtonian ($\alpha_{\rm{grav}} = 0$) and run a gradient ascent in the remaining model parameters to try and maximize $\ln P$ \citep{Fletcher_1963}. The gradient ascent usually takes about 100 iterations to converge. To further improve the fit, we supplement this by running a `line ascent' where $\ln P$ is maximized with respect to each parameter while holding all the other parameters fixed. We then repeat this extended version of gradient ascent for MOND ($\alpha_{\rm{grav}} = 1$). The parameters of the best-fitting models are shown in Table~\ref{Parameters_Newton_MOND}, whose last line shows the corresponding $\ln P$. This is much lower in the MOND case, with $\Delta \ln P = 175$. This can be thought of as equivalent to a statistical significance of $\sqrt{2 \Delta \ln P}$ standard deviations for a single Gaussian variable, implying that the WBT prefers Newtonian gravity over MOND at $19\sigma$ confidence. The high significance is in line with earlier forecasts that a few thousand systems should be more than sufficient for the WBT \citepalias[see the blue curves in figure~5 of][]{Banik_2018_Centauri}.

\renewcommand{\arraystretch}{1.2}
\begin{table}
    \centering
    \caption{The best-fitting model parameters with a fixed gravity law, found using gradient ascent. Notice the very substantial difference in $\ln P$ (Section~\ref{Binomial_statistics}), indicating that MOND provides a much poorer fit to the data.}
    \begin{tabular}{ccc}
    \hline
    & \multicolumn{2}{c}{Gravity law} \\
    Parameter & Newton & MOND \\ \hline
    $a_{\rm{break}}$ (kAU) & $~~5.10$ & $~~5.30$ \\
    $\beta$ & $-2.66$ & $-2.70$ \\
    $\gamma$ & $~~1.86$ & $~~1.96$ \\
    $f_{\rm{CB}}$ (\%) & $~~69.9$ & $~~65.7$ \\
    $k_{\rm{CB}}$ (\textperthousand) & $~~1.07$ & $~~2.10$ \\
    $f_{\rm{LOS}}$ (\%) & $~~1.45$ & $~~1.49$ \\
    $\alpha_{\rm{grav}}$ (fixed) & 0 & 1 \\ \hline
    $\ln P$ & $-1457.4$ & $-1632.6$ \\ \hline
    \end{tabular}
    \label{Parameters_Newton_MOND}
\end{table}
\renewcommand{\arraystretch}{1}

Our result broadly agrees with that of \citetalias{Pittordis_2023} if we consider their Newtonian and MOND models with a thermal eccentricity distribution ($\gamma = 1$), which is the closest distribution they consider to the slightly superthermal distribution preferred by our fits. Their analysis gives a difference in $\chi^2$ between the best Newtonian and MOND models of 525, implying that Newtonian gravity is preferred over MOND at a significance of $\sqrt{525}\sigma$ or about $23\sigma$. We expect that the actual significance would be somewhat lower because further exploration of the parameter space should improve the fit, with MOND benefitting far more due to its $2.6\times$ higher $\chi^2$. Even so, it is clear that the WBT favours Newtonian gravity over MOND at $\gg 5\sigma$ confidence. We will see later that this conclusion holds up in our more thorough exploration of the parameter space. Other attempts at the WBT are discussed in Section~\ref{Other_WBT_results}.

\begin{figure*}
    \centering
    \includegraphics[width=0.48\textwidth]{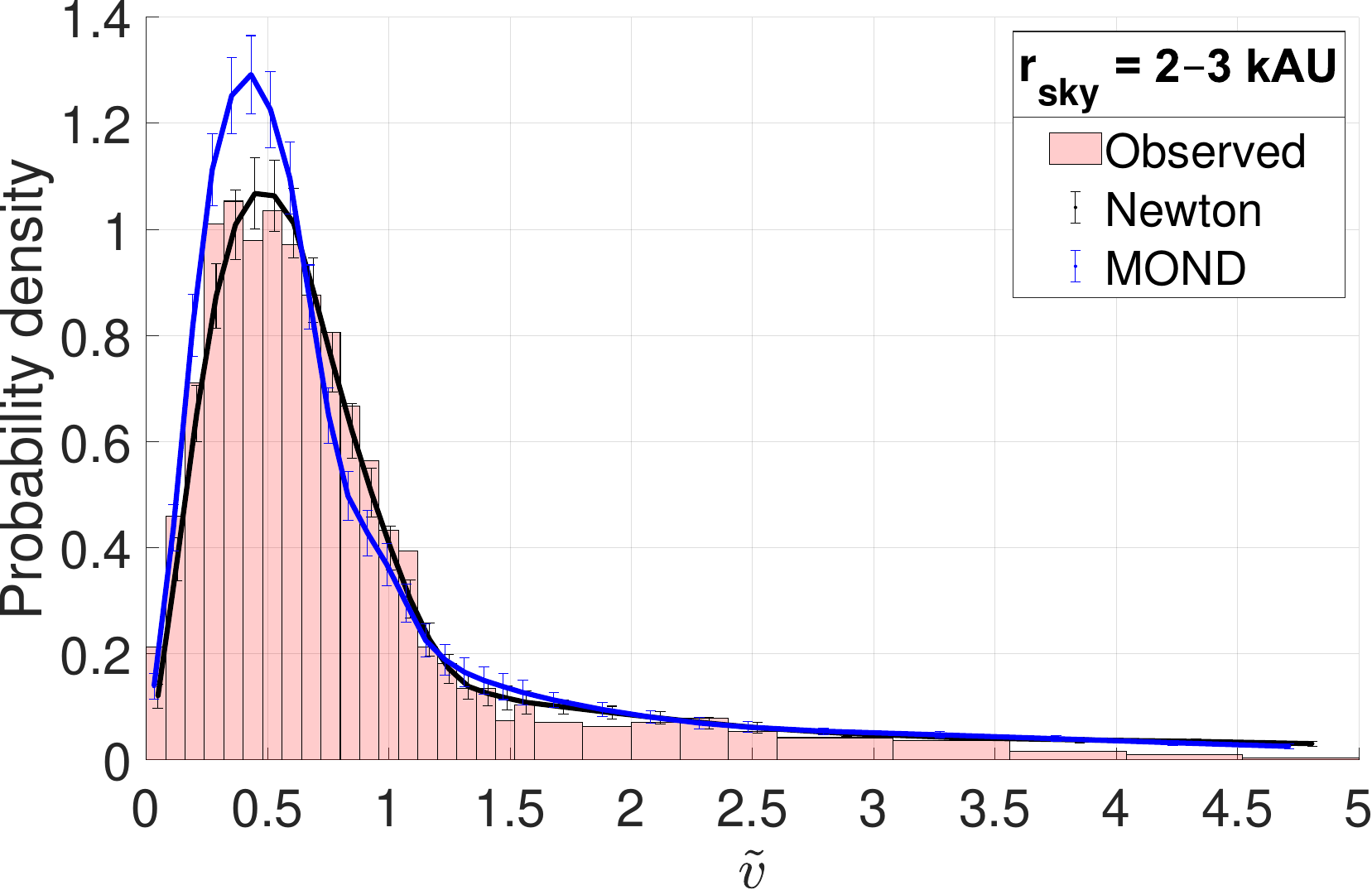}
    \hfill
    \includegraphics[width=0.48\textwidth]{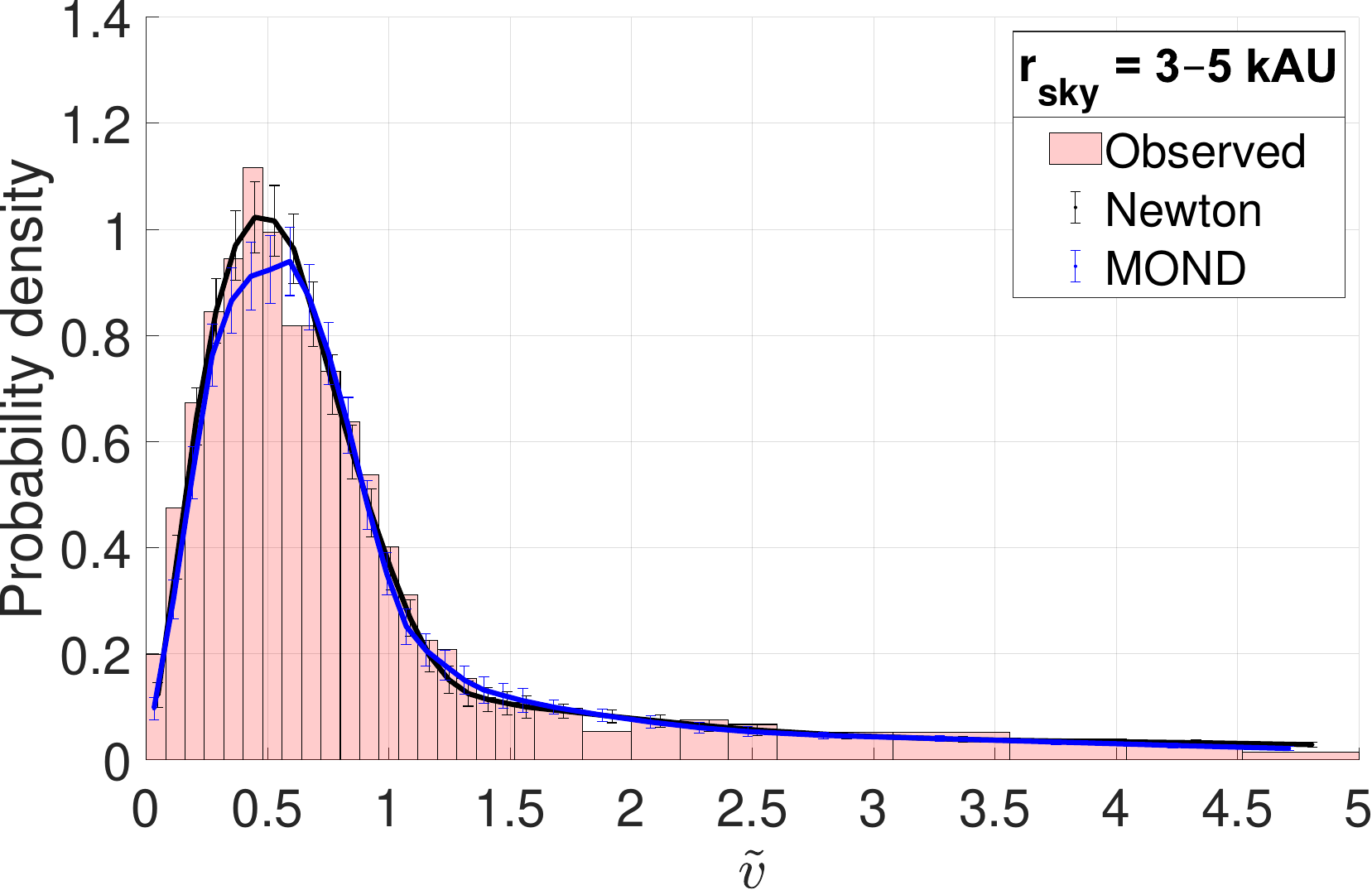}
    \includegraphics[width=0.48\textwidth]{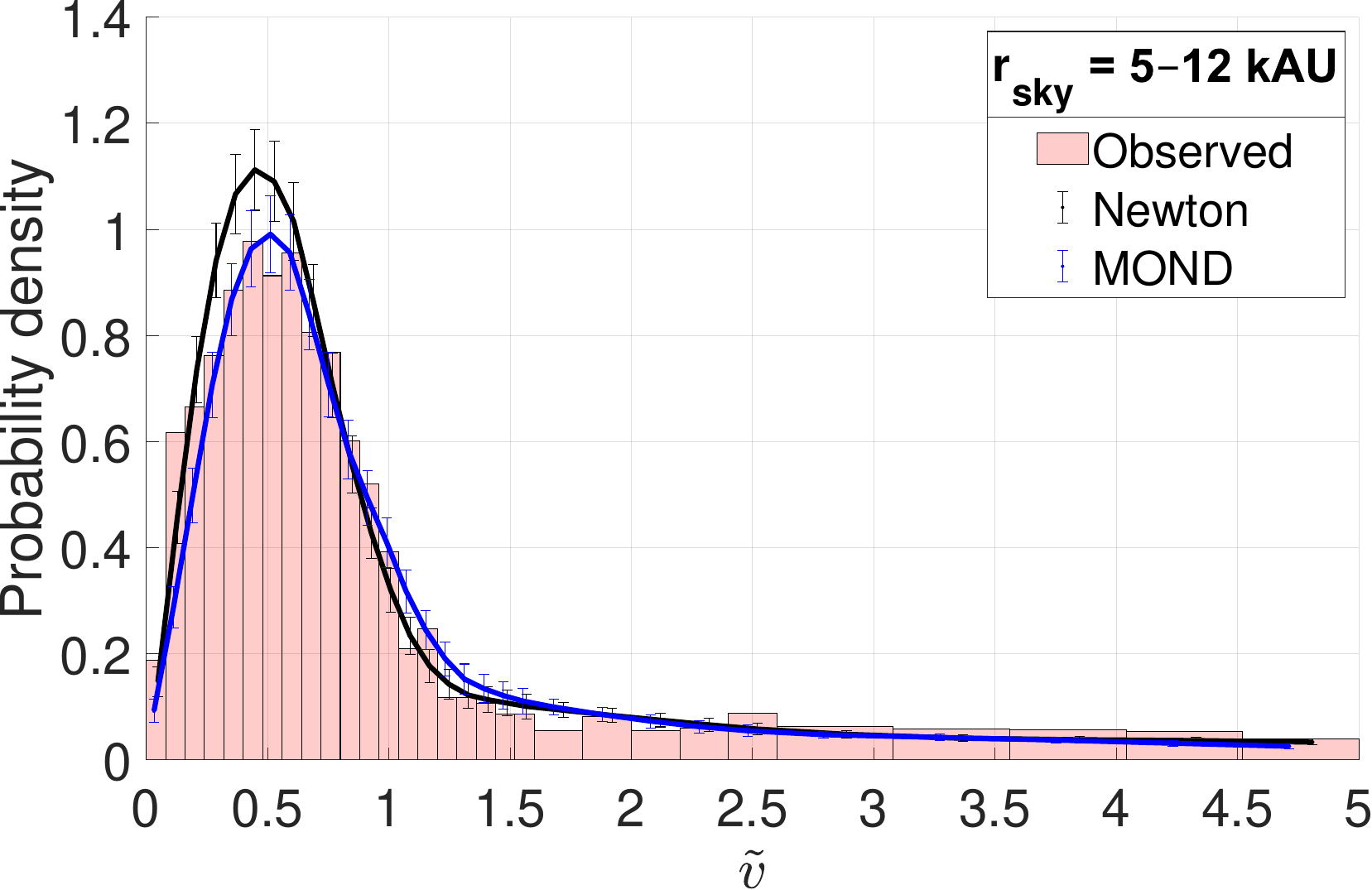}
    \hfill
    \includegraphics[width=0.48\textwidth]{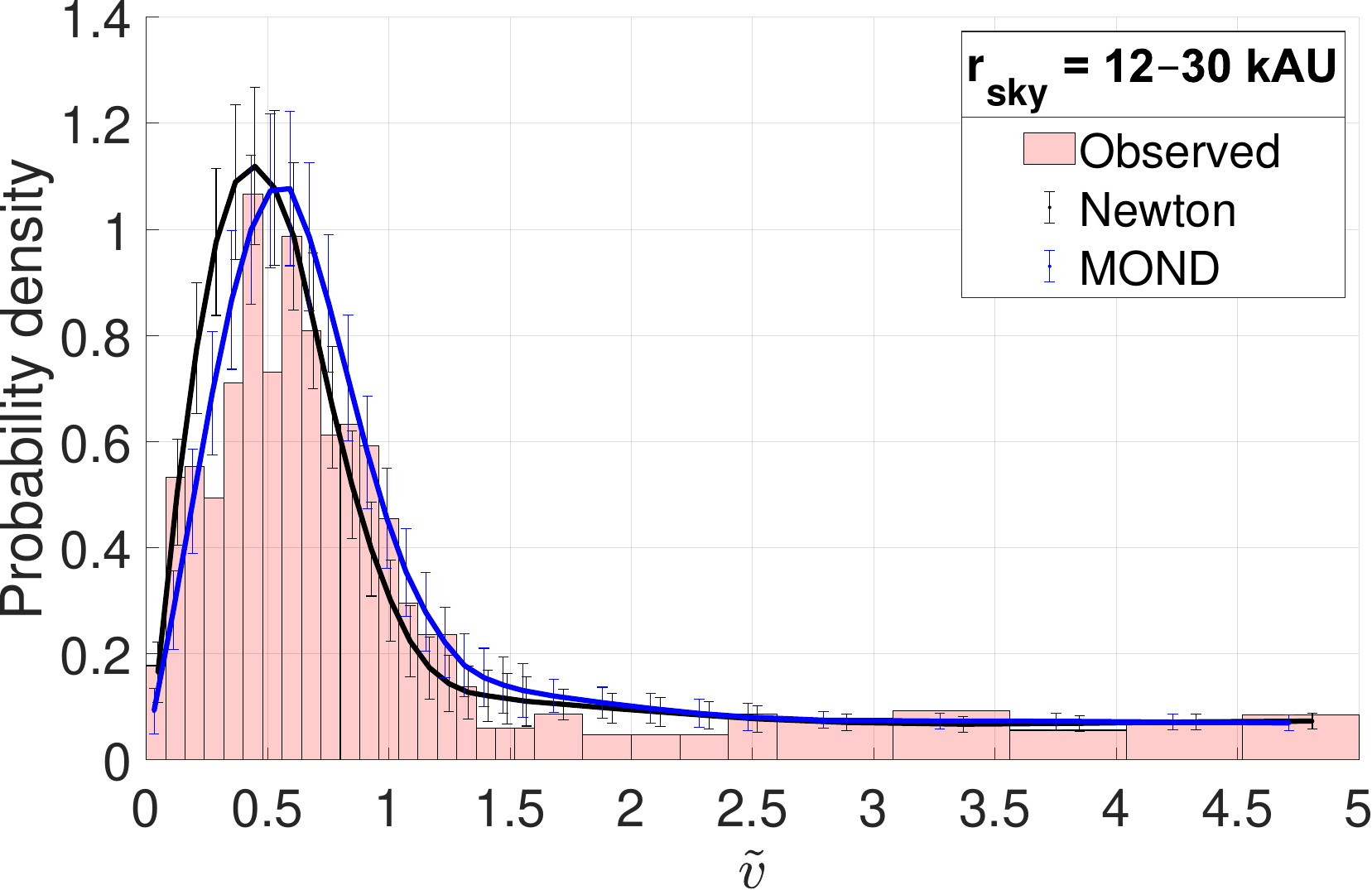}
    \caption{Comparison between the observed $\widetilde{v}$ distribution in four different $r_{\rm{sky}}$ ranges (different panels) with the prediction of our best Newtonian and Milgromian model (black and blue lines with error bars, respectively). Results with Newtonian (Milgromian) gravity are offset very slightly to the right (left) from the centre of each $\widetilde{v}$ bin for clarity in case the predictions are similar. Uncertainties are calculated using binomial statistics. The results shown here are normalized based on the observed WB distribution, so the simulated curves in each panel need not have an integral of 1.}
    \label{Photo_fixed_gravity}
\end{figure*}

To better understand why the Newtonian fit to the WB dataset is much better than the MOND fit, we use Figure~\ref{Photo_fixed_gravity} to plot the $\widetilde{v}$ distribution in four different $r_{\rm{sky}}$ ranges, normalized using the observed number of systems in each $r_{\rm{sky}}$ range. The error bars show binomial uncertainties in the model predictions. The top left panel has little direct sensitivity to the gravity law because of the low $r_{\rm{sky}}$, but there is an indirect dependence because the best model parameters differ depending on the gravity law. The lower $f_{\rm{CB}}$ in the best MOND fit (solid blue line) causes the $\widetilde{v}$ distribution to peak earlier at a much higher amplitude and then drop off more rapidly after the peak compared to the best Newtonian fit (solid black line). This causes MOND to severely disagree with the data, which may seem surprising as MOND is not directly expected to be very relevant at such low $r_{\rm{sky}}$. The disagreement must arise because of a tradeoff when simultaneously fitting the data over a wide $r_{\rm{sky}}$ range, so we turn to the situation at wider separations. As the gravity law gradually becomes more important, the predicted mode to the Milgromian $\widetilde{v}$ distribution eventually `overtakes' the Newtonian prediction.\footnote{Figure~3 of \citetalias{Banik_2018_Centauri} shows how the eccentricity distribution and the gravity law affect the $\widetilde{v}$ distribution, but here we focus on optimized fits for each gravity theory to enable a fairer comparison.} The Newtonian and MOND distributions become almost identical at $r_{\rm{sky}} = 3-5$~kAU (top right panel). The Newtonian model performs slightly better around the peak of the distribution because it predicts an earlier and higher peak despite the higher $f_{\rm{CB}}$, highlighting that the gravity law is now one of the dominant factors. This remains true at larger separations, where MOND always predicts a later peak to the $\widetilde{v}$ distribution. For $r_{\rm{sky}} = 5-12$~kAU (bottom left panel), MOND performs somewhat better in a handful of pixels around the peak region, though given the uncertainties, the Newtonian model is not that far off. The rapid decline after the peak and the very low number counts at $\widetilde{v} = 1 - 1.5$ work better in the Newtonian model. The situation is similar for $r_{\rm{sky}} = 12-30$~kAU (bottom right panel), where MOND again fits slightly better before the peak in the $\widetilde{v}$ distribution. The uncertainties are now discernibly larger due to the smaller proportion of the total WB sample expected (and observed) to be in this range compared to the other $r_{\rm{sky}}$ ranges mentioned above.\footnote{Binomial uncertainties in the model prediction for each pixel do not depend on how many WBs are observed in it.} Even so, the peak position is better reproduced in Newtonian gravity, which also provides a somewhat better match to the sharp observed decline afterwards over the range $0.5-1.5$ and the consequent rather low number counts over the range $1-1.5$. This range was expected to be critical for the WBT \citepalias[section~5 of][]{Banik_2018_Centauri} and is also a region where the model predictions depend almost entirely on the gravity law, which is obviously not the case near the peak of the distribution where other factors are also relevant (see their figure~3). Both models provide a very good fit to the almost flat $\widetilde{v}$ distribution over the range $2.5-5$, nicely demonstrating a balance between a declining trend from the extended CB tail and a rising trend from LOS contamination (Equation~\ref{LOS_contamination_pattern}). The rapid observed decline just beyond the peak of the $\widetilde{v}$ distribution when $r_{\rm{sky}} \ga 3$~kAU presumably forces the MOND fit to use a lower $f_{\rm{CB}}$, but this also causes the MOND model to provide a worse fit to precisely this range of $\widetilde{v}$ at very low $r_{\rm{sky}}$, as shown in the top left panel. Indeed, the combination of a too-rapid decline after the main peak at very low $r_{\rm{sky}}$ with a too-gradual decline at high $r_{\rm{sky}}$ is just what we would expect of an optimized Milgromian fit if the predicted change to the shape of the $\widetilde{v}$ distribution does not occur in the real data, as strongly suggested by Figure~\ref{Median_vtilde_values}.

\begin{figure*}
    \centering
    \includegraphics[width=0.48\textwidth]{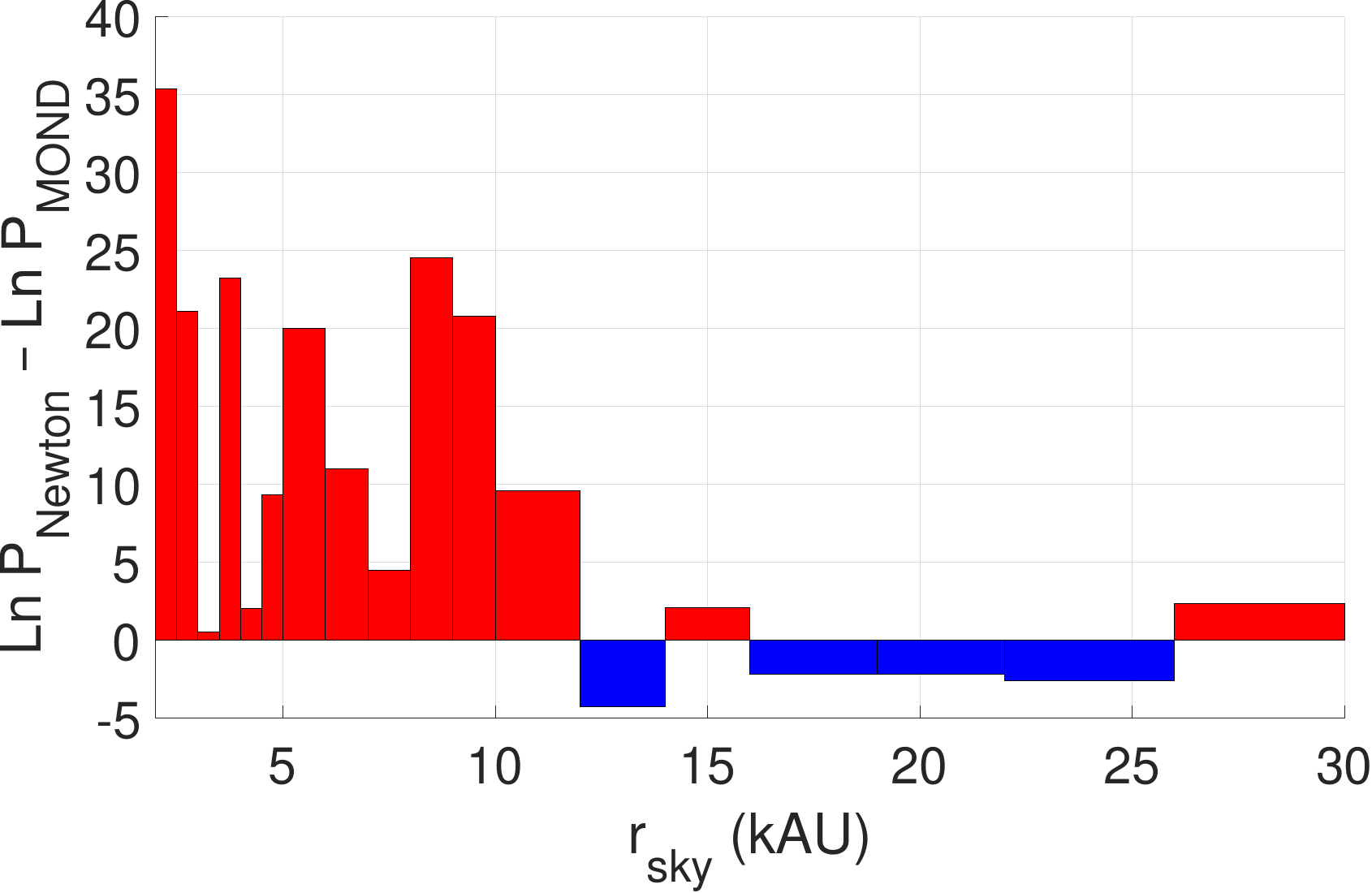}
    \hfill
    \includegraphics[width=0.48\textwidth]{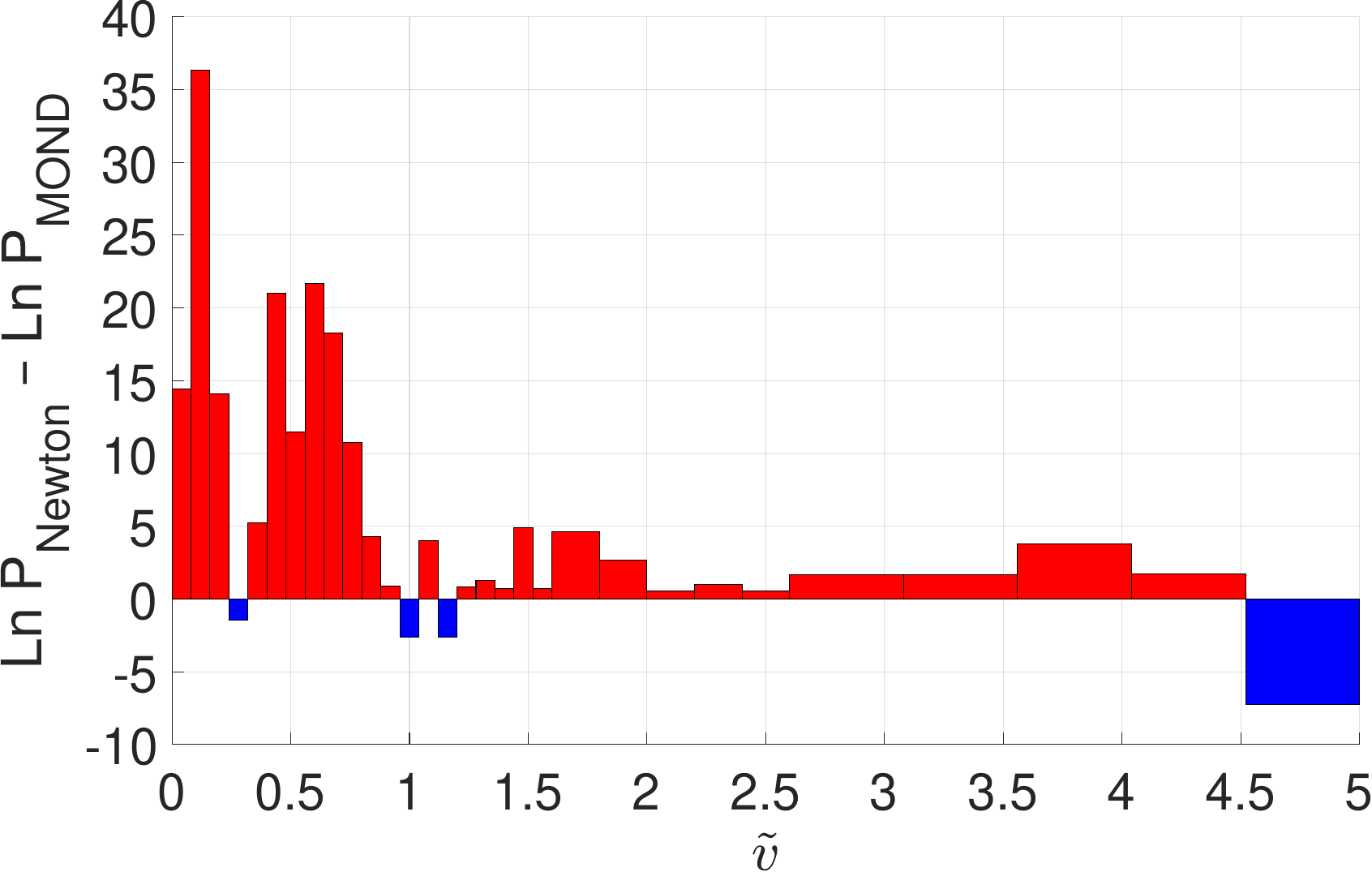}
    \caption{The difference in log-likelihood between our best Newtonian and Milgromian models, shown here after summing across all pixels in $\widetilde{v}$ at each $r_{\rm{sky}}$ (left panel) and vice versa (right panel). Blue bars indicate that MOND works better. Most bars are red, indicating a better fit in Newtonian gravity. Overall, $\ln P$ is higher in the Newtonian model by $\Delta \ln P = 175.2$, which suggests a preference for Newtonian gravity over MOND by $\sqrt{2 \Delta \ln P}$ standard deviations, i.e. at $18.7\sigma$ confidence. We use Figure~\ref{Ln_P_Newton_MOND} to show a pixel-by-pixel comparison between the log-likelihoods returned by each model.}
    \label{Ln_P_Newton_MOND_summed}
\end{figure*}

It is of course inevitable that Newtonian gravity would fit better in some pixels and MOND in others, so it is important to consider the overall goodness of fit (Equation~\ref{P_total}). To better pinpoint which regions of parameter space contribute to the much lower likelihood in MOND, we find the difference in $\ln P$ between the best-fitting Newtonian and MOND models for each pixel and sum the results across one of the dimensions. Figure~\ref{Ln_P_Newton_MOND_summed} shows the sum of the so-obtained $\Delta \ln P_{\rm{pixel}}$ values for all $\widetilde{v}$ pixels at fixed $r_{\rm{sky}}$ (left panel) and vice versa (right panel). The left panel shows that all $r_{\rm{sky}}$ pixels below 12~kAU show a preference for Newtonian gravity, which is sometimes very strongly preferred. In particular, the range $r_{\rm{sky}} = 5-12$~kAU provides the bulk of the evidence in favour of Newtonian gravity, even though we saw earlier that MOND works slightly better around the peak of the $\widetilde{v}$ distribution at these separations. Beyond 12~kAU, 4/6 of the $r_{\rm{sky}}$ pixels prefer MOND, but the preference is very weak in all cases. The right panel shows that nearly all $\widetilde{v}$ pixels prefer Newtonian gravity. The handful of $\widetilde{v}$ pixels which prefer MOND appear to be randomly distributed. The mild preference for MOND at very high $\widetilde{v}$ could be driven by a deficit of WBs here at very low $r_{\rm{sky}}$ due to our sample selection imposing that $v_{\rm{sky}} < 3$~km/s (Section~\ref{Gaia_DR3_sample}). At our lowest considered $r_{\rm{sky}}$ of 2~kAU and assuming $M = M_\odot$, this limit corresponds to $\widetilde{v} < 4.5$, which would cause only a very mild edge effect considering the very low number of WBs that we might reasonably expect in the affected pixels. This issue is further mitigated in a revised analysis where we use a narrower range of $M$ (Section~\ref{Restricted_mass_range}). We use Figure~\ref{Ln_P_Newton_MOND} to present the 2D version of the results shown here, with a colour scheme used to show all the $\Delta \ln P_{\rm{pixel}}$ values and small white circles to highlight pixels which work better in MOND, as sometimes occurs to a very limited extent.

\subsection{Markov Chain Monte Carlo (MCMC) analysis}
\label{MCMC_analysis}

Our main results are based on running an MCMC analysis starting from the optimal parameters identified by gradient ascent, which we start at $\alpha_{\rm{grav}} = 1/2$ to avoid biasing the results in favour of either gravity theory we are testing.\footnote{Using gradient ascent to initialize an MCMC analysis worked well in \citet{Asencio_2022}, though here we supplement the gradient ascent with a line ascent stage where only one parameter is varied at a time, with the gravity law being the last parameter to be optimized prior to the MCMC due to its importance.} MCMC is a standard statistical method that generates a sequence of parameter values whose frequency distribution matches the posterior inference on the model parameters. Starting from some initial guess for the parameters with likelihood $P$, the protocol calls for generating a proposal by adding Gaussian random perturbations to the parameters. The revised parameters lead to a model with likelihood $P_{\rm{next}}$. We follow the Metropolis-Hastings approach to MCMC in which the proposal is accepted if $P_{\rm{next}} > P$, but if not, then it is accepted with a likelihood of only $P_{\rm{next}}/P$ by using a random number generator. Every time the proposal is accepted, the parameter perturbations are applied and $P$ is updated. When the proposal is rejected, the previous parameters must be recorded again. For best results, the acceptance fraction should be close to 23.4\% \citep*{Gelman_1997}. We ensure it is always in the range $21\%-26\%$, which should ensure a well-mixed chain. This is achieved by setting the perturbation on each parameter to have a Gaussian dispersion similar to the estimated uncertainty, which we guess using our gradient ascent algorithm based on the curvature of $\ln P$ with respect to that parameter in the supplementary stage when all the other parameters are fixed (Section~\ref{Gradient_ascent_fixed_gravity}). We then ran a reduced length MCMC chain to estimate the parameter uncertainties more precisely and set the Gaussian dispersion of each parameter to its estimated uncertainty scaled by a common factor close to 2/3. This factor cannot be varied inside an MCMC chain, so we ran a few reduced length chains to find out that this is the appropriate factor to use in our nominal analysis (slightly different factors were needed in the variations discussed in Section~\ref{Analysis_variants}).\footnote{An alternative approach is to let the factor vary dynamically and then use the optimal value in a full length chain.} All the MCMC analyses presented in this contribution use $10^5$ trials.

\begin{figure*}
    \centering
    \includegraphics[width=0.95\textwidth]{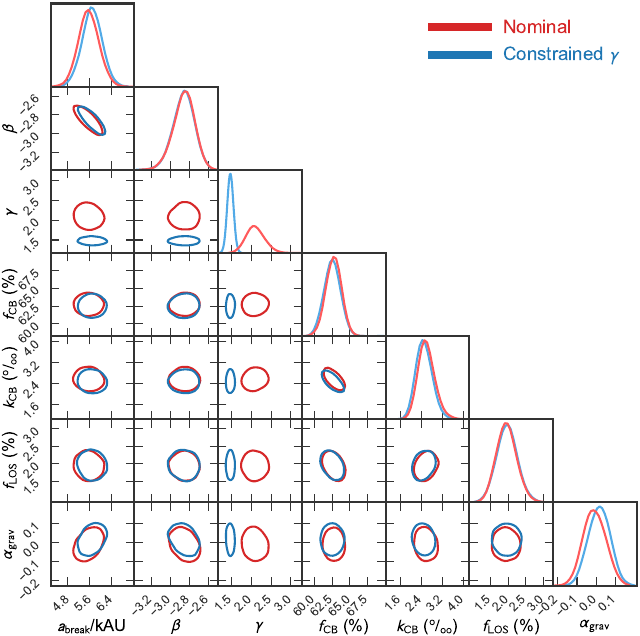}
    \caption{Triangle plot showing the posterior inference on each model parameter (top panels in each column) and the 68\% confidence region for every parameter pair (other panels). The red lines show our nominal analysis with a wide uninformative prior on $\gamma$ (Equation~\ref{gamma_definition}), while the blue lines impose a prior of $\gamma = 1.32 \pm 0.09$ based on the angle between the sky-projected separation and relative velocity of each WB \citep{Hwang_2022}. Notice how this constraint has little effect on the other model parameters, including especially the gravity law, which is clearly Newtonian (bottom right panel). All the triangle plots in this contribution were prepared using \textsc{pygtc} \citep{Bocquet_2016} and show the nominal analysis with solid red lines.}
    \label{Triangle_nominal_gamma_Hwang}
\end{figure*}

Figure~\ref{Triangle_nominal_gamma_Hwang} shows the result of our MCMC analysis as a `triangle plot' \citep{Bocquet_2016}, with the result for our nominal assumptions shown in red in all such plots. This assumes a uniform prior on the eccentricity index $\gamma$ (Equation~\ref{gamma_definition}) over the range $0-4$. The blue lines show the results assuming instead a Gaussian prior of $\gamma = 1.32 \pm 0.09$ \citep{Hwang_2022}. The top panel in each column shows the inference on just one model parameter marginalized over all others. The other panels show the inference on every pair of parameters, which clarifies whether there are any correlations in the uncertainties. There are only two noticeable cases of correlated errors, both of which are anti-correlations. The first of these is between the parameters governing the semi-major axis distribution of the WBs (Equation~\ref{a_cases}). This is similar to the correlation that can arise between the slope and intercept of a linear regression if the data are clustered about a point offset from the origin \citep[a similar correlation is evident between the parameters $r_{\rm{core}}$ and $\rm{Slope}_{P_r}$ in figure~10 of][]{Asencio_2022}. The other correlation evident from Figure~\ref{Triangle_nominal_gamma_Hwang} is between $f_{\rm{CB}}$ and $k_{\rm{CB}}$, which are the only two parameters that relate solely to the CB population. Reducing $k_{\rm{CB}}$ causes the CBs to orbit faster and have a larger impact on $\widetilde{v}$, spreading out the CB tail over a larger range in $\widetilde{v}$ and thus reducing its amplitude. To match the observed amplitude of the extended tail to the $\widetilde{v}$ distribution in the regime where neither genuine WB orbital motion nor LOS contamination should be significant, the algorithm is forced to increase the prevalence of CB companions.

Applying a restrictive prior on $\gamma$ has barely any effect on the other parameter inferences, though of course the inference on $\gamma$ is considerably tighter in this case. We see that in our nominal analysis where $\gamma$ is allowed to vary freely, it prefers a somewhat higher value than 1.32. This could indicate that the WB orbital eccentricity distribution is more super-thermal than reported by \citet{Hwang_2022}. We argued previously that this is quite possible given that CB and LOS contamination were not considered in their analysis but would tend to drive the distribution of the angle $\psi$ towards uniform, which is the expected angular distribution if $\gamma = 1$.

The preferred values of the parameters related to the WB semi-major axis distribution ($a_{\rm{break}}$ and $\beta$) are roughly in line with the results of \citet{Andrews_2017}, with the steeper decline at high $a$ likely due to our quality cuts selecting against large separation WBs where it can be harder to get tight constraints on $\widetilde{v}$. The occurrence rate of CBs (related to the parameter $f_{\rm{CB}}$) is slightly higher than the 50\% estimated by \citet{Clarke_2020}. However, stars do appear to have binary companions quite frequently, with observations indicating a fraction exceeding 40\% \citep*{Hartman_2022} and usually only placing a lower limit because the observations are only sensitive to some range of orbital parameters \citep{Riddle_2015}. After correcting for incompleteness, figure~1 of \citet{Offner_2023} suggests that a reasonable $f_{\rm{CB}}$ for our sample might be 50\% if most CB companions are not identified in our faint star catalogue (Section~\ref{Gaia_DR3_sample}) or through our cut on the parameter $\tt{ipd\_frac\_multi\_peak}$ (Section~\ref{ipd_cut}). We will see later that we can reduce our inferred $f_{\rm{CB}}$ from our nominal $63 \pm 1\%$ to about this much with some slight changes to the CB model that have little impact on the other parameters (Section~\ref{a_int_distribution}). The inferred $k_{\rm{CB}}$ of $2.5 \pm 0.3$\textperthousand\ is very much in line with our estimate based on the various selection effects at play (Section~\ref{aint_distribution}), even though our prior merely imposed that $k_{\rm{CB}} = 1-750$\textperthousand. Only a few percent of the WBs in our sample appear to be LOS contamination, which was to be expected from the WB distribution shown in Figure~\ref{Observed_photo_tiled} $-$ though it is also clear that we cannot adequately fit the data without this ingredient.

Since the non-gravitational parameters seem broadly in line with what one might expect, we now turn to the inferred gravity law. This is very consistent with Newtonian ($\alpha_{\rm{grav}} = 0$) and completely rules out MOND ($\alpha_{\rm{grav}} = 1$). A clear Newtonian result is not surprising in light of the model-independent Figure~\ref{Median_vtilde_values}, our discussion in Section~\ref{Gradient_ascent_fixed_gravity}, and the results of \citetalias{Pittordis_2023} using a somewhat different dataset and model for the WB population (comparisons with their study and other attempts at the WBT are discussed in Section~\ref{Other_WBT_results}).

\renewcommand{\arraystretch}{1.2}
\begin{table*}
    \centering
    \caption{The posterior inference on each of our model parameters, showing its mode and $1\sigma$ confidence interval. We provide a reference to the equation or section where the model parameter is defined. The first row with values shows the uniform prior on each parameter for our nominal analysis. Each subsequent row shows the result of an MCMC analysis, beginning with the nominal assumptions. Revised analyses are discussed further in Section~\ref{Analysis_variants}. The last column shows the model with the highest likelihood for each set of assumptions, considering both the MCMC chain and the final result of the gradient ascent used to initialize it.}
    \begin{tabular}{ccccccccc}
    \hline
    Altered & \multicolumn{7}{c}{Model parameter} & Best \\
    assumption & $a_{\rm{break}}$ (kAU) & $\beta$ & $\gamma$ & $f_{\rm{CB}}$ (\%) & $k_{\rm{CB}}$ (\textperthousand) & $f_{\rm{LOS}}$ (\%) & $\alpha_{\rm{grav}}$ & $\ln P$ \\ [2pt] \hline
    Definition & Eq.~\ref{a_cases} & Eq.~\ref{a_cases} & Eq.~\ref{gamma_definition} & Sec.~\ref{CB_WB_convolution} & Sec.~\ref{aint_distribution} & Sec.~\ref{Chance_alignments} & Sec.~\ref{Gravity_law_interpolation} & Sec.~\ref{Binomial_statistics} \\
    Prior & $\left( 1, 15 \right)$ & $\left( -15, -1 \right)$ & $\left( 0, 4 \right)$ & $\left( 0, 99 \right)$ & $\left( 1, 750 \right)$ & $\left( 0, 60 \right)$ & $\left( -2, 3.6 \right)$ & $-$ \\
    Nominal & $5.61^{+0.33}_{-0.42}$ & $-2.84^{+0.09}_{-0.12}$ & $2.04^{+0.27}_{-0.19}$ & $62.81^{+1.05}_{-1.20}$ & $2.52^{+0.32}_{-0.27}$ & $1.94^{+0.29}_{-0.27}$ & $-0.021^{+0.065}_{-0.045}$ & $-1438.98$ \\
    $\gamma = 1.32 \pm 0.09$ & $5.64^{+0.42}_{-0.32}$ & $-2.84^{+0.09}_{-0.13}$ & $1.47^{+0.08}_{-0.07}$ & $62.54^{+1.06}_{-1.12}$ & $2.37^{+0.37}_{-0.20}$ & $1.99^{+0.26}_{-0.31}$ & $0.024^{+0.042}_{-0.054}$ & $-1445.58$ \\
    $M = \left( 1 - 2 \right) M_\odot$ & $5.78^{+0.47}_{-0.32}$ & $-3.07^{+0.13}_{-0.14}$ & $1.86^{+0.30}_{-0.18}$ & $63.52^{+1.01}_{-1.45}$ & $2.43^{+0.26}_{-0.32}$ & $1.75^{+0.35}_{-0.31}$ & $-0.038^{+0.076}_{-0.052}$ & $-1329.82$ \\
    $M_\star > 0.5 \, M_\odot$ & $7.13^{+0.92}_{-0.79}$ & $-2.63^{+0.14}_{-0.24}$ & $1.49^{+0.36}_{-0.09}$ & $61.82^{+1.25}_{-1.27}$ & $3.93^{+0.62}_{-0.55}$ & $1.82^{+0.49}_{-0.33}$ & $-0.177^{+0.064}_{-0.081}$ & $-1263.68$ \\
    Flat $q$ for CBs & $5.53^{+0.37}_{-0.39}$ & $-2.84^{+0.11}_{-0.12}$ & $2.15^{+0.25}_{-0.25}$ & $62.94^{+0.98}_{-0.99}$ & $2.43^{+0.28}_{-0.29}$ & $2.00^{+0.29}_{-0.30}$ & $-0.033^{+0.074}_{-0.056}$ & $-1441.28$ \\
    Linear $q$ for CBs & $5.66^{+0.28}_{-0.47}$ & $-2.88^{+0.13}_{-0.07}$ & $2.06^{+0.24}_{-0.21}$ & $62.78^{+1.07}_{-1.10}$ & $2.47^{+0.27}_{-0.32}$ & $1.99^{+0.22}_{-0.37}$ & $-0.006^{+0.057}_{-0.056}$ & $-1441.18$ \\
    $k_{\rm{CB}} = 0.2$ & $3.84^{+0.22}_{-0.30}$ & $-2.73^{+0.07}_{-0.09}$ & $4.00^{+0.00}_{-0.14}$ & $67.08^{+1.52}_{-1.62}$ & 200 & $9.38^{+0.48}_{-0.36}$ & $-0.271^{+0.071}_{-0.090}$ & $-1994.27$ \\
    Tri 1.5~dex & $5.51^{+0.37}_{-0.37}$ & $-2.83^{+0.07}_{-0.12}$ & $2.12^{+0.24}_{-0.20}$ & $59.51^{+1.34}_{-1.21}$ & $3.40^{+0.45}_{-0.52}$ & $1.94^{+0.32}_{-0.28}$ & $-0.016^{+0.049}_{-0.061}$ & $-1434.16$ \\
    Flat 1~dex & $5.60^{+0.37}_{-0.35}$ & $-2.87^{+0.11}_{-0.10}$ & $1.99^{+0.19}_{-0.24}$ & $56.51^{+1.17}_{-1.01}$ & $2.05^{+0.26}_{-0.18}$ & $1.95^{+0.30}_{-0.28}$ & $0.017^{+0.041}_{-0.066}$ & $-1443.14$ \\
    $f_{\rm{CB}} = 0.3$ & $5.65^{+0.27}_{-0.38}$ & $-2.97^{+0.09}_{-0.13}$ & $1.05^{+0.09}_{-0.14}$ & 30 & $18.7^{+1.8}_{-0.9}$ & $7.33^{+0.47}_{-0.31}$ & $0.149^{+0.035}_{-0.048}$ & $-1852.05$ \\ \hline
    \end{tabular}
    \label{MCMC_inferences}
\end{table*}
\renewcommand{\arraystretch}{1}

Our prior for each model parameter is shown in Table~\ref{MCMC_inferences} along with the mode and $1\sigma$ confidence region of its posterior inference in all of our MCMC analyses (different rows). The alterations to the modelling assumptions are discussed in Section~\ref{Analysis_variants}. The last column shows the log-likelihood of the model where this is maximal, considering both the final result of the gradient ascent and the whole MCMC chain. Only the analysis with $f_{\rm{CB}}$ fixed at a rather low value prefers $\alpha_{\rm{grav}} > 0$ at any reasonable significance. However, this model has an extremely low log-likelihood compared to our nominal assumptions, suggesting a $29\sigma$ preference for the latter (the very severe problems with this model are discussed further in Appendix~\ref{Best_fit_fCB03}). Thus, it is not realistically possible for the WBT to give a Milgromian result. We explain later why constraints from galaxy rotation curves preclude altering the MOND interpolating function so as to evade our constraint from the WBT (Section~\ref{MOND_interpolating_function}).

\section{Discussion}
\label{Discussion}

Since the main goal of this contribution is to distinguish Newtonian gravity from MOND, our \emph{a priori} expectation was that $\alpha_{\rm{grav}}$ should be 0 or 1. Our results strongly prefer 0. If local WBs are actually Milgromian, there would have to be a confluence of observational errors and modelling deficiencies that shift $\alpha_{\rm{grav}}$ from a true value of $\approx 1$ down to almost exactly 0. Measurement errors on $\widetilde{v}$ are very small in our WB sample (Figure~\ref{v_tilde_error_distribution}) compared to the intrinsic dispersion of $\approx 0.5$ (Figure~\ref{Photo_fixed_gravity}). Moreover, uncertainties would tend to broaden the distribution, with the broadening occurring preferentially at high $r_{\rm{sky}}$ because the same velocity uncertainty in m/s translates into a larger uncertainty in $\widetilde{v}$ (Equation~\ref{v_tilde_definition}). Thus, measurement errors would if anything cause a Newtonian WB population to look somewhat Milgromian. This is the opposite of what it would take to reconcile our results with MOND.

The impact of modelling deficiencies is harder to assess, but \citetalias{Pittordis_2023} implemented the WBT with a somewhat different sample selection and model for the WB population. Those authors considered WBs with $r_{\rm{sky}} = 5-20$~kAU rather than our adopted range of $2 -30$~kAU, leading to a smaller proportion of systems deep into the Newtonian regime. To get a better handle on the CB properties, they considered systems with $\widetilde{v} < 7$ rather than $\widetilde{v} < 5$. In addition, they used a linear mass-luminosity relation rather than our cubic fit over the range most relevant to the WBT (Figure~\ref{Mass_luminosity_relation}). Despite these and other differences, they also found a very strong preference for Newtonian gravity over MOND, with the Newtonian model fitting the observed number counts rather well \citepalias[figure~12 of][]{Pittordis_2023}. Their analysis with a thermal eccentricity distribution ($\gamma = 1$) provides the best fit in both gravity theories and is also closest to the somewhat superthermal distribution preferred by our fits. It also better matches the distribution of the angle between the sky-projected separation and relative velocity of each WB \citep{Hwang_2022}. For the case $\gamma = 1$, the analysis of \citetalias{Pittordis_2023} prefers Newtonian gravity over MOND at about $23\sigma$ confidence. This is similar to our result in Section~\ref{Gradient_ascent_fixed_gravity} that the best Newtonian model is preferred over the best MOND model at $19\sigma$ confidence.

An important reason for the similar results is that the gravity law is mostly sensitive to the main peak region of the $\widetilde{v}$ distribution, so it is not necessary to have a very accurate model for the extended tail at $\widetilde{v} \ga 2$. This does not of course mean that we can completely neglect CB and LOS contamination, but it does mean that the manner in which these effects are included should have little effect on the inferred gravity law. Even differences in the eccentricity distribution have a less significant effect on the overall $\widetilde{v}$ distribution than whether gravity is Newtonian or Milgromian \citepalias[see figure~3 of][]{Banik_2018_Centauri}. This is also evident in Figure~\ref{Triangle_nominal_gamma_Hwang}, which shows that forcing $\gamma$ to a lower value than preferred in our nominal analysis only marginally affects the inferred $\alpha_{\rm{grav}}$ and its uncertainty. There is almost no correlation between these parameters in their joint inference.

We have assumed that $\gamma$ is independent of other parameters like the WB semi-major axis. A correlation here can masquerade as a correlation with the WB orbital acceleration because $g_{_N}$ depends more sensitively on the separation than it does on $M$ and the dynamic range in $r_{\rm{sky}}$ is larger than in the mass. It is possible that WBs with smaller separations mostly formed together, while more widely separated WBs formed via capture of field stars. This could lead to WBs with larger $a$ being on typically more eccentric orbits (higher $\gamma$ in Equation~\ref{gamma_definition}), which would reduce the median $\widetilde{v}$ because the WB would spend more time near apocentre \citepalias[see figure~3 of][]{Banik_2018_Centauri}. In principle, this could counteract the MOND prediction that WBs in the asymptotic regime have enhanced $\widetilde{v}$ due to the predicted enhancement to the gravity binding the WB. However, it seems unlikely that the cancellation would be sufficiently precise to yield the observed very flat trend of the median $\widetilde{v}$ with respect to $r_{\rm{sky}}/r_{_M}$ (Figure~\ref{Median_vtilde_values}).

\begin{figure}
    \centering
    \includegraphics[width=0.47\textwidth]{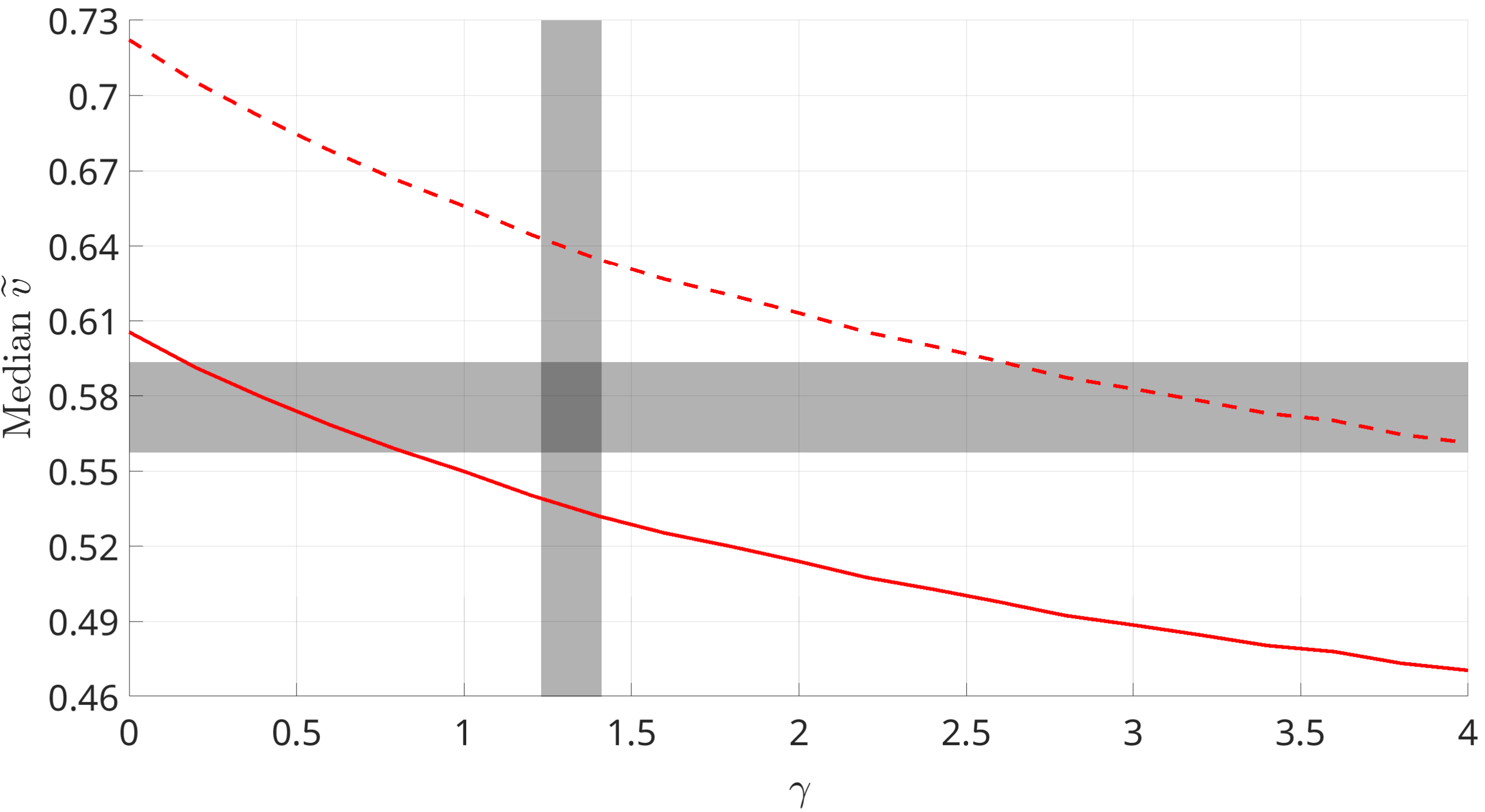}
    \caption{The median $\widetilde{v}$ for an ensemble of Keplerian orbits with a range of eccentricities viewed from a random viewing angle at a random orbital phase (solid red line). This result is scaled up by $\sqrt{\eta}$ to get the MOND prediction for local WBs with asymptotically large separation (dashed red line). The eccentricity distribution is governed by $\gamma$ (Equation~\ref{gamma_definition}). The vertical grey shaded band shows $\gamma = 1.32 \pm 0.09$ based on observations of WBs with separations similar to those we use for the WBT \citep[Section~\ref{WB_orbital_parameter_distribution}; see also][]{Hwang_2022}. The horizontal grey shaded band shows the median $\widetilde{v}$ of the WB population, with the range giving the uncertainty arising from whether we truncate the $\widetilde{v}$ distribution at 1.5, 2, or 2.5 when finding the median. The WBs are not further binned in $r_{\rm{sky}}/r_{_M}$ as this has little impact on the median $\widetilde{v}$ (Figure~\ref{Median_vtilde_values}).}
    \label{gamma_impact}
\end{figure}

To further explore this quite contrived possibility, we compute the $\widetilde{v}$ distribution for a grid of Keplerian orbits with different eccentricities. For each orbit, we consider all possible orbital phases and viewing angles.\footnote{The computational costs are reduced by only considering the period between pericentre and apocentre.} We then assume some eccentricity distribution governed by the parameter $\gamma$ (Equation~\ref{gamma_definition}). The median of the resulting $\widetilde{v}$ distribution depends on the assumed $\gamma$, as shown by the solid red line on Figure~\ref{gamma_impact}. The dashed red line shows this result rescaled by $\sqrt{\eta}$, which is what we would expect in MOND for local WBs with $r_{\rm{sky}} \gg r_{_M}$. The horizontal grey band shows the median $\widetilde{v}$ of the WB population, which is not that high (see also Figure~\ref{Median_vtilde_values}). While agreement with MOND is possible at very high $\gamma$, we must bear in mind that $\gamma$ affects the distribution of the angle $\psi$ between $\bm{r}_{\rm{rel}}$ and $\bm{v}_{\rm{rel}}$. This has been used to argue that $\gamma = 1.32 \pm 0.09$ for Newtonian WBs with separations of only a few kAU \citep[see figure~7 of][]{Hwang_2022}. We show this using the vertical grey band. The observed median $\widetilde{v}$ can be reconciled with MOND only for $\gamma \ga 2.5$, which would significantly affect the observed distribution of $\psi$ \citep[see figure~2 of][]{Hwang_2022}. Using this constraint, their figure~7 shows that there is no tendency for WBs at larger separations to prefer systematically higher $\gamma$ (the last three bins are most relevant for the WBT). Their figure~10 also shows that for a slightly superthermal eccentricity distribution ($\gamma$ slightly above 1), their method of determining $\gamma$ would preferentially overestimate it at large separations, making it even less likely that $\gamma \gg 1$ in this regime. Moreover, the actual distribution of $\psi$ in their figure~6 appears fairly symmetric with respect to $\psi \to \mathrm{\pi} - \psi$ for WBs with $r_{\rm{sky}} > 1$~kAU, providing an important sanity check. The observed approximate symmetry between systems heading towards and away from each other makes it unlikely that our results are contaminated much by recently ionized WBs that are currently dispersing. This is not too surprising given that unbound systems are expected to disperse quickly and that disruptive encounters with perturbers like passing stars would be rather infrequent \citepalias[see section~8.1 of][]{Banik_2018_Centauri}. It is therefore extremely unlikely that a rapidly rising trend in $\gamma$ with $a$ can explain the very clear Newtonian result obtained by our analysis and that of \citetalias{Pittordis_2023}. In a Newtonian context, our results suggest that CBs have slightly inflated the median $\widetilde{v}$ compared to the uncontaminated case shown in Figure~\ref{gamma_impact}. Since CBs introduce a hidden degree of freedom that pushes the $\psi$ distribution closer to uniform, they can also cause $\gamma$ to be somewhat underestimated in analyses like that of \citet{Hwang_2022} that do not account for CBs, perhaps explaining why our main analysis infers $\gamma \approx 2$ instead of their $1.32 \pm 0.09$. CBs can therefore allow a somewhat elevated $\gamma$ and $\widetilde{v}$ at the same time. Applying this argument to MOND reveals a major difficulty because it requires $\gamma \ga 3$, which can be reconciled with the \citet{Hwang_2022} result only with significant CB contamination, but the low observed median $\widetilde{v}$ leaves very little margin for inflation by CBs.

Another possible systematic effect is that WBs with larger separations could preferentially be younger because they are more fragile to disruption, which could systematically affect the mass-luminosity relation. However, it has been estimated that the half-life $t_{1/2}$ of a WB in the Solar neighbourhood is comparable to the 10~Gyr age of the Galactic disc \citep*{Knox_1999} only when $a = 31$~kAU \citep[equation~1 of][]{Jiang_2010}. Survival rates of WBs would be reduced modestly in MOND because although their binding energy is enhanced by the gravity boost factor $\eta$, the impulse from a passing star is also enhanced by the same factor, so the energy gained per interaction would be enhanced by a factor of $\eta^2$ if we treat impulses from different sources as contributing to a diffusive random walk \citep[heating rates from impulsive encounters in MOND are discussed extensively in section~3.3.2 of][]{Asencio_2022}. This would reduce $t_{1/2}$ by a factor of $\eta$, but since $t_{1/2} \propto a^{-2}$ \citep*[equation~4 of][]{Bahcall_1985}, we can maintain the value of $t_{1/2}$ by reducing $a$. This would entail reducing the above estimate of 31~kAU by a factor of $\sqrt{\eta}$, which would reduce it to 26~kAU. Even taking this into account, there should only be a mild tendency for WBs to preferentially be younger at larger separations when $r_{\rm{sky}} < 10$~kAU, which is the range that mostly contributes to the WBT preferring Newtonian gravity over MOND (left panel of Figure~\ref{Ln_P_Newton_MOND_summed}). WBs in the Solar neighbourhood must be fairly robust given the large number identified within 250~pc \citepalias{Pittordis_2023} and the larger number within 1~kpc \citep{Badry_2021}. In fact, the nearest star to the Sun is in a WB \citep{Kervella_2017}, with Monte Carlo simulations of this system indicating a 74\% survival probability over 5~Gyr \citep{Feng_2018}.

Even if there is some correlation between the separation of a WB and the ages of its stars, it is unclear how that would influence the estimated mass. Main sequence stars increase their brightness over time as their cores become denser and hotter due to the fusion of hydrogen into helium, so younger stars would tend to be less luminous \citep{Hejlesen_1980}. This would cause their mass to be underestimated and their $\widetilde{v}$ to be overestimated (Equation~\ref{v_tilde_definition}). If this occurs more commonly at large separations, it would enhance rather than hide a MOND signal in the data. However, stars that are not yet on the main sequence become fainter as they settle onto it, so younger stars would be brighter at this evolutionary stage \citep{Hayashi_1961a, Hayashi_1961b}. Depending on the time required to settle onto the main sequence, it could be that the effect is important at the low mass end, though we note that most of the stars in our WB sample have $M_\star > 0.3 \, M_\odot$ (blue bars in Figure~\ref{Mass_distribution}). We further mitigate this issue by explicitly considering the impact of removing all WBs where either star has $M_\star < 0.5 \, M_\odot$ (Section~\ref{Restricted_mass_range}).

We estimate the mass of each star from its apparent $G$-band magnitude and trigonometric parallax. While these should both be reliable, we also need to assume a relation between the absolute $G$-band magnitude and mass. For this, we fit a polynomial which is easily inverted, thereby smoothing over small-scale features (Section~\ref{ML_relation}). One such feature is the Jao gap \citep{Jao_2018}, which arises because stars with $M_\star \la 0.35 \, M_\odot$ have fully convective cores \citep{Chabrier_2000}. This leads to an unstable range of luminosities marked by the convective kissing instability \citep{Santana_2021}. However, the gap has a width of only 0.05 magnitudes \citep[section~2 of][]{Jao_2018}. This has a discernible impact on the luminosity function of stars \citep[figure~17 of][]{Gaia_2021}, but the impact on the mass-luminosity relation is very small because 0.05 magnitudes corresponds to only a 1\% difference in mass (Equation~\ref{Linear_mass_luminosity_relation}) and thus a 0.5\% difference in $\widetilde{v}$ (Equation~\ref{v_tilde_definition}). A more significant issue might be the broad hump evident in the mass-luminosity relation at $M_\star \approx 0.5 \, M_\odot$ \citep*{Kroupa_1993, Reid_2002}, which is probably the same feature evident at $M_G \approx 10$ in figure~16 of \citet{Gaia_2021}. While the mass-luminosity relation here is important to our analysis (blue bars in Figure~\ref{Mass_distribution}), a smooth polynomial fit to the $V$-band mass-luminosity relation only deviates from the data points by a few tenths of a magnitude \citep[figure~11 of][]{Reid_2002}. With accurate data, gaps in the mass-luminosity relation or sharp changes in its gradient can lead to sharp features in the distribution of absolute magnitudes (as with the Jao gap), but the steep mass-luminosity relation mitigates the impact on the estimated $M_\star$. Moreover, a polynomial fit to the mass-luminosity relation should not deviate systematically from the actual relation: if the fit overestimates the mass at some $M_G$, then it should underestimate the mass at a slightly different $M_G$. In any case, such modest deviations from our multi-part polynomial fit should be much less substantial than the difference between the cubic fit we use over most of the relevant range and the linear fit adopted by \citetalias{Pittordis_2023} in their equation~3. Since there is a very strong preference for Newtonian gravity over MOND in both studies and these even give a similar statistical significance (Section~\ref{Gradient_ascent_fixed_gravity}), it is unlikely that this preference is caused by issues with the assumed mass-luminosity relation, which moreover is calibrated with spectroscopic mass estimates from FLAME (Section~\ref{FLAME_calibration}).

Another important aspect of the WBT is the presence of undetected CB companions, which broaden the tail of the $\widetilde{v}$ distribution (Figure~\ref{Photo_buildup_MOND}). We have assumed that every star in our WB sample is equally likely to have such a companion. However, it is possible that $f_{\rm{CB}}$ depends on properties of the star and even its WB companion. For this to affect the WBT, $f_{\rm{CB}}$ would need to depend on the WB acceleration, which mostly depends on its separation. Counteracting the effect of MOND would require more widely separated WBs to have a smaller $f_{\rm{CB}}$, which could hold down the median $\widetilde{v}$ in the main peak region and hide the MOND signal. This seems unlikely because the CBs relevant to our analysis have a much smaller separation than the WB (Section~\ref{aint_distribution}), something that is also clear from the fact that our inferred $k_{\rm{CB}} \ll 1$. We would therefore expect $f_{\rm{CB}}$ to not depend very much on the separation of the WB \citep*[as also suggested by spectroscopic searches for binaries;][]{Tokovinin_2002, Tokovinin_2010}.

To test this expectation, we need a proxy for $f_{\rm{CB}}$. We develop one based on the fact that CBs affect the tail of the $\widetilde{v}$ distribution. Even in MOND gravity, we expect WBs to very rarely have $\widetilde{v} > 1.5$ in the absence of contaminating effects (Figure~\ref{Photo_buildup_MOND}). This motivates us to use the proportion of WBs with higher $\widetilde{v}$ as a measure of contamination. While this arises from both CBs and chance alignments, Figure~\ref{Observed_photo_tiled} shows that the latter is not very important if we restrict to $\widetilde{v} \la 2.5$. This is because LOS contamination becomes more important at high $\widetilde{v}$ (Equation~\ref{LOS_contamination_pattern}), opposite the situation for CB contamination \citepalias{Pittordis_2019}. We therefore argue that the incidence of CBs can be gauged by finding the likelihood $P_{\rm{tail}}$ that a WB with $\widetilde{v} < 2.5$ also has $\widetilde{v} > 1.5$.
\begin{eqnarray}
    P_{\rm{tail}} ~\equiv~ \left. P \left(\widetilde{v} > 1.5 \right| \widetilde{v} < 2.5 \right) .
    \label{P_tail_definition}
\end{eqnarray}
While $P_{\rm{tail}}$ cannot be equated with $f_{\rm{CB}}$, a higher incidence of CBs would increase $P_{\rm{tail}}$, making it a useful proxy for the prevalence of CB contamination. We therefore quantify $P_{\rm{tail}}$ and its uncertainty using binomial statistics. Assuming a uniform prior on $P_{\rm{tail}}$, its posterior distribution is characterized by \citep[equation~32 of][]{Asencio_2022}:
\begin{eqnarray}
    \label{Binomial_P_uncertainty}
    \textrm{mean} \!\!\! &=& \!\!\! \frac{N_{>1.5} + 1}{N_{<2.5} + 2} \, , \\
    \begin{matrix}
        \textrm{standard} \\
        \textrm{deviation}
     \end{matrix} \!\!\! &=& \!\!\! \frac{1}{N_{<2.5} + 2} \sqrt{\frac{ \left( N_{>1.5} + 1 \right) \left( N_{<2.5} - N_{>1.5} + 1 \right)}{N_{<2.5} + 3}} \, , \nonumber
\end{eqnarray}
where $N_{<2.5}$ is the number of WBs with $\widetilde{v} < 2.5$, of which $N_{>1.5}$ WBs also have $\widetilde{v} > 1.5$. If $N_{>1.5} = N_{<2.5} = 0$, the posterior distribution of $P_{\rm{tail}}$ is uniform over the range $0-1$ and we obtain the expected result for its mean and standard deviation.

\begin{figure}
    \centering
    \includegraphics[width=0.47\textwidth]{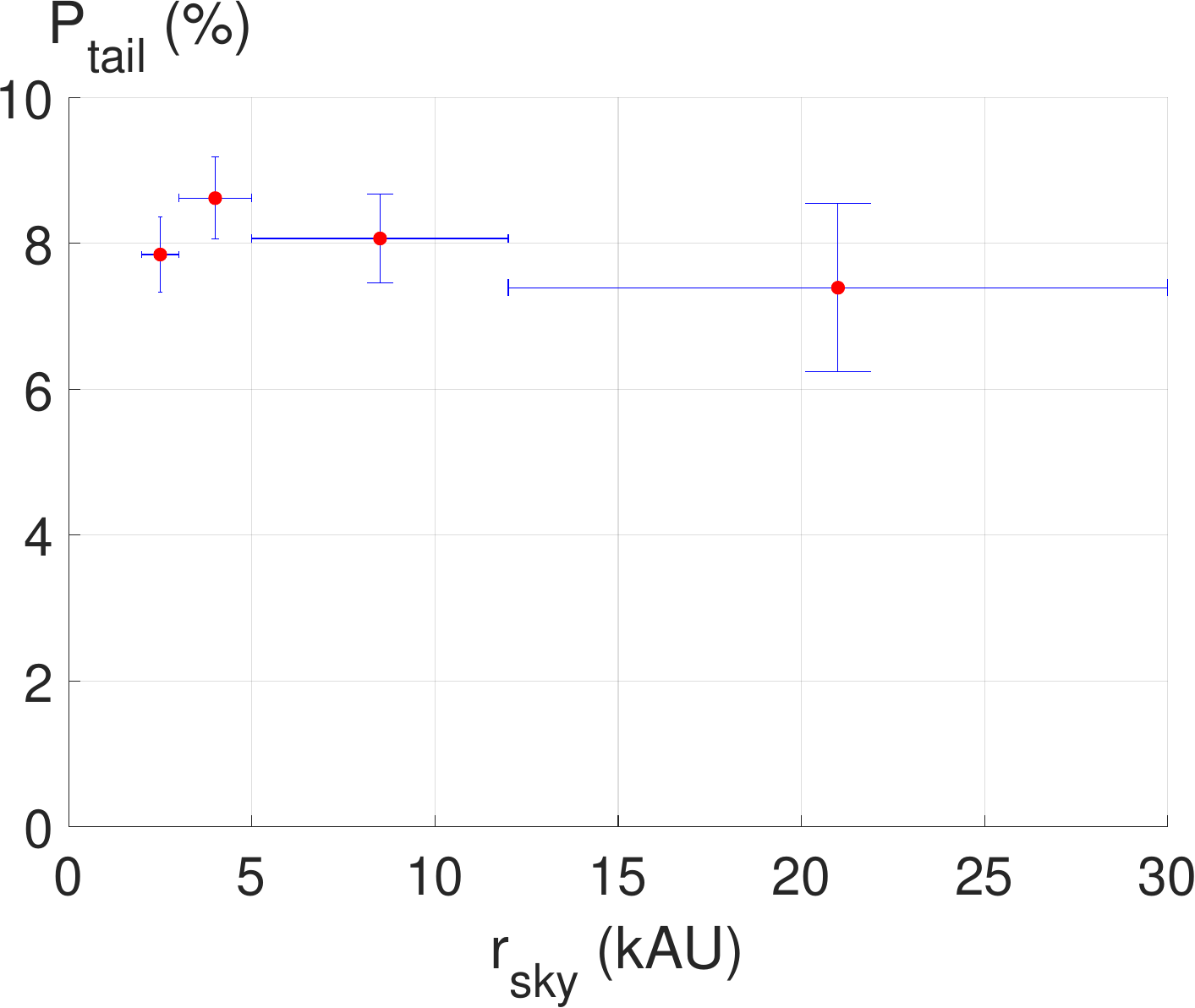}
    \caption{The impact of $r_{\rm{sky}}$ on $P_{\rm{tail}}$ (Equation~\ref{P_tail_definition}), which we argue in the text should correlate with the CB fraction and thus serves as a useful proxy for the latter. $P_{\rm{tail}}$ is shown for WBs in the same four $r_{\rm{sky}}$ bins as those used in Figure~\ref{Photo_fixed_gravity}. Error bars in $P_{\rm{tail}}$ show binomial uncertainties (Equation~\ref{Binomial_P_uncertainty}), while error bars in $r_{\rm{sky}}$ show the width of each bin.}
    \label{P_tail}
\end{figure}

To check if $P_{\rm{tail}}$ depends on the WB separation, we determine $P_{\rm{tail}}$ for the WBs in four $r_{\rm{sky}}$ intervals (the same as those used in Figure~\ref{Photo_fixed_gravity}). The result is shown in Figure~\ref{P_tail}. There is no discernible trend in $P_{\rm{tail}}$: a value of 8\% can adequately fit all four data points despite the rather small binomial uncertainties. This makes it very unlikely that our analysis mistakenly prefers Newtonian gravity due to trends in the likelihood of a star having a CB companion.

We test the robustness of our falsification of MOND by varying some of our modelling assumptions (Section~\ref{Analysis_variants}) and considering WBT results from other authors (Section~\ref{Other_WBT_results}). We then consider if uncertainties in the MOND interpolating function can plausibly allow WBs in the asymptotic regime to have orbital velocities that deviate only a few percent from the Newtonian prediction, as indicated by the WBT (Section~\ref{MOND_interpolating_function}). We will see that it is not possible to reconcile MOND with the WBT through plausible variations to the modelling assumptions or the interpolating function. We go on to discuss the broader implications of this result, focusing on whether the WB data and the tight observed RAR \citep{Li_2018, Desmond_2023} can be explained by further adjustments to MOND or by a completely new theory, bearing in mind other astrophysical constraints like Solar System ephemerides and the RAR in galaxy clusters (Section~\ref{Broader_implications}). A Newtonian result from the WBT causes severe difficulties for any theory which approximately reduces to the MOND field equation on kAU scales in the Solar neighbourhood, forcing us to consider if some other approach might be necessary.

\subsection{Variations to the nominal analysis}
\label{Analysis_variants}

In this section, we vary some of the modelling assumptions in our nominal analysis. The posterior inferences on the model parameters are summarized in Table~\ref{MCMC_inferences}.

\subsubsection{Restricted mass range for the WB stars}
\label{Restricted_mass_range}

Our model assumes that several quantities like the CB contamination fraction $f_{\rm{CB}}$ are independent of the WB mass $M$. It could be that some of these quantities do in fact depend on $M$. Moreover, higher mass stars are more luminous, allowing us to detect them out to a greater distance given the requirement that the apparent magnitude $m_G < 17$ (Section~\ref{Gaia_DR3_sample}). Since the total mass of a WB affects its MOND radius and thus the predicted enhancement to its Newtonian gravity, it is possible that our predicted $\widetilde{v}$ distribution in different $r_{\rm{sky}}$ bins is somewhat inaccurate due to using the overall distribution of $M$ as a prior in all $r_{\rm{sky}}$ bins.

\begin{figure*}
    \centering
    \includegraphics[width=0.48\textwidth]{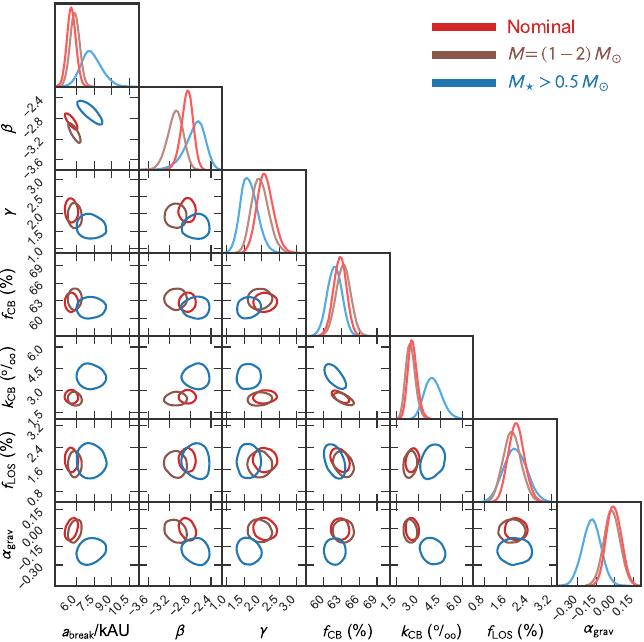}
    \hfill
    \includegraphics[width=0.48\textwidth]{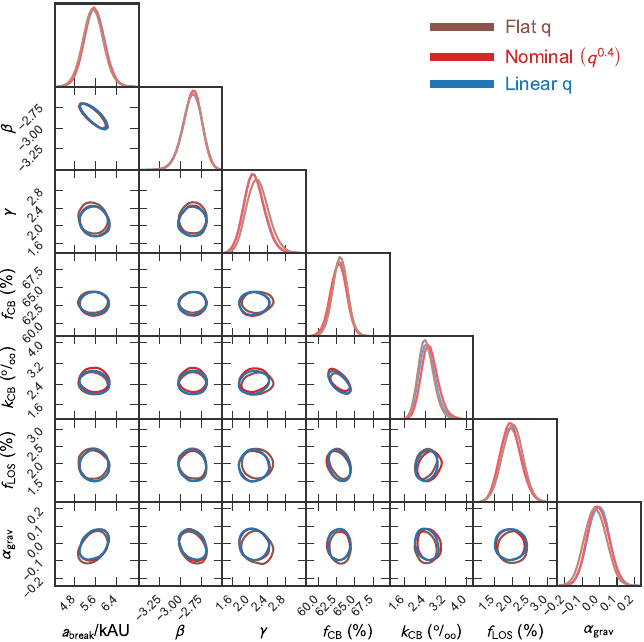}
    \includegraphics[width=0.48\textwidth]{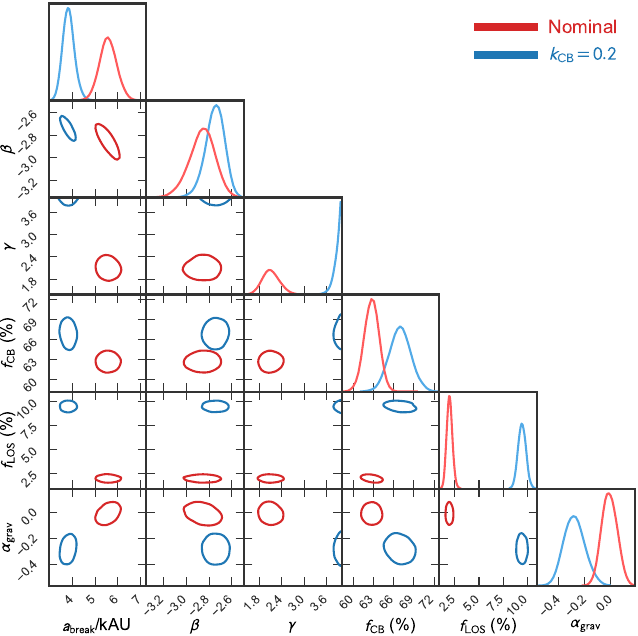}
    \hfill
    \includegraphics[width=0.48\textwidth]{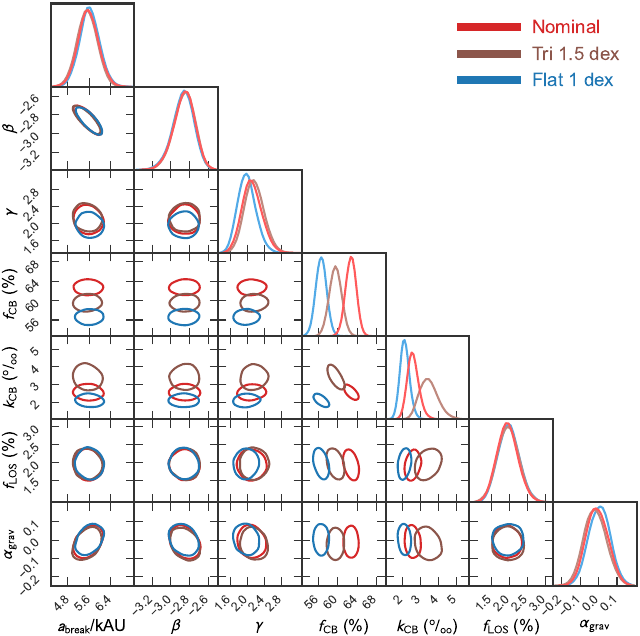}
    \caption{Triangle plot similar to Figure~\ref{Triangle_nominal_gamma_Hwang}, but now comparing the nominal analysis (solid red lines in all panels) to revised analyses where the total mass of each WB is restricted to the range $1.0 - 2.0 \, M_\odot$ or the individual stars in each WB are required to have $M_{\star} > 0.5 \, M_\odot$ (\emph{top left}; Section~\ref{Restricted_mass_range}), the CB mass ratio distribution is altered to bracket the uncertainties as illustrated in Figure~\ref{q_distribution_comparison} (\emph{top right}; Section~\ref{CB_mass_ratio_distribution}), the maximum semi-major axis of the CBs is fixed at 20\% of the WB semi-major axis (\emph{bottom left}; Section~\ref{a_int_distribution}), and the logarithm of the CB semi-major axis is assumed to have a triangular distribution of width 1.5~dex or a flat distribution of width 1~dex (\emph{bottom right}; Section~\ref{a_int_distribution}). The posterior inferences on the model parameters in these revised analyses are compared with the nominal analysis in Table~\ref{MCMC_inferences}.}
    \label{Triangles_revised}
\end{figure*}

While it is not feasible to use different $M$ distributions in different $r_{\rm{sky}}$ bins due to the increased complexity and computational cost, we can use a narrower mass range to mitigate modelling deficiencies of this kind. Since this reduces the sample size, we roughly centre our restricted mass range on the mode of our WB sample's mass distribution (red bars in Figure~\ref{Mass_distribution}). This leads us to consider only WBs with a mass of $1.0 - 2.0 \, M_\odot$. Since the impact of CBs depends on the distribution of $M_\star$ for the stars in our sample, we need to recalculate the CB velocity perturbations with revised $\Delta \widetilde{M}$ and $f_{\rm{pb}}$ (Equations~\ref{dM_tilde_definition} and \ref{f_pb_definition}, respectively). We also adjust the distribution of $M$ used to marginalize our WB orbit library and rerun our analysis in Appendix~\ref{Equal_mass_binaries}, which now gives $P_{\rm{eqm}} = 0.03$. The results are shown in brown in the top left panel of Figure~\ref{Triangles_revised}. The preferred gravity law is now even further from MOND, though it remains very consistent with the Newtonian prediction and differs only modestly from our nominal analysis. Shifts to the other parameter inferences are also small. Since the restricted mass range substantially reduces the scope for changes to the mass distribution across the parameter range used for the WBT, our results appear to be robust in this respect.

The WBT relies on having an accurate estimate of $M_\star$ for every star in our WB sample. Our $M_\star$ estimates are based on the mass-luminosity relation from \citet{Pecaut_2013}, but their accuracy has been improved slightly by calibrating against spectroscopic estimates from the \emph{Gaia} FLAME package (Section~\ref{FLAME_calibration}). This calibration is only possible when $M_\star > 0.5 \, M_\odot$ as there is essentially no data below this. We therefore consider restricting our WB sample to only those systems where both stars have $M_\star > 0.5 \, M_\odot$, which reduces $P_{\rm{eqm}}$ to 0.01. The results are shown in blue in the top left panel of Figure~\ref{Triangles_revised}. Apart from a $1.6\times$ increased $k_{\rm{CB}}$ which we discuss below, the most noteworthy change is a reduction in $\alpha_{\rm{grav}}$ to slightly negative values, causing almost $3\sigma$ tension with the Newtonian prediction ($\alpha_{\rm{grav}} = 0$). This shift in the inferred gravity law could indicate that the higher typical $M_\star$ of the sample increases the lowest $a_{\rm{int}}$ that we need to consider for the CBs. Our estimate for this in Appendix~\ref{Non_linear_orbital_motion} assumes that the $M_\star$ distribution easily extends down to $0.4 \, M_\odot$, which is correct for our nominal sample (blue bars in Figure~\ref{Mass_distribution}) but is no longer true if we require $M_\star > 0.5 \, M_\odot$. A higher mass would increase the separation corresponding to the same orbital period. Since this does not affect our argument in Section~\ref{aint_distribution} about the upper limit to the relevant range of $a_{\rm{int}}$, the dynamic range of $a_{\rm{int}}$ might be less than our nominal assumption of 1.5~dex. Keeping it fixed at 1.5~dex to minimize the changes could then force the analysis to raise the upper limit to $a_{\rm{int}}$ by increasing $k_{\rm{CB}}$, which does indeed happen (see Table~\ref{MCMC_inferences}). We note that in a different extension analysis, reducing the allowed range of $a_{\rm{int}}$ to 1~dex increases the inferred $\alpha_{\rm{grav}}$ by 0.04 (Section~\ref{a_int_distribution}). A similar change to the result of our $M_\star > 0.5 \, M_\odot$ analysis would reduce the discrepancy with Newtonian gravity to about $2\sigma$.


\subsubsection{The CB mass ratio distribution}
\label{CB_mass_ratio_distribution}

We assume that the CB mass ratio distribution $\propto q^{0.4} \left( q \leq 1 \right)$ based on \citet{Korntreff_2012}. We showed in Figure~\ref{q_distribution_comparison} that this provides a good match to our WB sample, especially when bearing in mind that selection effects can make very low $q$ systems hard to detect. Since the CBs are undetected, it would also be reasonable to consider a flat prior on $q$ or a linear prior, both of which seem marginally plausible. Considering these revised priors on $q$ should thus bracket the uncertainty in its distribution. In particular, a flat distribution roughly approximates the three-part power law recommended in figure~7 of \citet{Badry_2019_twin}.

The results of these two revised analyses are shown in the top right panel of Figure~\ref{Triangles_revised}, with the linear prior shown in blue and the flat prior shown in brown. While the actual distribution of $q$ is inevitably more complicated than the power-law forms considered in this study, the posterior inferences of all seven model parameters and the overall goodness of fit barely change in both analysis variants (Table~\ref{MCMC_inferences}). Our results thus seem robust to what exactly we assume here.

\subsubsection{The CB semi-major axis distribution}
\label{a_int_distribution}

The impact of CBs on the WB $\widetilde{v}$ depends on the distribution of $a_{\rm{int}}$, so we consider variations to our assumptions regarding it. We would generally expect any CB to become unstable if $k_{\rm{CB}} \ga 0.3$ \citepalias{Pittordis_2023}, but our analysis prefers a value of only 2.5\textperthousand. We argued that this is to be expected given the various selection effects at play as part of our strategy to minimize CB contamination and model the remaining CBs (Section~\ref{aint_distribution}). In particular, the faint star catalogue (Section~\ref{Gaia_DR3_sample}) and the use of the \emph{Gaia} catalogue parameter $\tt{ipd\_frac\_multi\_peak}$ (Section~\ref{ipd_cut}) should remove CBs based on their outright or marginal detection, respectively.

Even so, it is worth considering what happens if we fix $k_{\rm{CB}} = 0.2$ in Equation~\ref{k_CB_defintion}. The results are shown in the bottom left panel of Figure~\ref{Triangles_revised}, but with the rows and columns for $k_{\rm{CB}}$ omitted as this is now fixed. Compared to our nominal analysis where $k_{\rm{CB}}$ is a free parameter, the fits are pushed towards unrealistically high values of $\gamma$, contradicting the results of \citet{Hwang_2022}. This could be due to the much higher $k_{\rm{CB}}$ reducing the velocity perturbations from CBs and making it harder to fit the extended tail to the $\widetilde{v}$ distribution. This problem can be slightly alleviated by postulating higher orbital eccentricities, a higher fraction of CBs, and a higher fraction of chance-aligned WBs $-$ all of which are evident from the posteriors. The inferred $f_{\rm{LOS}}$ of 9.4\% is particularly high compared to the 1.9\% in our nominal analysis. It is hard to imagine that the small number of WBs towards the upper right of Figure~\ref{Observed_photo_tiled} is consistent with 9\% of the full sample of WBs being chance alignments, which would be less common at lower $\widetilde{v}$ (Equation~\ref{LOS_contamination_pattern}). These issues are reflected in the much poorer overall fit by a factor of $\exp \left( 555 \right)$ (last column in Table~\ref{MCMC_inferences}). Even if we leave aside these very serious issues, the inferred gravity law is further from Milgromian than in our nominal analysis to such an extent that it is now barely consistent with Newtonian, indicating that a higher $k_{\rm{CB}}$ cannot reconcile MOND with the WBT.

We also expect our results to depend somewhat on the width of the $a_{\rm{int}}$ distribution (Equation~\ref{k_CB_defintion}). We have assumed a flat logarithmic distribution for simplicity. We might expect the distribution to decline to zero more gradually at the edges. We therefore also consider a triangular logarithmic distribution in which the mode occurs at the centre of the considered range, i.e. 0.75~dex from either end. This extension analysis is labelled `Tri 1.5~dex' in Table~\ref{MCMC_inferences}, where we see that the fit improves slightly compared to our nominal analysis. The only noticeable shift to the parameter inferences is a slight decrease to $f_{\rm{CB}}$ and corresponding increase to $k_{\rm{CB}}$, roughly parallel to the direction along which their errors are anti-correlated (bottom right panel of Figure~\ref{Triangles_revised}). Importantly, the change to the inferred gravity law is about $10\times$ smaller than its uncertainty.

Our nominal analysis assumes a 1.5~dex width in $a_{\rm{int}}$ for reasons discussed in Section~\ref{aint_distribution}. To check how much this assumption affects our results, we now reduce this to 1~dex while continuing to assume a flat logarithmic distribution. This extension analysis is labelled `Flat 1~dex' in Table~\ref{MCMC_inferences}, where we see that the fit becomes slightly worse than in our nominal analysis. The only parameter shift that is much larger than the random uncertainty is a significant reduction in $f_{\rm{CB}}$, which is reduced from 63\% to 57\%. Despite significantly narrowing the allowed range of $a_{\rm{int}}$, there is only a modest reduction to the inferred $k_{\rm{CB}}$, bolstering our argument in Section~\ref{aint_distribution} that it should be about 2.5\textperthousand. The inferred gravity law also barely changes. On the other hand, the reduced incidence of CBs suggests that it is difficult for our model to reliably estimate just how common they actually are.

This difficulty probably involves CBs with low separations that cause large $\widetilde{v}$ perturbations, leading to a much broader low amplitude tail to the $\widetilde{v}$ distribution. Since we only consider WBs with $\widetilde{v} < 5$, it is possible that such tight CBs actually matter less for our analysis. In that scenario, there would be a degeneracy between the frequency of these very tight CBs and the overall CB fraction. Roughly speaking, suppose that we have a CB model where 50\% of the CBs create such large velocity perturbations that any WB containing such a subsystem would very likely have $\widetilde{v} > 5$ and thus be rejected from our sample, while the remaining CBs create more palatable velocity perturbations. If our analysis then needs 30\% of the WBs in our sample to contain a CB in order to adequately fit the extended tail at $\widetilde{v} \ga 1.5$, then we would have to assume that the likelihood of a WB containing a CB is about 46\%. This is because half of the contaminated systems would exit our sample altogether rather than contribute to the extended $\widetilde{v}$ tail, so the proportion of the remaining WBs which contain a CB would be about $23/\left( 100 - 23 \right) = 0.3$. However, we can envisage a different model in which none of the CBs are very tight, so any instance of CB contamination is very unlikely to cause the WB to have $\widetilde{v} > 5$. In this case, it would be possible to make do with a lower overall incidence of CBs while still adequately explaining the extended $\widetilde{v}$ tail. While this would inevitably affect the detailed shape of the distribution, this could be hard to tell given that our model has other degrees of freedom related to the WB population and the extent of LOS contamination. Moreover, our motivation for truncating the $a_{\rm{int}}$ distribution at the low end is related to the short orbital period (Appendix~\ref{Non_linear_orbital_motion}) rather than the velocity perturbations becoming so large that the contaminated WB is likely to exit our sample, an effect that is already included in our analysis. It is beyond the scope of our model to directly consider the astrometric oscillations that must arise with very tight CBs and how they would impact $\widetilde{v}$. These issues might be mitigated in future \emph{Gaia} data releases with a longer baseline, which would increase the minimum $a_{\rm{int}}$ that we need to consider and thus reduce the maximum possible perturbation to $\widetilde{v}$. Once this maximum is sufficiently small, it will be clear that WBs should remain in our sample despite receiving velocity perturbations from undetected CBs. This would correspond to a narrower range of $a_{\rm{int}}$ and most likely reduce $f_{\rm{CB}}$.

These difficulties could explain why changing from a flat to a triangular logarithmic distribution of $a_{\rm{int}}$ reduces $f_{\rm{CB}}$ by 0.033 and narrowing the width from 1.5~dex to 1~dex reduces the inferred $f_{\rm{CB}}$ by 0.063. One can envisage that combining these changes would reduce $f_{\rm{CB}}$ by 0.1 from its nominal value of 0.63. Further fine-tuning of the $a_{\rm{int}}$ distribution could perhaps reduce $f_{\rm{CB}}$ to about 0.5. It seems unlikely that $f_{\rm{CB}}$ could be much smaller than this (see Appendix~\ref{Best_fit_fCB03}). Even so, it is clear that unlike the inferred gravity law, the inferred incidence of CB companions is not as robust within its formal uncertainty. We therefore suggest that one should not read too much into the somewhat high $f_{\rm{CB}}$ inferred by our analysis.



\subsubsection{The frequency of CB companions}
\label{Frequency_CBs}

Given that LOS contamination would cause a rising $\widetilde{v}$ distribution (Equation~\ref{LOS_contamination_pattern}) and that $\widetilde{v}$ uncertainties are very small (Figure~\ref{v_tilde_error_distribution}), our model must explain the extended declining tail towards high $\widetilde{v}$ (Figure~\ref{Photo_fixed_gravity}) using mostly undetected CBs. This was also suggested by several other workers (Section~\ref{Undetected_companions}) and expected \emph{a priori} \citepalias{Banik_2018_Centauri}. We argued previously that trends in the CB fraction with the WB separation are unlikely to mask the MOND signal (Figure~\ref{P_tail}). However, this still leaves open the possibility that $f_{\rm{CB}}$ could have a different value to the $63 \pm 1\%$ inferred by our analysis. Since the properties of the extended tail are to some extent degenerate with the inferred gravity law, we consider whether we can push this towards MOND by fixing $f_{\rm{CB}} = 0.3$, which is close to half the somewhat high value inferred by our nominal analysis. The idea is that the model cannot easily fit the extended tail using CBs, so it might try to do so by changing the gravity law.

\begin{figure}
    \centering
    \includegraphics[width=0.47\textwidth]{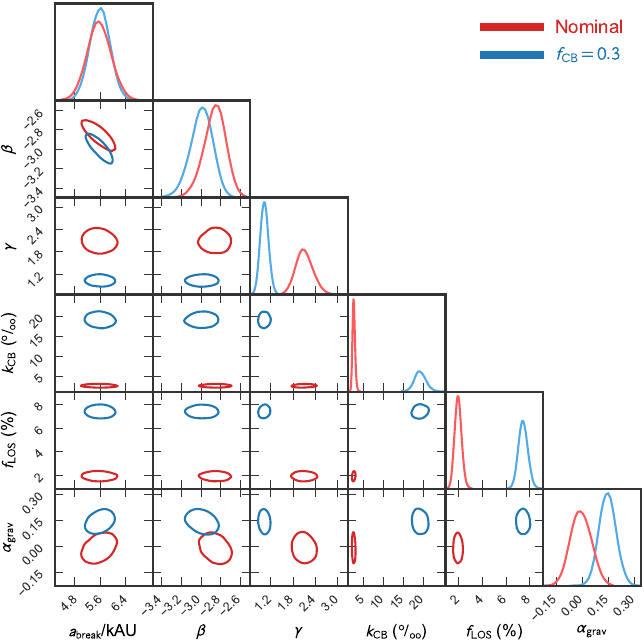}
    \caption{Triangle plot comparing the nominal analysis (red) to a revised analysis in which the likelihood of a star in our sample having an undetected CB companion is fixed at 30\% (Section~\ref{Frequency_CBs}).}
    \label{Triangle_fCB}
\end{figure}

The results of this exercise are shown in Figure~\ref{Triangle_fCB}. The inferred gravity law shifts moderately towards MOND, but it still remains much more consistent with the Newtonian prediction, which now faces just over $3\sigma$ tension. Since this reanalysis is the only one to substantially raise the inferred $\alpha_{\rm{grav}}$, we consider this model in more detail in Appendix~\ref{Best_fit_fCB03}. We argue that since the likelihood of the best overall fit is lower than in our nominal analysis by a factor of $\exp \left( 413 \right)$, there is a $29\sigma$ preference for the latter. We trace this to a visible failure to match nearly all aspects of the WB dataset when $f_{\rm{CB}} = 0.3$ (Figure~\ref{Photo_best_nominal_fCB03}).

\subsection{Comparison with prior WBT results}
\label{Other_WBT_results}

Our approach to the WBT is very similar to \citetalias{Pittordis_2023}, who also considered the distribution of $\left( r_{\rm{sky}}, \widetilde{v} \right)$. Our stellar mass estimates should be more reliable because we use a cubic rather than linear mass-luminosity relation and calibrate the resulting mass estimates with spectroscopic FLAME masses (Equation~\ref{FLAME_adjustment}). We calculate the relative velocity of each WB much more carefully, including an allowance for perspective effects due to the systemic RV. Our sample selection is also much stricter in several ways (Section~\ref{Refined_quality_cuts}), including through a quality cut on the carefully calculated $\widetilde{v}$ uncertainty. On the theoretical side, we improve upon their WB model by using orbits integrated in a rigorously calculated gravitational field \citepalias{Banik_2018_Centauri} and allow the eccentricity distribution to be superthermal, which seems very likely observationally \citep{Hwang_2022}. Our CB model is also an improvement because we allow both stars in a WB to have an undetected CB companion and model the CBs themselves in a somewhat different way, paying particular attention to the relevant range of CB orbit sizes and the inflation of the WB system mass due to undetected companion(s). Our treatment of chance alignments with unbound stars is simpler for computational reasons (Section~\ref{Chance_alignments}), though it should still be adequate given that the inferred $f_{\rm{LOS}}$ is only about 2\% (Table~\ref{MCMC_inferences}). Our exploration of the parameter space is vastly more thorough thanks to an optimized algorithm \citep{Banik_2021_plan}. In addition, we conduct a much broader range of extension analyses where we vary the sample selection and model assumptions, thus demonstrating the robustness of our results to a much greater extent (Section~\ref{Analysis_variants}). Despite these changes, our overall conclusions are very similar with regards to how strongly Newtonian gravity is preferred over MOND (Section~\ref{Gradient_ascent_fixed_gravity}).

While this study was under review, two additional publications appeared which argue that the WBT prefers MOND \citep{CHAE_2023, Hernandez_2023}. The study of \citet{Hernandez_2023} focused on building a very clean sample of WBs where CB contamination should be very small. While this could simplify the interpretation of the results, the sample size is unfortunately also rather small (see its figure~8). In the following, we therefore focus on the work of \citet{CHAE_2023}. The particular issue that we identify could also affect the analysis of \citet{Hernandez_2023}, though we have not checked this in detail.

The WB analysis of \citet{CHAE_2023} uses the WB sample of \citet{Badry_2021}, but it attempts to infer the 3D separation and relative velocity from the available projected information instead of modelling into the space of the observables. Various complicated procedures are used to deal with the fact that WBs are expected to have a wide range of eccentricities and to be viewed from a random direction at a random orbital phase. The eccentricity is constrained using the angle between the projected separation and relative velocity, making heavy use of the approach pioneered by \citet{Hwang_2022}. The overall result is that MOND is preferred at a confidence of $10\sigma$ using the nominal sample of 26615 WBs. This sample is not very well suited to the WBT due to a mass-dependent cutoff to the allowed $\widetilde{v}$, but as this seems to have only a modest impact, we defer discussion of this issue to Appendix~\ref{vtilde_limit_Kyu_impact}.

\citet{CHAE_2023} follow equation~4 of \citet{Badry_2021} in neglecting the systemic RV and setting
\begin{eqnarray}
    \label{vsky_simple}
    v_{\rm{sky}} ~=~ \overline{d} \mu_{\rm{rel}} \, , \quad \bm{\mu}_{\rm{rel}} ~\equiv~ \begin{bmatrix} \mu_{\alpha,*}^{\rm{rel}} \\ \mu_{\delta}^{\rm{rel}} \end{bmatrix} \, ,
\end{eqnarray}
where $\overline{d}$ is the inverse variance weighted mean of the parallax distances to the stars A and B that make up the WB, $\mu_{\rm{rel}} \equiv \left| \bm{\mu}_{\rm{rel}} \right|$ is the relative proper motion, $\mu_{\alpha,*}$ is the proper motion along the East-West direction, $\mu_\delta$ is the proper motion along the North-South direction, and `rel' superscripts denote relative values found by subtracting the value for star~B from the value for star~A (e.g. $\mu_{\delta}^{\rm{rel}} \equiv \mu_{\delta}^A - \mu_{\delta}^B$). The proper motion $\mu_{\delta} \equiv \dot{\delta}$, where $\delta$ is the declination and an overdot denotes a time derivative. To account for the spherical geometry of quantities on the sky, the proper motion component $\mu_{\alpha,*} \equiv \dot{\alpha} \cos \delta$, where $\alpha$ is the right ascension. Both stars are assumed to be at the same distance, which should be correct on average if we assume that uncertainties in trigonometric distances are much larger than actual separations along the LOS. Neglecting the perspective effect due to the systemic RV should also be reasonable given the typically small angular separations of WBs (Section~\ref{Systemic_RV}).

\begin{figure*}
    \centering
    \includegraphics[width=0.48\textwidth]{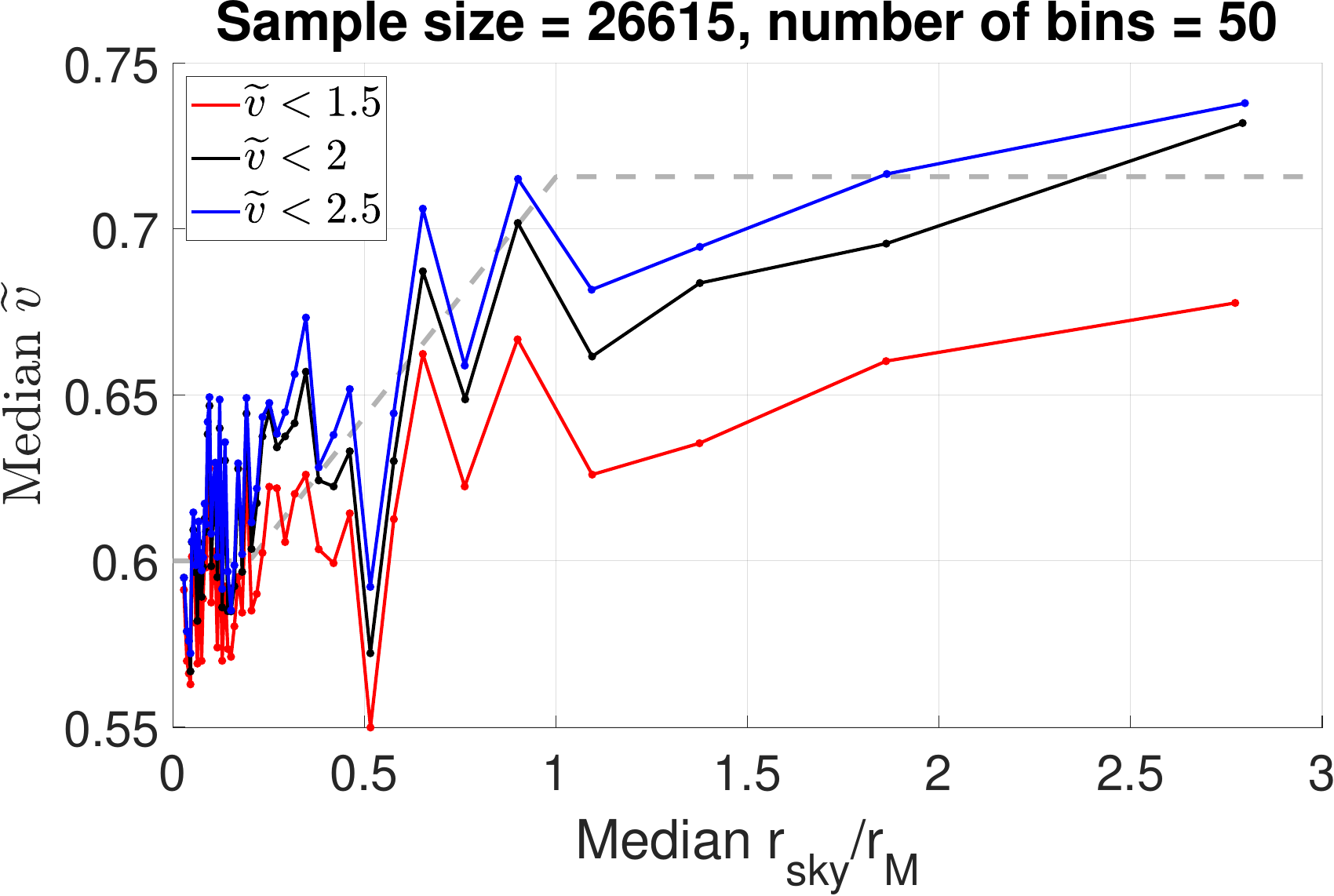}
    \hfill
    \includegraphics[width=0.48\textwidth]{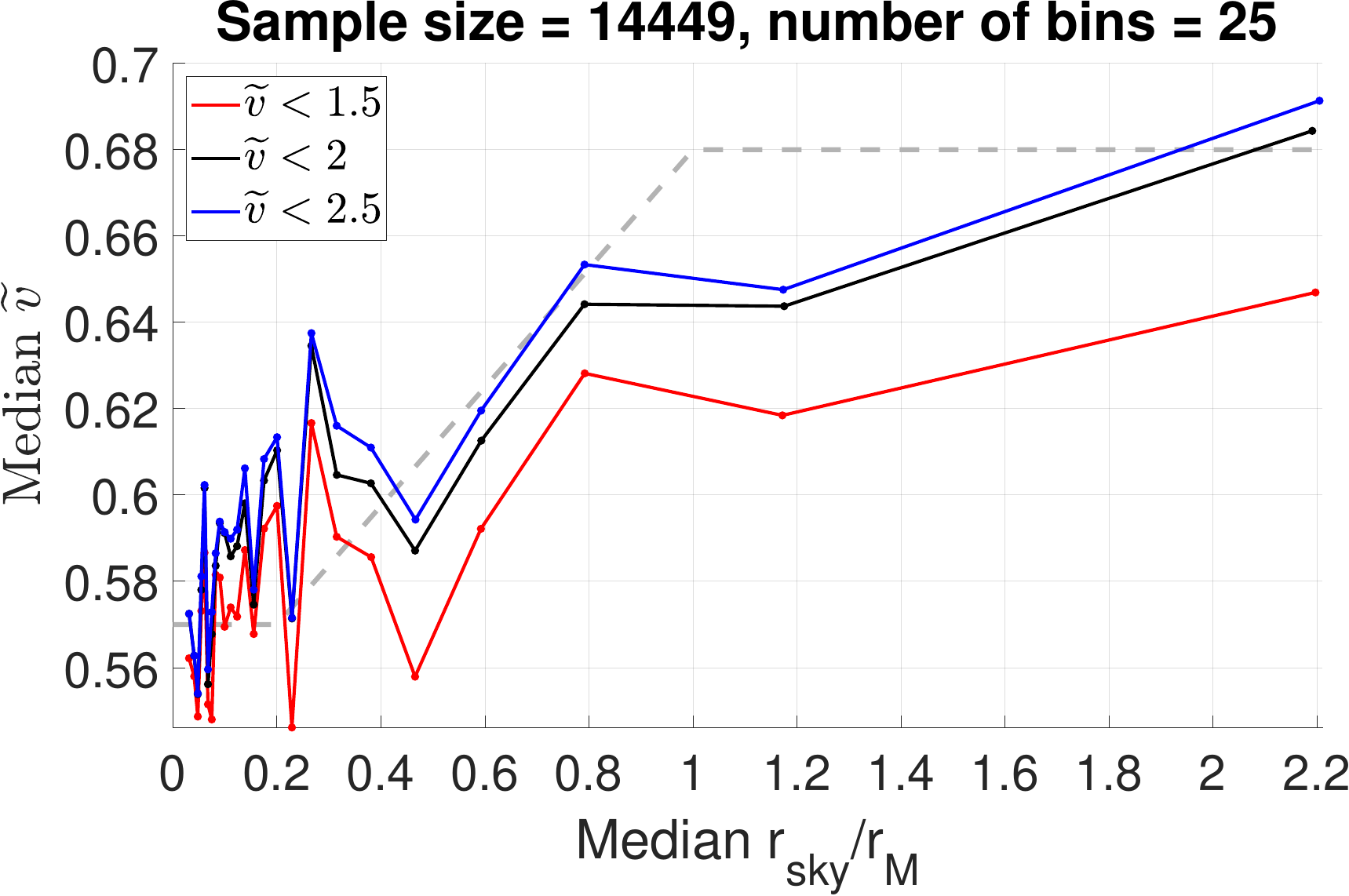}
    \caption{The median $\widetilde{v}$ as a function of the median $r_{\rm{sky}}/r_{_M}$ for the nominal sample of \citet{CHAE_2023}, which requires both proper motion components of both stars in each WB to be more precise than 1\% (left panel). This is tightened to 0.3\% in the right panel, as done in their appendix~B and \citet{Chae_2024}. The sample size and number of bins are shown at the top. Similarly to Figure~\ref{Median_vtilde_values}, each point corresponds to the same number of WBs apart from the last point, which has a slightly different number due to rounding.}
    \label{Kyu_medians_all}
\end{figure*}

We now try to reproduce the claimed detection of MOND by \citet{CHAE_2023}. Without going into the details of the deprojection algorithm or the model for the CB and WB populations, a simple way to consider the WBT is to find the median $\widetilde{v}$ in different ranges of $r_{\rm{sky}}/r_{_M}$. Bearing in mind the above small caveat about perspective effects and the need to accurately estimate the stellar masses, these quantities can all be unambiguously calculated with no assumptions about how exactly WBs behave, how eccentric their orbits typically are, etc. The only assumption we need to make is that in the theories under consideration, WBs would very rarely have $\widetilde{v} > 1.5$, so values much beyond this are indicative of some sort of contamination. It is therefore important to focus on only those WBs with $\widetilde{v} \la 2$ when calculating the median, since otherwise trends in the level of contamination could create the appearance of an acceleration-dependent trend in the WB dynamics (magenta line in Figure~\ref{Median_vtilde_values}). For the nominal sample of 26615 WBs used by \citet{CHAE_2023}, the result of this exercise is shown in the left panel of Figure~\ref{Kyu_medians_all} based on masses and $v_{\rm{sky}}$ values from that study. Due to the larger sample size compared to our study, we are able to use 50 bins while keeping the scatter low. A rising trend is immediately apparent, especially for the subsamples with $\widetilde{v} < 2$ or 2.5. This could arguably indicate that a limit of 1.5 is too restrictive and loses some of the MOND signal, which is possible given that the upper limit to $\widetilde{v}$ for a bound Milgromian WB is about 1.7 \citepalias{Banik_2018_Centauri}. The results agree very well with the dashed grey line, which shows the shape that we expect in MOND (the normalization has been adjusted to fit the results at high accelerations or low $r_{\rm{sky}}/r_{_M}$). The median $\widetilde{v}$ undergoes some scatter but with little trend when $r_{\rm{sky}}/r_{_M} < 0.2$. It then rises rapidly at about the expected rate before flattening out. While we do not expect the observed rising trend at $r_{\rm{sky}}/r_{_M} \ga 1.5$ (see Figure~\ref{Radial_gravity_boost}), the median $\widetilde{v}$ is only rising quite slowly by this point and is broadly in agreement with the expected flat trend. All this seems to provide compelling evidence in favour of local WBs following Milgromian gravity.

A critical consideration for the WBT is the accuracy of $v_{\rm{sky}}$, which is the main source of uncertainty in $\widetilde{v}$. We quantify its uncertainty very carefully in order to limit the $\widetilde{v}$ uncertainty of the WBs in our sample (Section~\ref{v_tilde_uncertainty}). As argued there, since the main peak to the $\widetilde{v}$ distribution lies at $\widetilde{v} \approx 0.5$ (Figure~\ref{Photo_fixed_gravity}) and $\widetilde{v}$ uncertainties would add in quadrature to the intrinsic width of the $\widetilde{v}$ distribution, we need the uncertainty in $\widetilde{v}$ to be $\la 0.1$ in the regime most relevant to the WBT. An uncertainty of 0.1 would inflate the dispersion from 0.5 to $\sqrt{0.5^2 + 0.1^2} = 0.51$, implying only a 2\% broadening $-$ too little to affect the inferred gravity law. While this approach might seem somewhat conservative, we have seen that it still leaves a reasonable sample size, partly because we impose a less strict limit to the $\widetilde{v}$ uncertainty outside the main peak region (Equation~\ref{Max_vtilde_uncertainty}).

Unfortunately, \citet{CHAE_2023} do not estimate the uncertainty in $\widetilde{v}$ or in $v_{\rm{sky}}$. Their approach is to require a maximum 1\% uncertainty on the heliocentric proper motions $\mu_{\alpha,*}^A$, $\mu_{\alpha,*}^B$, $\mu_{\delta}^A$, and $\mu_{\delta}^B$ of the stars in each WB. To see why this approach is not sufficient, suppose that both stars have a heliocentric velocity of 30~km/s in the East-West and North-South directions (the small difference due to WB orbital motion is not relevant here). The sky-projected heliocentric velocity of each star could then have an uncertainty of 300~m/s along each direction. Since we need to combine four proper motion components to calculate $v_{\rm{sky}}$ (Equation~\ref{vsky_simple}), this has an uncertainty of $\approx 300\sqrt{4}$~m/s or 600~m/s. However, the Newtonian $v_c$ of two Sun-like stars with a 10~kAU separation is only 420~m/s, implying a very poorly constrained $\widetilde{v}$. This completely undermines the WBT because its main idea is that some typical measure of $\widetilde{v} \propto \sqrt{g/g_{_N}}$, making it absolutely essential to have precise constraints on $\widetilde{v}$.

To try and address this issue, \citet{CHAE_2023} conduct an extension analysis in their appendix~B where the 1\% requirement mentioned above is tightened to 0.3\%. This is unlikely to be enough because in the previous argument, we would have a maximum allowed $\sigma \left( v_{\rm{sky}} \right)$ of 180~m/s rather than 600~m/s, where $\sigma \left( Q \right)$ is the uncertainty on any quantity $Q$. However, if the Newtonian $v_c$ is 420~m/s and the mode of the $\widetilde{v}$ distribution is at 0.5, the intrinsic velocity dispersion is 210~m/s, which is not much larger than the upper limit to $\sigma \left( v_{\rm{sky}} \right)$. Alternatively, the plausible scenario described above would lead to $\sigma \left( \widetilde{v} \right) = 180/420 = 0.43$, which is still too large for the WBT. Even so, we repeat the median $\widetilde{v}$ analysis with all four relevant heliocentric proper motions required to have an accuracy of at least 0.3\%. The sample size is almost halved by this much stricter cut, so we use only half as many bins to reduce the scatter. Despite this, we still achieve a good range in $r_{\rm{sky}}/r_{_M}$ and probe well into the asymptotic regime (right panel of Figure~\ref{Kyu_medians_all}). A rising trend is evident here, though the agreement with the MOND expectation is somewhat less good. Importantly, it is not really possible to fit the results with a flat line, as would be expected in Newtonian gravity.

A more rigorous but still very simple way to estimate the uncertainty in $v_{\rm{sky}}$ is provided by equation~5 of \citet{Badry_2021}. If we neglect other sources of error like perspective effects and the small error in $\overline{d}$, we can approximate that the only uncertain term is $\mu_{\rm{rel}}$. Its uncertainty is
\begin{eqnarray}
    \label{sigma_mu_rel}
    \sigma \left( \mu_{\rm{rel}} \right) ~&=&~ \sqrt{ \sigma^2 
 \left( \mu_{\alpha,*}^{\rm{rel}} \right) + \sigma^2 \left( \mu_{\delta}^{\rm{rel}} \right)} \, , \\
    \sigma \left( \mu_{\alpha,*}^{\rm{rel}} \right) ~&=&~ \sqrt{ \sigma^2 \left( \mu_{\alpha,*}^A \right) + \sigma^2 \left(\mu_{\alpha,*}^B \right)} \, , \\
    \sigma \left( \mu_\delta^{\rm{rel}} \right) ~&=&~ \sqrt{ \sigma^2 \left( \mu_\delta^A \right) + \sigma^2 \left( \mu_\delta^B \right)} \, .
\end{eqnarray}
This result is based on the fact that at first order in $\sigma \left( \mu_{\rm{rel}} \right)/\mu_{\rm{rel}}$, the uncertainty in $\mu_{\rm{rel}}$ arises only from the proper motion uncertainty along the direction $\left( \mu_{\alpha,*}^{\rm{rel}}, \mu_\delta^{\rm{rel}} \right)$ because uncertainty in $\bm{\mu}_{\rm{rel}}$ in the orthogonal direction mainly affects its direction rather than its magnitude. This approximation becomes inaccurate if $\sigma \left( \mu_{\rm{rel}} \right) \ga \mu_{\rm{rel}}$, so a Monte Carlo approach might be preferable even if we could assume that the uncertainty arises entirely from uncorrelated astrometric errors in the proper motions.

With the above approximations, the uncertainty in $v_{\rm{sky}}$ is simply
\begin{eqnarray}
    \sigma \left( v_{\rm{sky}} \right) ~=~ \overline{d} \sigma \left( \mu_{\rm{rel}} \right) \, .
    \label{sigma_v_sky}
\end{eqnarray}
Dividing this by the Newtonian $v_c$ then yields an estimated $\widetilde{v}$ uncertainty. We suggest that such an approach would capture the major uncertainties affecting $\widetilde{v}$ because the mass and systemic distance should both have small uncertainties, while perspective effects should likewise have little effect (Section~\ref{Systemic_RV}).

\begin{figure}
    \centering
    \includegraphics[width=0.47\textwidth]{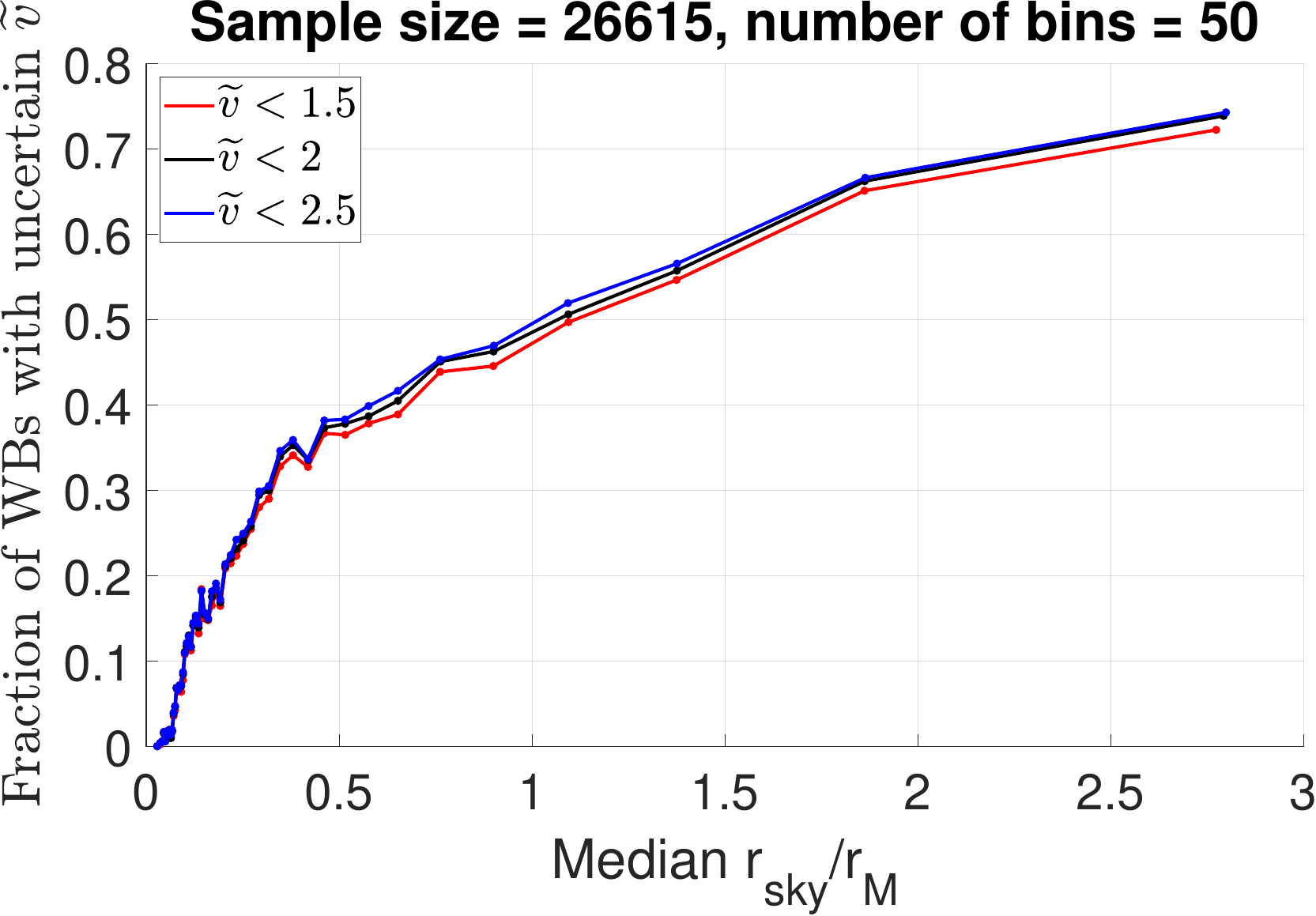}
    \caption{The fraction of WBs in each $r_{\rm{sky}}/r_{_M}$ bin of the nominal \citet{CHAE_2023} sample with $\sigma \left( \widetilde{v} \right) > 0.1 \max \left( 1, \widetilde{v}/2 \right)$. This applies to most of the WBs in the asymptotic regime critical to testing MOND.}
    \label{Kyu_highvterr_fraction}
\end{figure}

We can now check whether the result of \citet{CHAE_2023} is robust against a quality cut based on the estimated $\widetilde{v}$ uncertainty of each WB. We begin by quantifying the proportion of WBs in each $r_{\rm{sky}}/r_{_M}$ bin with $\sigma \left( \widetilde{v} \right) > 0.1 \max \left( 1, \widetilde{v}/2 \right)$, the cut imposed in our own analysis (Equation~\ref{Max_vtilde_uncertainty}). The result is shown in Figure~\ref{Kyu_highvterr_fraction}. It is clear that most WBs in the asymptotic regime fail to pass our quality cut on $\sigma \left( \widetilde{v} \right)$. This raises the possibility that the MOND-like trend in Figure~\ref{Kyu_medians_all} is caused by measurement errors broadening the $\widetilde{v}$ distribution at low accelerations.

\begin{figure}
    \centering
    \includegraphics[width=0.47\textwidth]{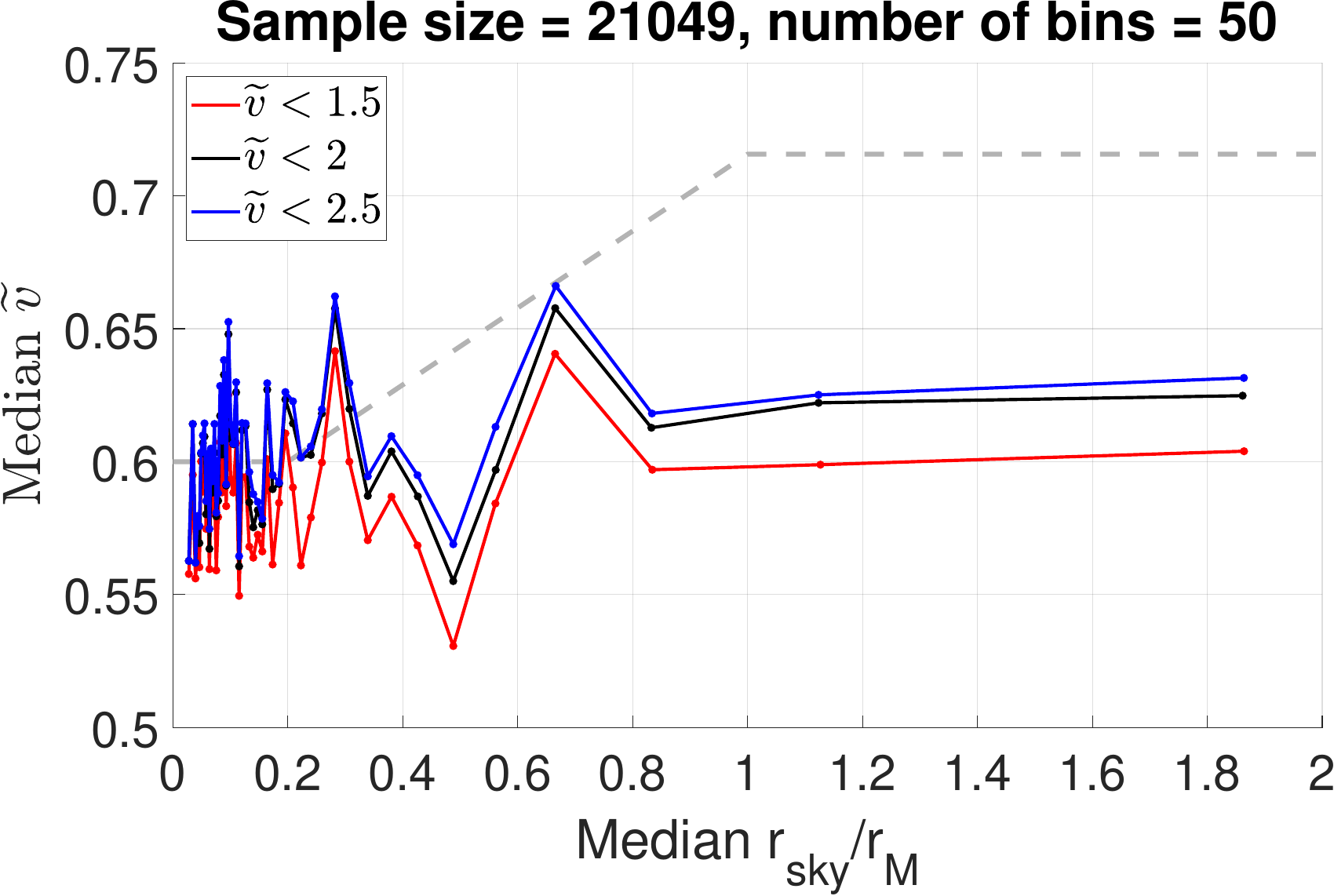}
    \caption{Similar to the left panel of Figure~\ref{Kyu_medians_all}, but with the \citet{CHAE_2023} nominal sample restricted to WBs with an estimated $\sigma \left( \widetilde{v} \right) < 0.1 \max \left( 1, \widetilde{v}/2 \right)$ according to Equations~\ref{sigma_mu_rel} and \ref{sigma_v_sky}. The median $\widetilde{v}$ is now flat with respect to our proxy for the WB acceleration thanks to this cut on the estimated $\widetilde{v}$ uncertainty, which is critical to reliably conducting the WBT.}
    \label{Kyu_medians_v01}
\end{figure}

To analyse the dataset of \citet{CHAE_2023} more robustly, we restrict their nominal sample to the same limiting $\sigma \left( \widetilde{v} \right)$ as we used in our own analysis (Equation~\ref{Max_vtilde_uncertainty}). The result is shown in Figure~\ref{Kyu_medians_v01}. The sample size only decreases by about 1/5, allowing us to continue using 50 bins in $r_{\rm{sky}}/r_{_M}$, albeit with a somewhat reduced range given that the sample is reduced by half at the MOND radius and slightly more beyond that. Even so, we still have many WBs well into the asymptotic regime. These show that the median $\widetilde{v}$ is now almost completely flat with respect to $r_{\rm{sky}}/r_{_M}$. There is clearly a major disagreement with the dashed grey line showing the MOND expectation calibrated to the high-acceleration end. This dramatic difference is caused by removing only 1/5 of the WBs, indicating that the issue we identified is indeed extremely serious.

The MOND-like rising median $\widetilde{v}$ trend in the nominal sample of \citet{CHAE_2023} persists despite an apparently tight 0.3\% cut on the proper motion errors, which leads to an even smaller sample size than in Figure~\ref{Kyu_medians_v01}. This highlights that the important quantity for the WBT is the relative velocity of the stars in each WB rather than the heliocentric velocity of each star. While one can envisage that a sufficiently restrictive cut on the latter would be sufficient, this is simply not targeting the relevant quantity, so the sample size is likely to become insufficient before the cut has been adequately tightened for the purpose of the WBT. As an exercise in this direction, we found that requiring a 0.1\% precision reduces the sample to only 4977 WBs and makes the median $\widetilde{v}$ nearly flat with respect to $r_{\rm{sky}}/r_{_M}$, so perhaps this cut is tight enough. However, the sample size can be much larger while still maintaining sufficient accuracy on $\widetilde{v}$ provided that the quality cut is targeted at $\widetilde{v}$, or at least at $\mu_{\rm{rel}}$.

The importance of this issue stems from the fact that for the same $\sigma \left( v_{\rm{sky}} \right)$ in m/s, $\sigma \left( \widetilde{v} \right) \propto \sqrt{r_{\rm{sky}}}$ because the Newtonian $v_c \propto 1/\sqrt{r_{\rm{sky}}}$, making it increasingly difficult to pass a strict quality cut on $\sigma \left( \widetilde{v} \right)$ (Figure~\ref{Kyu_highvterr_fraction}). Without such a cut, as we get to large $r_{\rm{sky}}$ and thus also $r_{\rm{sky}}/r_{_M}$, the $\widetilde{v}$ distribution might be broadened by measurement errors to a greater extent, pushing up the median $\widetilde{v}$ and creating a fake Milgromian signal. This raises the more general issue that any systematic trend with $r_{\rm{sky}}$ can appear to be an acceleration-dependent trend. In principle, one could disentangle the two because our proxy for the WB internal acceleration depends on both $r_{\rm{sky}}$ and $M$. However, the former is by far the main factor because of its large range and the fact that $r_{_M}$ scales only as $\sqrt{M}$, making the WB mass relatively less important. There is also an issue with the dataset of \citet{CHAE_2023} pertaining to the masses that makes it difficult to use a broad range (Appendix~\ref{vtilde_limit_Kyu_impact}). One therefore has to be cautious before concluding that any trend discovered in the WB data is best understood as causally related to their internal acceleration.

Given our results in this section, it seems extremely unlikely that local WBs prefer Milgromian over Newtonian gravity once the WB sample is selected carefully to ensure that $\widetilde{v}$ is accurately known. It should be straightforward to alter the quality cut related to the astrometric accuracy by focusing on how precisely we know the relative velocity within each WB, since this is what enters into the calculation of the physically relevant $\widetilde{v}$ parameter.\footnote{This would allow to relax the requirement for the heliocentric proper motion components to have a 1\% precision, increasing the sample size somewhat.} We stress that even if $\widetilde{v}$ is not calculated explicitly, the WBT is ultimately about the ratio between the relative velocity and the Newtonian circular orbit prediction because the WB orbits are too long to directly measure the acceleration \citep{Banik_2019_Proxima}. The uncertainty on this velocity ratio is therefore of critical importance to the WBT and must be quantified, with systems rejected if the uncertainty is too large.

This concern likely also applies to the analysis of \citet{Hernandez_2023}: its section~4 states that $v_{\rm{sky}}$ should exceed $1.5\times$ its uncertainty. This translates to a fractional $\widetilde{v}$ accuracy of 2/3, so a system with $\widetilde{v} = 1$ might have $\sigma \left( \widetilde{v} \right) = 0.6$ \citep[this concern also applies to][]{Hernandez_2024}. Addressing this issue similarly to our study would reduce the already small sample size. However, our work suggests that the loss might be modest, especially given the high overall quality of the WB sample used in their study.

\subsection{The MOND interpolating function}
\label{MOND_interpolating_function}

Our results for the WBT depend on the assumed MOND interpolating function because local WBs are subject to the Galactic external field of magnitude $1.8 \, a_{_0}$ \citepalias[Section~\ref{WB_population}; see also section~7.1 of][]{Banik_2018_Centauri}. Thus, our falsification of MOND could in principle be avoided with a sufficiently rapid transition to Newtonian behaviour when $g_{_N} > a_{_0}$. However, this is in strong tension with rotation curve constraints. This is because the enhancement to $g_{_N}$ needed to fit the host galaxy rotation curve is very similar to the predicted enhancement to the gravity binding a WB in the asymptotic regime. This remains true in both AQUAL and QUMOND, which give almost the same boost for the same interpolating function (Figure~\ref{Radial_gravity_boost}). We clarify this by briefly giving the main analytic results relevant to EFE-dominated WBs and thereby estimate the predicted $\alpha_{\rm{grav}}$ with different MOND formulations and interpolating functions \citepalias[for further details, see sections~2.2 and 7.1 of][]{Banik_2018_Centauri}.

The WBT is mostly sensitive to $\eta$ (Equation~\ref{eta_definition}) in the asymptotic regime. In QUMOND, equation~16 of \citetalias{Banik_2018_Centauri} tells us that
\begin{eqnarray}
    \eta ~=~ \nu_e \left( 1 + \frac{K_e}{3} \right) , \quad K_e ~\equiv~ \frac{\partial \ln \nu_e}{\partial \ln g_{_{N,e}}} \, ,
    \label{eta_QUMOND}
\end{eqnarray}
where $\nu_e$ is the QUMOND interpolating function if we set its argument to the Newtonian-equivalent Galactic gravity $g_{_{N,e}}$ (Equation~\ref{gNe}). This result matches equation~57 of \citet{QUMOND}.

AQUAL does not directly use $\nu$ but instead uses the interpolating function $\mu$, which is defined so that $\mu \bm{g} = \bm{g}_{_N}$ in spherical symmetry. We can talk of AQUAL and QUMOND theories as having the `same' interpolating function if they give the same relation between $\bm{g}$ and $\bm{g}_{_N}$ in spherical symmetry, which requires that $\mu \left( g \right) \nu \left( g_{_N} \right) = 1$. According to equation~35 of \citetalias{Banik_2018_Centauri}, the AQUAL version of Equation~\ref{eta_QUMOND} is
\begin{eqnarray}
    \eta ~=~ \frac{\tan^{-1} \sqrt{L_e}}{\mu_e \sqrt{L_e}} \, , \quad L_e ~\equiv~ \frac{\partial \ln \mu_e}{\partial \ln g_e} \, ,
    \label{eta_AQUAL}
\end{eqnarray}
where $L_e$ is the AQUAL interpolating function considering only the Galactic gravity $g_e$. We can find $L_e$ from the QUMOND $K_e$ using the relation $\left( 1 + L_e \right) \left( 1 + K_e \right) = 1$ \citepalias[equation~38 of][]{Banik_2018_Centauri}. The angle-averaged enhancement to $g_{_N}$ shown above matches that stated in equation~65 of \citet{QUMOND} once we use the relation $\sin^2 \theta = \tan^2 \theta / \left( \tan^2 \theta + 1 \right)$.

The observable in the WBT is the relative velocity and this is related to the circular velocity, which scales as $\sqrt{\eta}$. Moreover, the parameter $\alpha_{\rm{grav}}$ only captures deviations from the Newtonian result because it is defined to be 0 in the Newtonian case, when $\eta \equiv 1$. Thus, we can approximate that the predicted $\alpha_{\rm{grav}} ~\propto~ \sqrt{\eta} - 1$. Since the normalization must be chosen such that $\alpha_{\rm{grav}} = 1$ if we use QUMOND with the simple interpolating function,
\begin{eqnarray}
    \alpha_{\rm{grav}} ~=~ \frac{\sqrt{\eta} - 1}{0.193} \, .
    \label{alpha_grav_est}
\end{eqnarray}

Table~\ref{alpha_grav_values} shows $\eta$ and the estimated $\alpha_{\rm{grav}}$ for different MOND formulations and interpolating functions. Differences between AQUAL and QUMOND are very small, as stated in section~5.1 of \citet{QUMOND} and explained further in section~7.2 of \citetalias{Banik_2018_Centauri}, which is devoted to a comparison of the two MOND formulations. The simple and MLS \citep[$\nu^{-1} = 1 - \exp \left( - \sqrt{g_{_N}/a_{_0}} \right)$;][]{McGaugh_2008, Famaey_McGaugh_2012, McGaugh_Lelli_2016} interpolating functions give similar predictions for the WBT because they are numerically very similar. The main difference is that the MLS function converges to Newtonian gravity much faster when $g \ga 10 \, a_{_0}$ due to the exponential term, but this is not important for the WBT because WB observations are sensitive to the gravity binding it rather than merely how much that deviates from $g_{_N}$.\footnote{The simple interpolating function provides a good match to the velocity dispersion profiles of elliptical galaxies up to almost $100 \, a_{_0}$ \citep{Chae_2015, Chae_2019, Chae_2020_elliptical}, beyond which the MOND enhancement to $g_{_N}$ must be very small.} Compared to the simple and MLS cases, the standard interpolating function significantly reduces the predicted signal for the WBT, though even then the predicted $\alpha_{\rm{grav}} = 0.18$ is still much higher than inferred by our nominal MCMC analysis (bottom right panel of Figure~\ref{Triangle_nominal_gamma_Hwang}). But since some variations to the modelling assumptions do increase the inferred $\alpha_{\rm{grav}}$ somewhat (Table~\ref{MCMC_inferences}), it can be argued that the WBT is marginally consistent with the standard interpolating function
\begin{eqnarray}
    \nu^2 ~=~ \frac{1}{2} + \sqrt{\frac{1}{4} + \left( \frac{a_{_0}}{g_{_N}} \right)^2} \, .
    \label{Standard_interpolating_function}
\end{eqnarray}
However, this is in tension with rotation curve constraints \citepalias[section~7.1 of][]{Banik_2018_Centauri}. We revisit their arguments in more detail below.

\begin{table}
    \centering
    \caption{The enhancement factor $\eta$ to the radial gravity (Equation~\ref{eta_definition}) and our corresponding estimate for $\alpha_{\rm{grav}}$ (Equation~\ref{alpha_grav_est}) with different MOND formulations and interpolating functions. Results differ little between AQUAL and QUMOND and between the MLS and simple functions.}
    \begin{tabular}{ccccc}
    \hline
    Interpolating & \multicolumn{2}{c}{AQUAL} & \multicolumn{2}{c}{QUMOND} \\
    function & $\eta$ & $\alpha_{\rm{grav}}$ & $\eta$ & $\alpha_{\rm{grav}}$ \\ \hline
    Simple & 1.4056 & 0.96 & 1.4228 & 1 \\
    MLS & 1.3508 & 0.84 & 1.3692 & 0.88 \\
    Standard & 1.0661 & 0.17 & 1.0726 & 0.18 \\
    Sharp & 1 & 0 & 1 & 0 \\ \hline
    \end{tabular}
    \label{alpha_grav_values}
\end{table}

Equations~\ref{eta_QUMOND} and \ref{eta_AQUAL} show that $\eta \approx \nu_e = 1/\mu_e$ up to a modest correction for azimuthal averaging. This is because $\nu_e$ can be arbitrarily large, but $K_e$ is always between $-1/2$ and 0 while $L_e$ is between 1 and 2, with low $K_e$ and high $L_e$ corresponding to the deep-MOND limit ($K_e = -0.26$ in the Solar neighbourhood). As a result, the enhancement to the orbital velocities of WBs in the asymptotic regime should be only slightly less than the enhancement to the Newtonian baryonic rotation curve of the parent galaxy in the vicinity of the WBs. This means a very stringent null detection of MOND effects in the WBT can only be reconciled with MOND if the Galactic rotation curve at the Solar circle has almost the same amplitude as it would do in Newtonian gravity with baryons alone.

\begin{figure}
    \centering
    \includegraphics[width=0.47\textwidth]{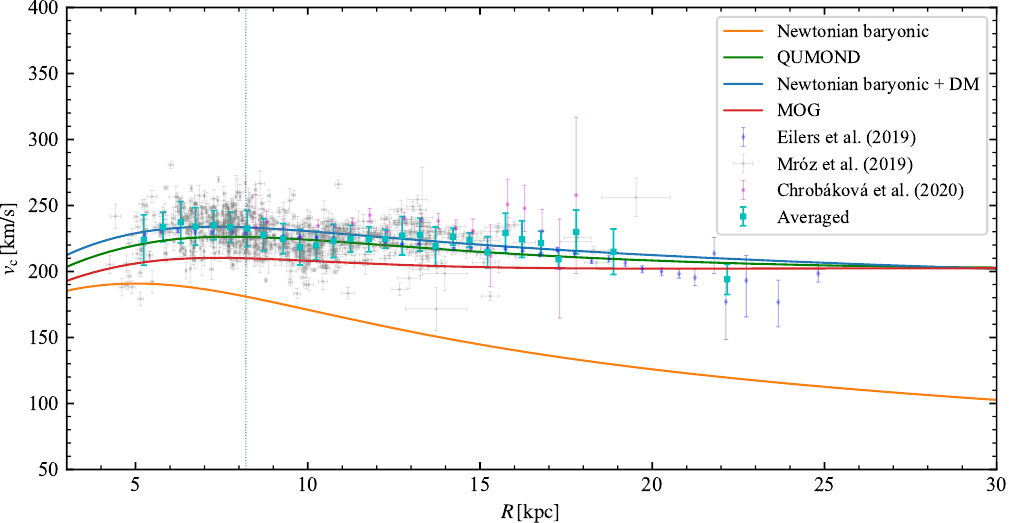}
    \caption{The MW rotation curve as measured by averaging different observational determinations \citep[cyan squares with error bars;][]{Eilers_2019, Mroz_2019, Chrobakova_2020} and as predicted in different gravity theories using the baryonic mass distribution from \citet*{Wang_2022}. Notice that the Newtonian model with only baryons (orange line) falls well short of the observed rotation curve at the Solar circle radius of 8.2~kpc (dotted blue vertical line). This deficiency can be rectified by adding dark matter (blue line), but the halo parameters must be inferred from the data and are not predicted \emph{a priori}. An alternative solution with very little flexibility is QUMOND with the simple interpolating function (green line; Equation~\ref{Simple_interpolating_function}), but this only works because this function predicts a significant enhancement to Newtonian gravity at the relevant acceleration (compare the simple and sharp interpolating functions in Table~\ref{alpha_grav_values}). This enhancement should also be detectable in local WBs. Reproduced from figure~1 of \citet{Zhu_2023}, modified by Haixia Ma.}
    \label{Galactic_rotation_curve}
\end{figure}

We test this in Figure~\ref{Galactic_rotation_curve}, where the cyan squares with error bars show binned observational results and the dotted blue vertical line shows the Solar circle radius of 8.2~kpc \citep[the uncertainty on this is very small;][]{Gravity_2019}. The largest uncertainty in the kinematics comes from the peculiar (non-circular) velocity of the Sun and hence on the actual value of $v_{c,\odot}$, which normalizes the entire curve. However, the resulting uncertainty has been substantially mitigated by the direct measurement of the Solar System's acceleration relative to distant quasars \citep{Klioner_2021}, which is a direct probe of $\bm{g}_e$. Besides the kinematics, there is also some systematic uncertainty in the Newtonian rotation curve of the Galactic disc because we do not perfectly know its actual mass-to-light ratio and gas content. With these small caveats in mind, the Newtonian rotation curve of the baryons alone (orange line) is clearly below the actual rotation curve at the Solar circle, which is also much flatter. Since scaling the Newtonian rotation curve (e.g. to raise the mass-to-light ratio) would make the predicted decline even steeper, it is clear that MOND can only work with an interpolating function that significantly enhances Newtonian gravity at the Solar circle. Indeed, the solid green line shows that QUMOND with the simple interpolating function fits rather well $-$ even the bumps and wiggles apparent in the observations are reproduced nicely if we consider similar features in the Galactic surface density profile \citep{McGaugh_2018}. However, the required significant Milgromian enhancement to the local amplitude of the rotation curve contradicts our stringent null detection of MOND effects in the WBT, which implies an enhancement by at most a few percent. This contradiction is underpinned by the directly measured acceleration of the Solar System relative to distant quasars \citep{Klioner_2021}, which precludes substantial deviations from the kinematically deduced $\bm{g}_e$ (Section~\ref{WB_population}). If we very conservatively assume that $v_{c, \odot}$ is at most 10~km/s higher than our adopted 232.8~km/s, then the stronger EFE would only reduce the QUMOND $\alpha_{\rm{grav}}$ for the simple interpolating function from 1 to 0.92, while an even higher $v_{c, \odot} = 250$~km/s would give $\alpha_{\rm{grav}} = 0.87$. Such a large deviation of $v_{c,\odot}$ from our adopted value is almost inconceivable given the many decades of research on this issue. Turning instead to the possibility of a much sharper interpolating function, the required enhancement to the baryonic surface density would need to be truly substantial, which seems rather unlikely.

Another interesting aspect of Figure~\ref{Galactic_rotation_curve} is the failure of Moffat gravity \citep[MOG;][]{Moffat_2006}, whose prediction is shown as the red line. MOG modifies gravity only beyond a certain distance, so it passes the WBT \citep[see section~2.3 of][]{Roshan_2021_disc_stability}. However, MOG underpredicts the Galactic rotation curve \citep[as shown previously by][]{Negrelli_2018}. Moreover, a joint fit to the velocity dispersion profile and star formation history of Dragonfly~44 rules out MOG at $5.5\sigma$ confidence \citep{Haghi_2019_DF44}. This shows how difficult it is for a modified gravity theory to remain consistent with the WBT and simultaneously explain galaxy dynamics, even if we neglect extragalactic data.

\begin{figure}
    \centering
    \includegraphics[width=0.47\textwidth]{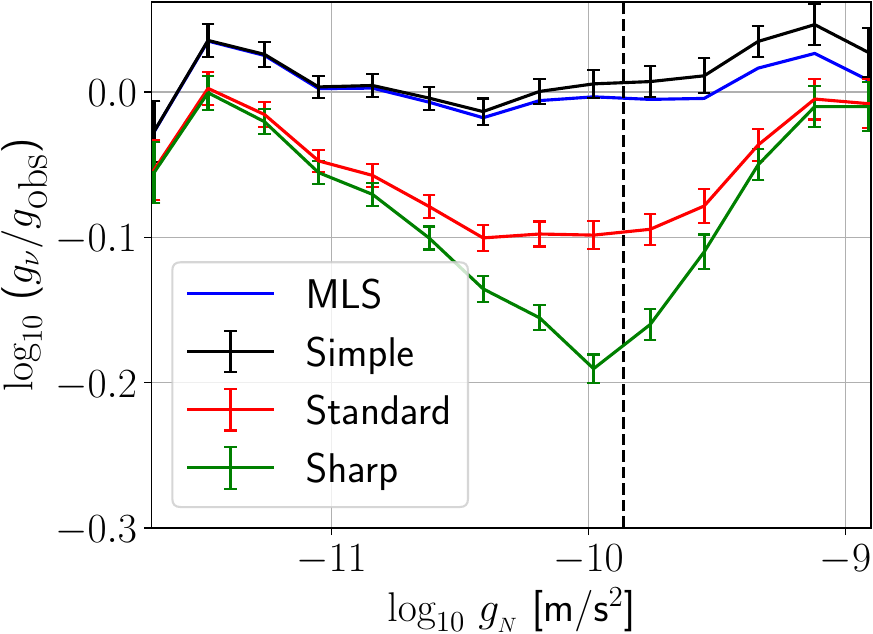}
    \caption{The ratio between the observed $g$ at fixed baryonic $g_{_N}$ \citep{McGaugh_Lelli_2016} and the prediction $g_\nu$ according to different MOND interpolating functions for the canonical $a_{_0} = 1.2 \times 10^{-10}$~m/s\textsuperscript{2} \citep{Begeman_1991, Gentile_2011, Lelli_2017}. Observational uncertainties are found by dividing the dispersion in $\log_{10} g$ at each $g_{_N}$ by $\sqrt{N}$, where $N$ is the number of data points in the bin. Due to the numerical similarity of the simple and MLS functions (black and blue lines, respectively); error bars are omitted for the latter, but they are shown on the standard interpolating function (red; Equation~\ref{Standard_interpolating_function}) and for the case of an infinitely sharp transition between the Newtonian and MOND regimes (green; Equation~\ref{Sharp_interpolating_function}). The dashed vertical line shows the Solar circle $g_{_N}$ assuming the simple interpolating function adopted for this study (Equation~\ref{Simple_interpolating_function}).}
    \label{g_nu}
\end{figure}

Of course, one should also consider the rather precise constraints on MOND available nowadays from extragalactic rotation curves. In particular, the Spitzer Photometry and Accurate Rotation Curves \citep[SPARC;][]{SPARC} catalogue reveals a tight RAR, which we show in Figure~\ref{g_nu} as the ratio between the observed $g$ at fixed $g_{_N}$ and the prediction $g_\nu$ with different interpolating functions.\footnote{A unique relation is not expected in modified gravity theories because disc galaxies are not spherically symmetric, but deviations from the spherically symmetric relation should be very small \citep{Jones_2018, Chae_2022}.} The error bars show the logarithmic dispersion in $g_{_N}$ about a smooth relation scaled down by $\sqrt{N}$, where $N$ is the number of data points in the $g_{_N}$ bin. It is clear that the MLS and simple functions are numerically rather similar and both match the observations fairly well. However, the standard interpolating function (Equation~\ref{Standard_interpolating_function}) deviates very substantially from observations \citep[this is also apparent in figure~1 of][]{Tian_2019}. As with other analyses, the crucial assumption is the stellar mass-to-light ratio from stellar population synthesis models \citep*{Schombert_2022}, which would need substantial modifications to allow the standard interpolating function to match the data.

This discrepancy is worsened by using an infinitely sharp transition between the Newtonian and deep-MOND regimes, i.e.
\begin{eqnarray}
    \nu ~=~ \max \left( 1, \, \sqrt{\frac{a_{_0}}{g_{_N}}} \right) \, .
    \label{Sharp_interpolating_function}
\end{eqnarray}
This yields no enhancement to gravity at the Solar circle ($\alpha_{\rm{grav}} = 0$), so the results of the WBT work best with this function given that our posterior inference on $\alpha_{\rm{grav}} = -0.021^{+0.065}_{-0.045}$. Even the modest enhancement predicted by the standard interpolating function causes $3\sigma$ tension with our WB results. At the same time, this enhancement is too little to explain extragalactic rotation curves across almost the full range of $g_{_N}$ that they probe (Figure~\ref{g_nu}). Similar conclusions can be drawn from the Galactic rotation curve, which barring major systematics shows that $g > g_{_N}$ by about 50\% at the Solar circle even though both slightly exceed $a_{_0}$ (Figure~\ref{Galactic_rotation_curve}). Therefore, it seems impossible for MOND to simultaneously match galaxy dynamics and \emph{Gaia} data on local WBs. This falsifies classical modified gravity versions of MOND at high significance.

\subsection{Broader implications}
\label{Broader_implications}

So far, we have assumed that MOND should be understood as a modification to Newtonian gravity in the weak-field regime (Equation~\ref{g_cases}). Given the failure of this approach in local WBs, it is helpful to consider MOND as a modification to inertia at low accelerations while preserving Newtonian gravity \citep{Milgrom_1994, Milgrom_2011, Milgrom_2022}. This may avoid significant MOND effects in local WBs because their internal orbital motion has a much higher frequency than their Galactocentric orbit \citepalias[e.g. see figure~14 of][]{Banik_2018_Centauri}. We can consider this in terms of an effective Newtonian external field $g_{_{N,e}}^{\rm{eff}}$, defined such that the boost to the gravity binding a WB is still given by Equation~\ref{eta_QUMOND}, but with argument $g_{_{N,e}}^{\rm{eff}}$ instead of the Newtonian gravity $g_{_{N,e}}$ sourced by the rest of the Galaxy. Since the MOND model considered in the main part of our study predicts a 42\% enhancement to the average radial gravity (Table~\ref{alpha_grav_values}) and our results show that $\alpha_{\rm{grav}} < 0.2$ at quite high confidence (Table~\ref{MCMC_inferences}), we can interpret the WBT as constraining the average gravity binding a WB to within 8\% of the Newtonian prediction in the asymptotic regime where the Galactic gravity dominates over the internal gravity. The corresponding lower limit on $g_{_{N,e}}^{\rm{eff}}$ is $7.8 \, a_{_0}$ ($4.6 \, a_{_0}$) for the simple (MLS) interpolating function. In both cases, this is several times larger than the expected value of $\approx 1.2 \, a_{_0}$ (Section~\ref{WB_orbit_modelling}). The high $g_{_{N,e}}^{\rm{eff}}$ could indicate that the Galactic EFE suppresses the gravity of a local WB by much more than we have assumed, making it very nearly Newtonian.

Modified inertia theories of MOND are still at an early stage of development, so it is not yet possible to conduct detailed simulations of situations like interacting galaxies with a low degree of symmetry. Even so, focusing on systems where the predictions are clear led to the first attempt to distinguish whether MOND is better understood as a modification to gravity or inertia \citep{Petersen_2020}. The results mildly prefer modified gravity at $\approx 1.5\sigma$ confidence. Modified inertia theories make the very strong prediction that test particles on circular orbits will have a unique relation between their kinematic acceleration and the gravity $\bm{g}_{_N}$ that they experience. As a result, the inner and outer parts of galaxy rotation curves should fall on the same RAR. A difference is expected in modified gravity, essentially because the vertical gravity just outside the disc plane is higher nearer the galactic centre, reducing the MOND enhancement \citep*{Banik_2018_Toomre}. Recently, a stacked analysis of SPARC rotation curves identified a $6.9\sigma$ difference between the RAR traced by data points from the inner and outer parts of galaxy rotation curves if the observational uncertainties are taken at face value \citep{Chae_2022}. The difference in the RAR evident in their figure~6 is similar to that expected in modified gravity formulations of MOND. This could be a coincidence caused by unknown systematic errors, but this seems unlikely because almost the same signal is detected when excluding galaxies with a significant bulge or with a high luminosity, which could be signs of a greater degree of pressure support and thus possibly mean a larger uncertainty on the data points at low radii. This suggests that we should take seriously the observed difference in the RAR traced by the inner and outer parts of rotation curves. The high formal significance of the difference evident in their figure~7 is \emph{a priori} not expected in the modified inertia interpretation of MOND. One caveat is that gas motions are not perfectly circular, so until we have a fully fledged formulation of modified inertia theories capable of handling somewhat eccentric orbits, it is impossible to fully conclude on whether the results of \citet{Chae_2022} indeed rule out such theories. But taking their results at face value and given also that the WBT rules out MOND as modified gravity \citep[though see][]{Milgrom_2023, Milgrom_2023b}, we most likely need to fundamentally change MOND if it is to survive the latest constraints, which questions its validity on all scales.

WBs are not the only bound systems that challenge MOND. Its predicted gravity in galaxy clusters generally falls short of the observed value, though one can assume a non-baryonic dark mass component that makes up about half of the gravitating mass \citep*[e.g.][]{Sanders_2003, Angus_2010}. This discrepancy was recently illustrated in terms of the cluster RAR, as shown in figure~8 of \citet{Eckert_2022_cluster_RAR} and figure~5 of \citet{Li_2023}. The RAR traced by galaxy clusters lies about 0.3~dex ($2\times$) above the RAR traced by galaxies when $g_{_N} \ga 0.1 \, a_{_0}$ \citep[][and references therein]{McGaugh_2020}. This gap narrows further and almost vanishes at the low acceleration end. The discrepancy with the Newtonian expectation without DM is of course larger at about 0.8~dex or $6\times$, which matches the ratio between total and baryonic mass in $\Lambda$CDM fits to the power spectrum of anisotropies in the cosmic microwave background radiation \citep{Planck_2020}.

The WBT tests MOND on a much smaller scale than the traditional galaxy-scale tests with rotation curves, velocity dispersions, and weak lensing \citep[section~3 of][and references therein]{Banik_Zhao_2022}. This raises the possibility of suppressing MOND effects at short range, for instance with a length-dependent cutoff to any MOND enhancement to gravity below a scale of 0.1~pc. \citet*{Babichev_2011} proposed an extended version of the Vainshtein screening mechanism in massive gravity \citep*{Vainshtein_1972, Babichev_2010} in which MOND effects are suppressed below a length $r_{_B} \propto M^{1/4}$ around a point mass $M$. Since $r_{_B}$ rises slower than the MOND radius ($r_{_M} \propto M^{1/2}$), we get that in a sufficiently massive system, the suppression of MOND effects arises only in the Newtonian regime and thus has no effect. But in a very low mass system, a new regime appears where the distance $r$ lies in the range $r_{_M} < r < r_{_B}$. In this regime, MOND as classically formulated predicts a departure from Newtonian gravity, but the behaviour becomes Newtonian with the extended screening mechanism \citep{Babichev_2011}.

Another way to think of this is in terms of the phantom dark matter (PDM) density $\rho_{\rm{pdm}}$ generated by MOND, which is defined such that
\begin{eqnarray}
    \nabla \cdot \bm{g} ~\equiv~ -4 \mathrm{\pi} G \left( \rho + \rho_{\rm{pdm}} \right) \, ,
    \label{rho_pdm_definition}
\end{eqnarray}
where $\bm{g}$ must be found by solving Equation~\ref{QUMOND_governing_equation} and $\rho$ is the physical mass density. Since $g \propto \sqrt{M}/r$ in the deep-MOND limit (Equation~\ref{g_cases}), $\rho_{\rm{pdm}} \propto \sqrt{M}/r^2$. If we focus on the PDM density at $r = r_{_B} \propto M^{1/4}$, we get that $\rho_{\rm{pdm}}$ is a constant. Thus, we can think of the model as providing a maximum limit to $\left| \rho_{\rm{pdm}} \right|$ \citep[the modulus is needed because the PDM density can be negative in more complicated geometries; see][]{Milgrom_1986_negative, Oria_2021}. Such an upper limit arises in some attempts to unify the acceleration discrepancies in galaxies with the late-time accelerated expansion of the Universe \citep{Zhao_2007}. We can use the WBT to estimate an upper bound on $\rho_{_0}$, the maximum possible $\left| \rho_{\rm{pdm}} \right|$. Since our analysis indicates that $\alpha_{\rm{grav}} < 0.2$ at rather high confidence (Table~\ref{MCMC_inferences}), we can assume that there is at most a 4\% enhancement to the orbital velocities of WBs in the asymptotic regime. This corresponds to an enclosed phantom mass equal to 8\% of the actual mass in the stars. Assuming our results are sensitive down to $M = 1 \, M_\odot$ and up to separations of $r = 20$~kAU or 0.1~pc, we can estimate that
\begin{eqnarray}
    \rho_{_0} ~<~ 0.08 \, M_\odot \div \frac{4 \mathrm{\pi} r^3}{3} \, = \, 20 \, M_\odot/\rm{pc}^3 \, .
    \label{rho_0_estimate}
\end{eqnarray}
This is three orders of magnitude above the local DM density inferred from a Newtonian dynamical analysis of Galactic data \citep{Read_2014, Hagen_2018, Salomon_2020, Salas_2021}. While we expect $\rho_{\rm{pdm}}$ to be somewhat higher in a dwarf galaxy due to its lower mass, it is clear that the application of MOND to galaxies would be unaffected by a maximum limit to $\left| \rho_{\rm{pdm}} \right|$ of order 20~$M_\odot$/pc\textsuperscript{3}. The impact on larger scales would be even less significant, leaving open the issue of how MOND might be reconciled with observations of galaxy clusters and larger structures.

A maximum limit to $\left| \rho_{\rm{pdm}} \right|$ capable of reconciling MOND with the WBT would substantially reduce MOND effects in the Solar System, invalidating attempts to explain some peculiarities of the Kuiper Belt using MOND \citep{Brown_2023, Migaszewski_2023} but improving the agreement with the rather tight bound on non-standard effects provided by Cassini radio tracking measurements of the Earth-Saturn range \citep{Hees_2014, Hees_2016}. This bound is already so tight that it formally rules out even the exponentially truncated MLS interpolating function at $8\sigma$ confidence \citep[see equation~C15 of][]{Brown_2023}, so it could be difficult to reconcile Solar System ephemerides with galaxy dynamics in classical modified gravity theories of MOND (Desmond et al., in preparation). Since the PDM density at the MOND radius of an isolated point mass $M$ scales as $\rho_{\rm{pdm}} \propto 1/\sqrt{M}$ and we are roughly at the MOND radius of the MW with $M = 7 \times 10^{10} \, M_\odot$ \citep{Banik_2018_escape}, adding $\rho_0$ as a new constant of nature would suppress MOND effects only in systems up to six orders of magnitude less massive than the MW, i.e. up to about $M = 10^5 \, M_\odot$. This is in the regime of massive star clusters and globular clusters, which might help to explain why the internal kinematics of the outer halo globular cluster NGC~2419 are consistent with Newtonian expectations despite feeling little gravity from the Galaxy by virtue of its distance \citep{Ibata_2011a, Ibata_2011b}. While it has been argued that the observations are consistent with MOND as classically understood \citep{Sanders_2012a, Sanders_2012b}, this requires one to invoke observational systematics like a radially varying polytropic equation of state. Since the mass of NGC~2419 is $9 \times 10^5 \, M_\odot$, a clear detection or exclusion of MOND effects here would help to constrain how MOND effects must be suppressed on small scales to ensure consistency with the WBT. Interestingly, MOND correctly predicts the velocity dispersions of isolated LG dwarf galaxies down to $10^6 \, M_\odot$ \citep[figure~3 of][]{McGaugh_2021}, while tidal stability considerations of Fornax Cluster dwarfs reach down to about the same mass \citep[figure~7 of][]{Asencio_2022}. It is therefore possible that MOND as classically formulated breaks down at lower masses and that hints of this are already apparent in NGC~2419. A problem with this scenario is the asymmetric tidal tails of star clusters in the Solar neighbourhood, which have been argued to favour MOND \citep{Kroupa_2022, Jan_2023}. Further investigation is needed to see if the results can be explained in Newtonian gravity with a more complicated Galactic model that includes bars and spiral arms, though recent work indicates that the bar is not sufficient by itself for any pattern speed \citep{Thomas_2023}. If MOND effects persist down to order $10^3 \, M_\odot$ but are not apparent in the WBT, then this would give a tight constraint on where classical MOND breaks down.

A limit to the PDM density might also limit the predicted Milgromian enhancement to the vertical gravity of the Galactic disc, which for now is neither confirmed nor rejected \citep{Zhu_2023}. If it becomes clear that the enhancement is smaller than predicted, this would be interesting because the Galactic rotation curve works very well in MOND \citep{McGaugh_2018, Zhu_2023} and probes scales which are not much larger than the disc thickness, thereby pinning down the boundary between where MOND works and where additional physics must be invoked to suppress MOND effects. While such tests would be useful, for MOND to be valid on kpc scales and ultimately provide the correct explanation for the RAR, we have to invoke at least two new fundamental constants ($a_{_0}$ and $\rho_{_0}$ or a fundamental length). This seems rather contrived and not in line with Occam's Razor, especially as $\rho_0$ is not obviously related to any cosmological density scale (unlike $a_{_0}$) and cannot be substantially reduced below our estimate in Equation~\ref{rho_0_estimate} if we are to preserve the successes of MOND in dwarf galaxies \citep{McGaugh_2021, Asencio_2022}. We therefore discuss other theories that may be able to explain galaxy dynamics while passing the WBT.

Since we do not have a fully fledged quantum gravity theory, the idea of emergent gravity \citep[EG;][]{Verlinde_2017} is that gravity is an entropic force arising from underlying microscopic degrees of freedom that are currently not understood $-$ and do not always need to be, much like how a very detailed understanding of molecules is not needed to understand the ideal gas law. EG faces major issues on galaxy cluster scales \citep{Tamosiunas_2019} and fails in galaxies because it predicts hook-shaped deviations from a single universal RAR that cannot be hidden within the uncertainties \citep*{Lelli_2017_EG}. While its prediction for the WBT is unclear, EG as presently formulated is strongly inconsistent with Solar System ephemerides \citep*{Hees_2017, Chan_2023}. These precise constraints also rule out gravitational dipoles \citep{Hajdukovic_2020, Banik_2020_dipole} and scale-invariant dynamics \citep{Maeder_2020, Banik_2020_SID}. The WBT greatly exacerbates the already severe difficulties faced by modified gravity theories in explaining the RAR without DM while leaving no discernible trace in highly precise Solar System ephemerides, which our work effectively extends to 20~kAU at lower precision. We thus explore whether some form of DM on galaxy scales might provide a tight RAR \citep{Famaey_McGaugh_2012} while lessening the very severe difficulties encountered by the standard CDM approach in other respects \citep[][and references therein]{Kroupa_2010, Kroupa_2012, Kroupa_2015, Valentino_2021, Banik_Zhao_2022}. We stress that the failures of $\Lambda$CDM identified in those works are unrelated to the validity of any alternatives proposed as solutions.

A hybrid MOND-DM model is provided by superfluid dark matter \citep[SFDM;][]{Berezhiani_2015}. Its basic idea is that galaxies are embedded in DM haloes, but their total mass within the region traced by rotation curves is not significant compared to the baryonic mass. Instead, the flat rotation curve problem is solved by postulating additional non-gravitational interactions between the baryons. These are mediated by phonons propagating in the DM halo, which is possible in its central superfluid core. On larger scales, SFDM reduces to $\Lambda$CDM as the superfluid phase only arises at low temperature and high density, so it arises in galaxies but not in galaxy clusters. SFDM should alleviate the problem faced by $\Lambda$CDM with the fast observed rotation speeds of galaxy bars \citep{Roshan_2021_bar_speed} thanks to reduced dynamical friction on subsonic flows like a rotating galaxy bar \citep*{Berezhiani_2019}. However, the LG satellite planes are still very difficult to understand in SFDM because their anisotropy strongly suggests a tidal origin but their internal velocity dispersions imply some enhancement to the forces binding the satellites, which is hard to understand as the more distant satellites in these structures would be outside the superfluid portion of the halo \citep[see section~5.6 of][]{Roshan_2021_disc_stability}. Another major issue is that the phonon-mediated forces at the heart of how SFDM reproduces the MOND phenomenology only enhance the forces on baryons, so it is difficult to understand why the RAR inferred from rotation curves is also evident in strong lensing \citep*{Mistele_2022_lensing} and weak lensing data down to $g_{_N} \approx 10^{-5} \, a_{_0}$ \citep{Brouwer_2021} $-$ the latter are particularly problematic for SFDM \citep*{Mistele_2023}. These observational difficulties need to be considered alongside theoretical difficulties in making a stable covariant theory of DM superfluidity that reduces to MOND in galaxies \citep*{Hertzberg_2021} and avoids significant orbital decay of stars due to Cherenkov-like radiation caused by their orbital velocity exceeding the local sound speed in the superfluid \citep{Mistele_2022_Cherenkov}. If these difficulties are ultimately overcome, a positive aspect of SFDM is that the superfluid phase reduces to a normal phase near a star due to the steep potential gradient, avoiding any anomalous effects in the Solar System. However, this screening mechanism only works within a few hundred AU of a Sun-like star, beyond which MOND-like behaviour would be recovered \citep[equation~86 of][]{Berezhiani_2015}. Therefore, the WBT falsifies SFDM as presently understood. To pass the WBT, the force binding each WB should differ from the Newtonian expectation by $\la 8\%$ out to the separation limit of our WB sample. In the context of SFDM, this would require the normal phase bubble around every local Sun-like star to extend out to $\ga 20$~kAU. This might be difficult to achieve by tuning the model parameters given constraints from galaxies and galaxy clusters \citep{Hodson_2017_SFDM_clusters}.

Thus, the extra forces needed to explain flat galaxy rotation curves might come from the mass of the DM rather than through it mediating a MOND-like interaction between the baryons. While the severe difficulties encountered by $\Lambda$CDM make this approach very unlikely \citep[e.g.][and references therein]{Banik_Zhao_2022}, the hypothetical DM particles might interact with each other $-$ the standard assumption that they only interact gravitationally for all practical purposes is only made for simplicity. Self-interacting dark matter \citep[SIDM;][]{Spergel_2000} offers a promising explanation for galaxies following a tight RAR \citep{Ren_2019}. The required self-interaction cross-section is much larger than the upper limit imposed by the Newtonian dynamical mass profiles of galaxy clusters \citep{Eckert_2022_SIDM}, but this could indicate that the cross-section depends strongly on the velocity. SIDM would limit the central DM density in galaxies, reducing dynamical friction on galaxy bars and thus possibly alleviating the $13\sigma$ fast bar tension faced by $\Lambda$CDM \citep{Roshan_2021_bar_speed}. SIDM would reduce to $\Lambda$CDM on large scales and thus still suffer the same issues with regards to the KBC void and Hubble tension \citep{Haslbauer_2020} and the early formation of galaxies \citep{Haslbauer_2022_JWST} and galaxy clusters \citep{Asencio_2021, Asencio_2023}. In addition, SIDM does not provide an obvious explanation for the LG satellite planes because DM self-interactions do not have any obvious effect on the positions and velocities of satellite galaxies around their host $-$ though the frequency of satellite planes in SIDM should be checked. Moreover, it is unclear how SIDM can explain the observed signs of tidal disturbance in Fornax Cluster dwarf galaxies and the lack of low surface brightness dwarfs towards the cluster centre \citep{Asencio_2022}. Reducing the central DM density would make the situation less problematic, but the DM fraction within the baryonic extent of the Fornax dwarfs would need to be less than for isolated LG dwarfs with accurately measured internal velocity dispersions $-$ these require a considerable amount of DM in a Newtonian context. Thus, SIDM models face some challenges but also appear to hold some promise, especially as they may explain some of the MOND phenomenology while being consistent with the WBT and Solar System ephemerides due to the lack of any change in the gravity sector. The same is true if we postulate a long-range non-gravitational interaction between baryons and dark matter beyond their self-interactions \citep{Famaey_2020}.

\subsection{Future prospects}
\label{Future_prospects}

The significant challenge to MOND presented by the WBT is an important result which should be confirmed with additional data. There are good prospects for future improvement: the next \emph{Gaia} data release (DR4) is anticipated at the end of 2025. DR4 will double the time baseline analysed in DR3, giving an $\approx 2.8\times$ improvement in proper motion precision and reducing the impact of CBs at AU separations (Appendix~\ref{Non_linear_orbital_motion}). We estimate that using DR4 with a similar selection to our main analysis but extended to $m_G < 17.6$ and $d < 400$~pc would approximately double the search volume and almost triple the final sample size, bearing in mind that more precise proper motions will lead to more systems passing our cut on the $\widetilde{v}$ precision (Equation~\ref{Max_vtilde_uncertainty}).

Systems rejected here due to a lack of either star's RV can also be recovered from ground-based RV observations at a moderate precision of $\approx 1$~km/s. This would increase the sample size by about 10\%, but the more useful gain could be that the contamination rate can be reduced if the RV is known for both stars.

There are also interesting prospects for directly constraining the triple population to reduce the degrees of freedom inherent in our modelling \citep{Manchanda_2023}. Their simulations have shown that we should be able to detect nearly all triples with a single main sequence star orbited by a CB whose inner period $\ga 3$~years, which is long enough to give a substantial time-averaged perturbation to the photocentre velocity (Appendix~\ref{Non_linear_orbital_motion}). Such hierarchical systems can be found either by \emph{Gaia} astrometric accelerations in the 10-year final dataset (for separations $\la 25$~AU); or by direct, speckle, or coronagraphic imaging (for separations $\ga 20$~AU). The required imaging observations are much less time-consuming than high-precision RV measurements, so they may be feasible for a substantial fraction of the WB sample, especially for the high $\widetilde{v}$ tail. This would provide an important check on our estimated CB contamination fraction.

In principle, it is possible to refine the WBT via the addition of binary RV differences. This would allow the use of 3D velocities and 2D projected separations, which would reduce the scatter in the $\widetilde{v}$ distribution inherent from projected velocities. However, this would require high precision (order 0.05~km/s) RVs for all stars, which would be very costly in observing time and require corrections for gravitational redshift \citep{Loeb_2022} and convective blueshift \citep{Liebing_2021}. A fully 3D version of the WBT would also require much more precise astrometry to obtain kAU-level constraints on the relative heliocentric distances to the stars in each WB. This is not envisaged in the foreseeable future for a statistically large sample of WBs. More precise information is available for some very nearby systems \citep{Kervella_2016, Kervella_2017}, but an individual WB cannot test MOND using velocities alone even if $\widetilde{v} > \sqrt{2}$ because a single snapshot would not prove that the WB is bound \citepalias[though a chance flyby is unlikely; see section~8.1 of][]{Banik_2018_Centauri}. A test is possible if the acceleration can be directly measured, which may become feasible with future observatories \citep{Banik_2019_Proxima}.

Our work has assumed that the potential of a star at kAU distances must be traced by another star. It is possible for a spacecraft to serve as the tracer instead \citep{Banik_2019_spacecraft}. The predicted aspherical shell of PDM concentrated around the Solar MOND radius would also affect Solar System ephemerides \citep{Hees_2014, Hees_2016, Brown_2023}.

\section{Summary and conclusions}
\label{Conclusions}

MOND has enjoyed unparalleled predictive success with regards to galaxy dynamics \citep[][and references therein]{Famaey_McGaugh_2012, Banik_Zhao_2022}. Its central postulate is that the dynamics of a system deviates from Newtonian expectations when the gravity $g \la a_{_0}$. This could in principle occur at any distance from a system provided it has a sufficiently low mass (Equation~\ref{MOND_radius}). Since the MOND radius of a star is $7\sqrt{M/M_\odot}$~kAU, MOND ought to have detectable effects on local WBs with kAU separations. Indeed, simple analytic estimates backed up by detailed calculations show a 20\% enhancement to the orbital velocity over the Newtonian expectation \citepalias{Banik_2018_Centauri}, as also evident in $N$-body simulations of local star clusters for much the same reason \citep{Kroupa_2022}.

We test this prediction using an observational sample of 8611 WBs from \emph{Gaia}~DR3 \citep{Gaia_2023} where the uncertainty on $\widetilde{v}$ (Equation~\ref{v_tilde_definition}) is very small and other quality cuts have been applied (Sections~\ref{Gaia_DR3_sample} and \ref{Refined_quality_cuts}). Dividing our sample into ten equally sized subsamples shows no trend in the median $\widetilde{v}$ with respect to $r_{\rm{sky}}/r_{_M}$ once we restrict our attention to $\widetilde{v} \la 2$, which should significantly reduce contaminating effects while preserving a genuine MOND signal (Figure~\ref{Median_vtilde_values}). Observational uncertainties would be expected to broaden the $\widetilde{v}$ distribution preferentially at large separations because the same uncertainty on the relative velocity implies a larger uncertainty in $\widetilde{v}$. This would if anything lead to a broader $\widetilde{v}$ distribution at low accelerations. As a result, the flat behaviour of the median $\widetilde{v}$ with respect to our proxy for the internal WB acceleration strongly suggests that local WBs are Newtonian.

We then present the most detailed statistical hypothesis test of MOND to date using local WBs. Our model includes a rigorous calculation of the MOND gravitational field under an EFE of strength $g_e = 1.8 \, a_{_0}$ (Section~\ref{WB_population}), an allowance for an undetected CB companion to one or both of the stars in each WB (Section~\ref{Undetected_companions}), and LOS contamination (Section~\ref{Chance_alignments}). The procedure was fixed in advance as much as possible to minimize biases that could arise when conducting the WBT \citep{Banik_2021_plan}. The model parameters relate to those of:
\begin{enumerate}
    \item The WBs, for which we need the $\alpha_{\rm{grav}}$ parameter interpolating between different gravity laws (see Equation~\ref{alpha_grav_est} for its relation to the gravity law) and the distribution of semi-major axes ($a_{\rm{break}}$ and $\beta$; see Equation~\ref{a_cases}) and orbital eccentricities ($\gamma$; see Equation~\ref{gamma_definition});
    \item The CBs, for which we assume the same $\gamma$ parameter but also need the fraction $f_{\rm{CB}}$ of stars in our WB sample with an undetected CB companion (Section~\ref{CB_WB_convolution}) and the upper limit to the CB semi-major axis, which is defined by $k_{\rm{CB}}$ (see Equation~\ref{k_CB_defintion}); and
    \item The LOS contamination fraction $f_{\rm{LOS}}$ due to chance alignments (see Section~\ref{Chance_alignments}).
\end{enumerate}
These seven parameters are allowed to vary freely so as to best match the observed distribution of $\left( r_{\rm{sky}}, \widetilde{v} \right)$ in 540 pixels (Table~\ref{Pixellation_decision}).

Our best-fitting Newtonian model significantly outperforms our best-fitting MOND model: the likelihood ratio of $\exp \left( 175 \right)$ implies a preference for Newtonian gravity at $19\sigma$ confidence according to our analyses with fixed $\alpha_{\rm{grav}}$. This is in line with the result of \citetalias{Pittordis_2023}, who conducted a less detailed version of the WBT using a somewhat different $r_{\rm{sky}}$ range of $5-20$~kAU and considered WBs with $\widetilde{v} < 7$. Their analysis also differs from ours in several other respects, including their use of a linear rather than cubic mass-luminosity relation (Figure~\ref{Mass_luminosity_relation}), their CB model, and their much more limited exploration of the parameter space. Those authors found that the $\chi^2$ of their best Newtonian model is smaller than that of their best MOND model by 525, suggesting a $23\sigma$ preference for Newtonian gravity. While our best-fitting Newtonian model is not a perfect representation of the WB dataset, it is clear that the data do not show the predicted broadening to the $\widetilde{v}$ distribution at larger separations (Figure~\ref{Photo_fixed_gravity}).

In our main analysis, we allow the gravity law to interpolate between Newtonian and Milgromian. For this, we use the gravity law parameter $\alpha_{\rm{grav}}$, which is 0 in Newtonian gravity and 1 in MOND. Using $10^5$ MCMC trials to explore the parameter space, we find that the inferred $\alpha_{\rm{grav}} = -0.021^{+0.065}_{-0.045}$. This is very consistent with Newtonian dynamics but rules out MOND at $16\sigma$ confidence, which is in line with the results of \citetalias{Pittordis_2023} and our results for fixed $\alpha_{\rm{grav}}$. Our result is robust to various changes in the modelling assumptions and the sample selection (Table~\ref{MCMC_inferences}), including when we use a narrower mass range to minimize possible trends in the mass distribution across the parameter range used for the WBT (Section~\ref{Restricted_mass_range}). We also explain in some detail why, without a drastic change to our understanding of baryonic surface densities, the MOND interpolating function cannot be chosen to simultaneously pass the WBT and constraints from the rotation curves of galaxies, including our own (Section~\ref{MOND_interpolating_function}).

Our conclusion disagrees with two studies that were published while this work was under review \citep{CHAE_2023, Hernandez_2023}. In Section~\ref{Other_WBT_results}, we focus on the study of \citet{CHAE_2023} due to its much larger sample size and its claim to have detected the predicted MOND enhancement to Newtonian gravity at $10\sigma$ confidence. We find that the nominal sample of 26615 WBs used in that study does indeed show a clear signal that closely resembles the MOND expectation and appears to rule out Newtonian dynamics (Figure~\ref{Kyu_medians_all}). However, we then identify a major deficiency with the handling of astrometric uncertainties. Despite the importance of the relative velocity between the stars in each WB, the uncertainty on this quantity is not estimated in \citet{CHAE_2023}. Instead, the focus is on ensuring that the heliocentric velocity of each star is very precise. This is insufficient to ensure a precisely known relative velocity and thus a reliable $\widetilde{v}$, which however is essential to conducting the WBT. We therefore use equation~5 of \citet{Badry_2021} to estimate the uncertainty in $v_{\rm{sky}}$ and thus in $\widetilde{v}$. We then impose an additional condition on the $\widetilde{v}$ uncertainty matching that used in our own analysis (Equation~\ref{Max_vtilde_uncertainty}). This completely removes the apparent MOND signal despite only reducing the sample size by about 1/5 (see Figure~\ref{Kyu_medians_v01}). We argue that a similar issue may well have affected the analysis of \citet{Hernandez_2023} based on some statements in that paper, though we do not study its WB sample in detail. We also find that the WB sample used by \citet{CHAE_2023} artificially imposes that $\widetilde{v} < \sqrt{5 \, M_\odot/M}$, making it not very well suited to the WBT (Appendix~\ref{vtilde_limit_Kyu_impact}).

We conclude that the gravity law inferred from our analysis of local WBs is consistent with Newtonian expectations but rules out MOND as modified gravity at $\gg 5\sigma$ confidence in both our nominal analysis and a considerable range of variations to it. This conclusion is of course reliant on our modelling approach, which is not a perfect match to the data (Figure~\ref{Photo_best_nominal_fCB03}).\footnote{This is similar to the rotation curve predictions in MOND, which often do not match the data within formal uncertainties but visually provide a good fit, which is typically considered sufficient given the inevitable modelling deficiencies \citep{Kroupa_2018, Cameron_2020}.} Even so, neither our main analysis nor any of the variations considered can significantly improve the fit by using a gravity law different to Newtonian. This is also evident from the model-independent Figure~\ref{Median_vtilde_values}, which considers the median $\widetilde{v}$ in subsamples with different $r_{\rm{sky}}/r_{_M}$ as a proxy for the WB internal acceleration. It has been argued that modified inertia interpretations of MOND could reduce the predicted enhancement to the Newtonian acceleration and thus remain consistent with the WBT \citep{Milgrom_1994, Milgrom_2011, Milgrom_2022}. This would lead to the inner and outer parts of galaxy rotation curves following the same RAR, but a difference is detected at $6.9\sigma$ confidence \citep{Chae_2022}. Although the predictions of modified inertia for slightly non-circular orbits are not known, the magnitude of the above-mentioned difference is consistent with expectations of MOND as modified gravity $-$ but this is ruled out by the WBT. Further modifications to MOND could preserve its successes on galaxy scales, especially with regards to issues like the LG satellite planes, which are difficult to understand any other way \citep{Pawlowski_2021_Nature, Pawlowski_2021, Banik_2022_satellite_plane}. This entails at least one new fundamental scale beyond $a_{_0}$, marking the end of MOND as a purely acceleration-dependent modification to standard physics. Limiting the MOND phantom density to $\left| \rho_{\rm{pdm}} \right| \la 20 \, M_\odot$/pc\textsuperscript{3} would yield consistency with the WBT and also alleviate tensions related to the null detection of MOND effects in Cassini radio tracking data from Saturn \citep{Hees_2014, Hees_2016, Brown_2023}. It might also improve the agreement with observations of globular clusters and the vertical gravity from the Galactic disc. Limiting $\left| \rho_{\rm{pdm}} \right|$ would have little effect on larger scales, similarly to the Vainshtein mechanism used to screen modified gravity effects in the Solar System \citep{Babichev_2011}. This would preserve MOND's successes in galaxies but leave open the issue of galaxy clusters and large-scale structure, for which the hybrid $\nu$HDM paradigm may be a promising approach \citep[section~9.2 of][and references therein]{Banik_Zhao_2022}. While our results falsify MOND as currently understood, given the many problems for $\Lambda$CDM discussed in that work, our results cannot be used to argue that it is the correct model either $-$ both models are clearly incomplete. Hybrid models like SFDM struggle to explain galactic-scale observations like lensing \citep{Mistele_2022_lensing, Mistele_2023} and the WBT, while non-MOND modifications to gravity like EG and MOG usually fail in galaxies (Section~\ref{Broader_implications}). SIDM may be a promising approach but it is very similar to $\Lambda$CDM on large scales, thus encountering the same difficulties with the KBC void \citep{Keenan_2013, Haslbauer_2020, Wong_2022} and Hubble tension \citep{Valentino_2021}. This significant anomaly for standard cosmology appears to persist in the JWST era \citep{Yuan_2022} and must be solved consistently with the ages of the oldest stars \citep[][and references therein]{Cimatti_2023}. We hope that our results from local WBs motivate the development of a more complete theory, which is likely to borrow some elements from both $\Lambda$CDM and MOND given their successes in different domains.

\section*{Acknowledgements}
\label{Acknowledgements}

IB is supported by Science and Technology Facilities Council grant ST/V000861/1, which also partially supports HZ. IB acknowledges support from a ``Pathways to Research'' fellowship from the University of Bonn, during which the primary statistical analysis was largely coded. BF and RI acknowledge funding from the Agence Nationale de la Recherche (ANR projects ANR-18-CE31-0006 and ANR-19-CE31-0017) and from the European Research Council (ERC) under the European Union’s Horizon 2020 Framework programme (grant agreement number 834148).

The authors thank the referee for highlighting some important deficiencies with an earlier version of this work regarding the modelling of undetected companions. They also thank Pavel Kroupa for comments on the article and for lending computational resources from the Stellar Populations and Dynamics Research Group at the University of Bonn, where all the statistical analyses were conducted. IB is very grateful to Elena Asencio for helping to check every line of the statistical analysis code and for lending algorithms used for making several figures in this paper, especially the triangle plots made using \textsc{pygtc} \citep{Bocquet_2016}. He thanks Haixia Ma for providing Figure~\ref{Galactic_rotation_curve}. The authors are grateful for a masters project by Nathan Findlay which gave a much better understanding of how to analyse the dataset and what quality cuts would be appropriate. IB acknowledges significant computational resources provided by Kyu-Hyun Chae to simulate the impact of undetected companions. Most of the figures were prepared using \textsc{matlab}.

\section*{Data availability}

The WB dataset used in this contribution will be made available upon reasonable request, showing \emph{Gaia} catalogue information for the sample of 19786 WBs described in the text and an extra flag to indicate if each WB is one of the 8611 systems used in our main analysis. The WB dataset used in the nominal analysis of \citet{CHAE_2023} was provided by Kyu-Hyun Chae. \emph{Gaia}~DR3 is publicly available.

\bibliographystyle{mnras}
\bibliography{WBT_bbl}

\begin{appendix}

\section{Exactly equal mass binaries}
\label{Equal_mass_binaries}

Although we might expect the mass ratio distribution of binaries to be smooth, observations indicate a population of exactly equal mass WBs \citep{Badry_2019_twin}. We therefore include an allowance for a $\delta$-function in the $\widetilde{q}$ distribution at $1/2$, where $\widetilde{q}$ is the fraction of the CB mass in the less massive star. To find the likelihood $P_{\rm{eqm}}$ of a binary having exactly equal mass stars, we plot the cumulative distribution of $0.5 - \widetilde{q}$ for the WBs in our sample, which we restrict to $\widetilde{v} < 1$ to try and mitigate CB contamination. The idea is to extrapolate the distribution down to zero and look for a positive intercept. Since a $\delta$-function would be smeared somewhat by observational uncertainties, we need to start our fit at some slightly positive value of $0.5 - \widetilde{q}$. We also need to avoid extending our fit to very high $0.5 - \widetilde{q}$ because our polynomial fitting function might become inaccurate.

\begin{figure}
    \centering
    \includegraphics[width=0.47\textwidth]{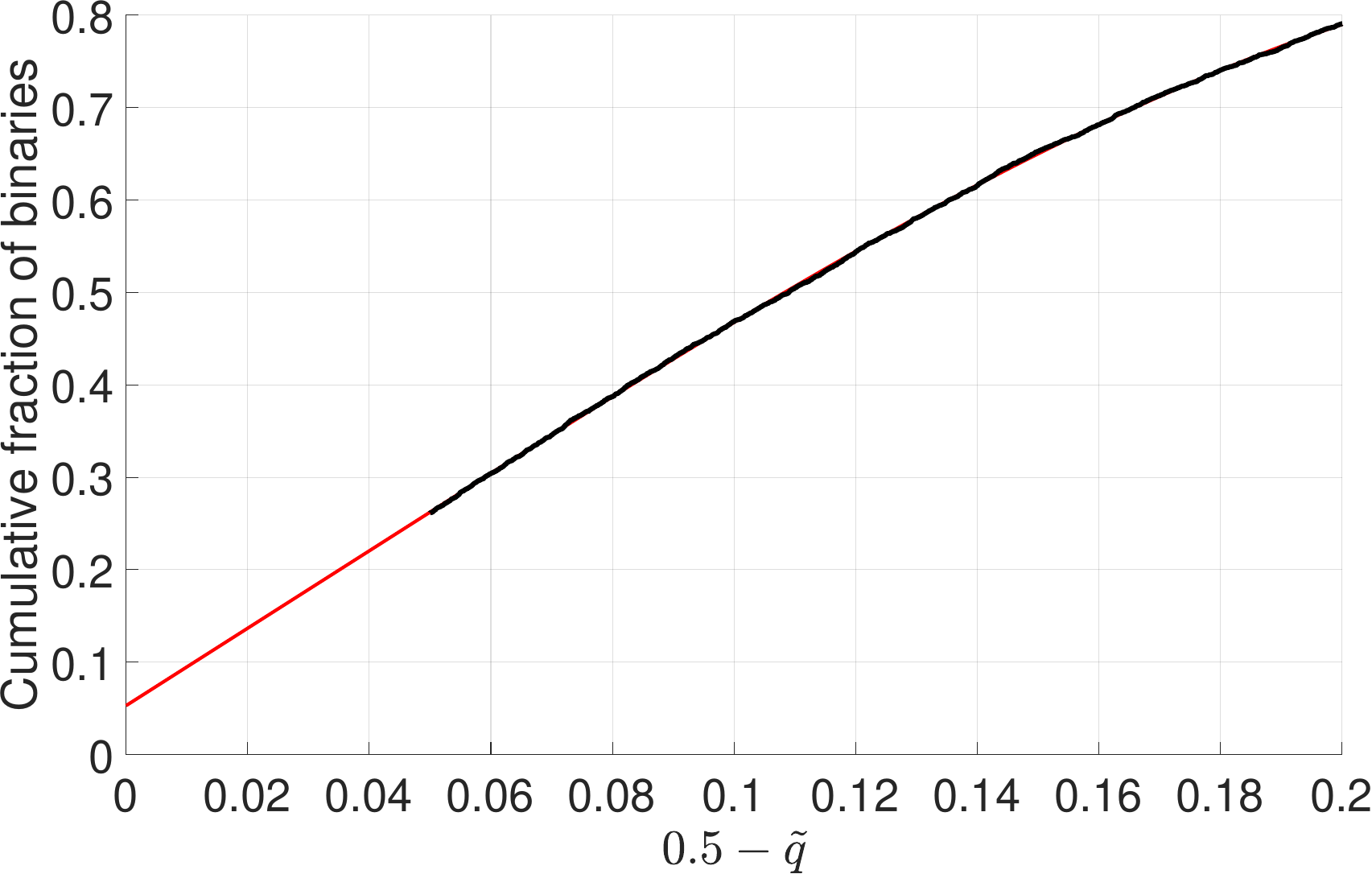}
    \includegraphics[width=0.47\textwidth]{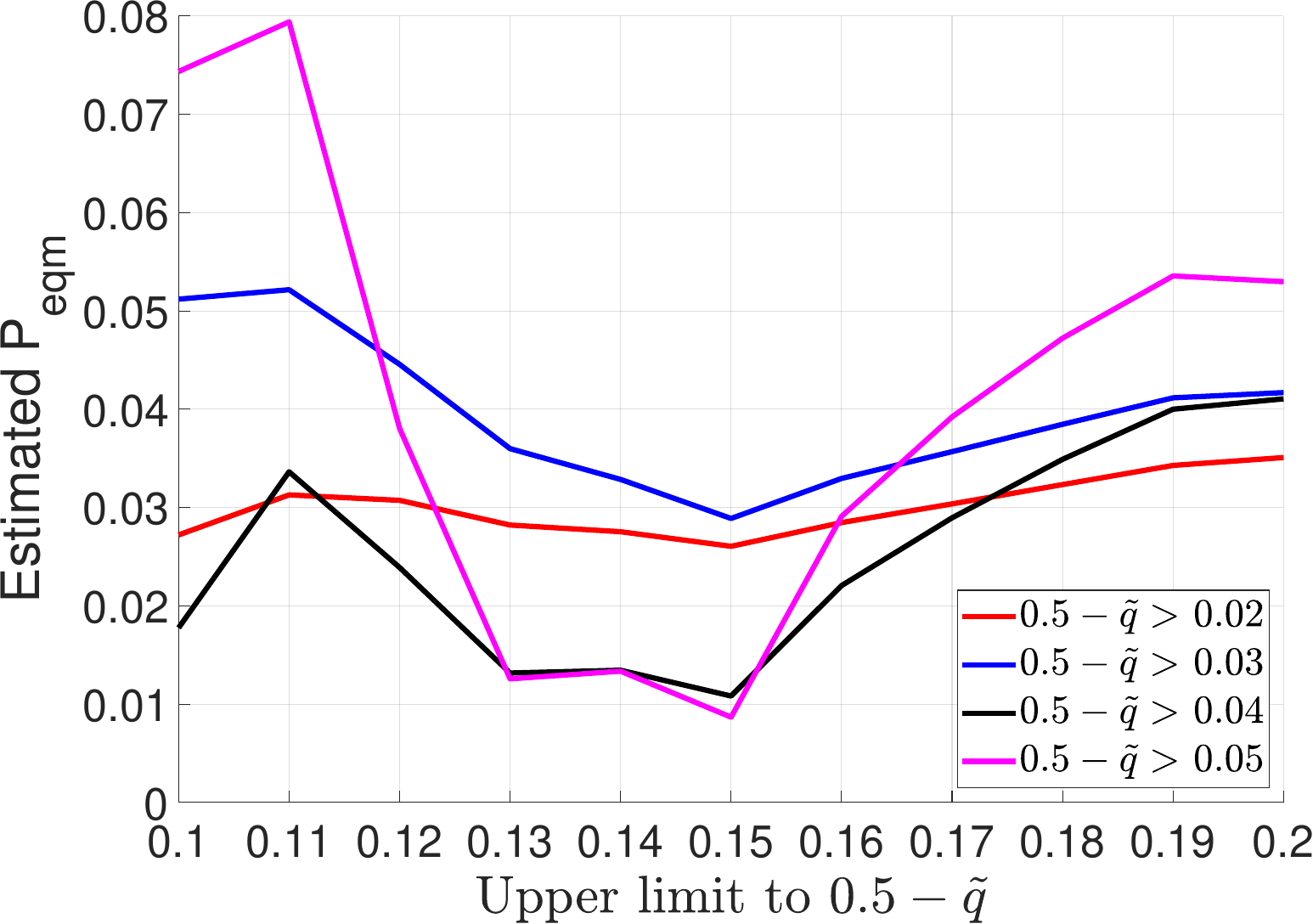}
    \caption{\emph{Top}: The cumulative distribution of $0.5 - \widetilde{q}$ for WBs with $\widetilde{v} < 1$ (black line). We only show results over the range $0.05-0.2$ because a $\delta$-function in the probability distribution at zero would be smeared somewhat by observational uncertainties. The red line shows a cubic fit over this range, but also extrapolates it down to zero. Notice that the intercept is slightly positive, indicating the presence of a population of WBs with an exactly equal mass \citep[consistent with][]{Badry_2019_twin}. \emph{Bottom}: The above-mentioned intercept is plotted as a function of the upper limit to the fitting range. Different lines show results with different lower limits, as indicated in the legend. Notice that for a wide range of fitting ranges, the results converge around 0.04, which we assume is the likelihood $P_{\rm{eqm}}$ of a binary having exactly equal mass components.}
    \label{P_eqm_justification}
\end{figure}

The results in the top panel of Figure~\ref{P_eqm_justification} show real data and a cubic fit over the range $0.05-0.2$. The data are shown in black, while the cubic fit is shown in red $-$ a cubic is the lowest degree polynomial which provides an accurate match to the data. The bottom panel of Figure~\ref{P_eqm_justification} shows the intercept as a function of the upper limit to $0.5 - \widetilde{q}$, with each coloured line used to show results for a different lower limit. All analysis variants show a positive intercept, whose value converges at close to 0.04 for a wide range of lower and upper limits to $0.5 - \widetilde{q}$. We therefore adopt $P_{\rm{eqm}} = 0.04$ for the CB model in our main analysis. This assumption is not relevant to modelling uncontaminated WBs, whose actual mass ratio distribution is used when needed (Section~\ref{CB_WB_convolution}). An equal mass fraction of a few percent is in line with the double main sequence apparent in the colour-magnitude diagram of our WB sample (Figure~\ref{Colour_magnitude_diagram}).

\section{When orbital motion becomes non-linear}
\label{Non_linear_orbital_motion}

We assume uniform rectilinear motion for the WB and any undetected CB companions, implicitly assuming that the orbital periods are very long. This should be a very good approximation for the WB given the kAU separations of even the tightest WBs we consider. However, CB orbital periods can be much shorter. This can substantially reduce the mean motion of the CB over the \emph{Gaia}~DR3 observing baseline of $t_G = 34$~months. In this section, we estimate the CB orbital period below which we can no longer safely assume a very long orbital period $P$. This is used to estimate a minimum CB semi-major axis $a_{\rm{int}}$ when setting up its distribution in Section~\ref{aint_distribution}.

For simplicity, we assume that the CB is on a circular orbit and that the line connecting its components rotates by some angle $\theta$ over the \emph{Gaia} observing baseline. We consider the impact of increasing $\theta$ by reducing the CB separation while not altering the other CB orbital parameters. The induced recoil velocity on the star detected as part of the WB depends on the CB orbital velocity $v \propto 1/\sqrt{a_{\rm{int}}}$. Bearing in mind Kepler's Third Law that $P \propto a_{\rm{int}}^{3/2}$, we get that $v \propto P^{-1/3}$. Since the CB rotation angle $\theta$ over a fixed duration of time varies as $\theta \propto 1/P$, we get that
\begin{eqnarray}
    v ~\propto~ \theta^{1/3} \, .
    \label{v_theta_linear}
\end{eqnarray}
The positive exponent captures the fact that larger $\theta$ is associated with a tighter CB on a faster orbit.

However, the mean velocity $\overline{v}$ will in general involve some additional shape factor $s \left( \theta \right)$ that accounts for the extent to which the orbital arc over the \emph{Gaia} observing baseline deviates from a straight line. Basic trigonometry tells us that the linear distance between two points on a circle is smaller than the distance between them along the circumference by a factor of $\sinc \left( \theta/2 \right)$. Putting this extra factor into Equation~\ref{v_theta_linear} tells us that
\begin{eqnarray}
    \overline{v} ~\propto~ \theta^{-2/3} \left| \sin \left( \frac{\theta}{2} \right) \right| \, .
    \label{v_theta}
\end{eqnarray}
Treating this as a function of $u \equiv \theta/2$, we find that the maximum occurs when $\tan u = 3u/2$. The first non-trivial solution is $u = 0.97$ or $\theta = 111^\circ$. Rotation by this angle over $t_G$ implies $P = 3.25 \, t_G$, so we expect our linear motion approximation to break down when $P < 110$~months or 9.2~years.

The corresponding orbital separation depends on the mass of the CB. For a low-mass undetected companion, the CB total mass would only slightly exceed the mass of the contaminated star, which is just one of the two stars detected as a WB. The blue bars in Figure~\ref{Mass_distribution} show that the stars in our WB sample easily reach down to $0.4 \, M_\odot$. At this mass, an orbital period of 9.2~years corresponds to $a_{\rm{int}} = 3.2$~AU. While CBs with even smaller separations would still affect the inferred WB orbital velocity, the impact rapidly becomes smaller than implied by Equation~\ref{v_theta_linear}.

We note that 3.2~AU is much larger than the minimum $a_{\rm{int}}$ of 0.1~AU considered in the orbit integrations of \citetalias{Pittordis_2023} (see their section~3.2). CBs with a separation of 0.1~AU would complete many orbits over $t_G$, so the effect of such tight CBs may have been overestimated in their analysis.

\section{Comparing the best Newtonian and MOND models}

\begin{figure}
    \centering
    \includegraphics[width=0.47\textwidth]{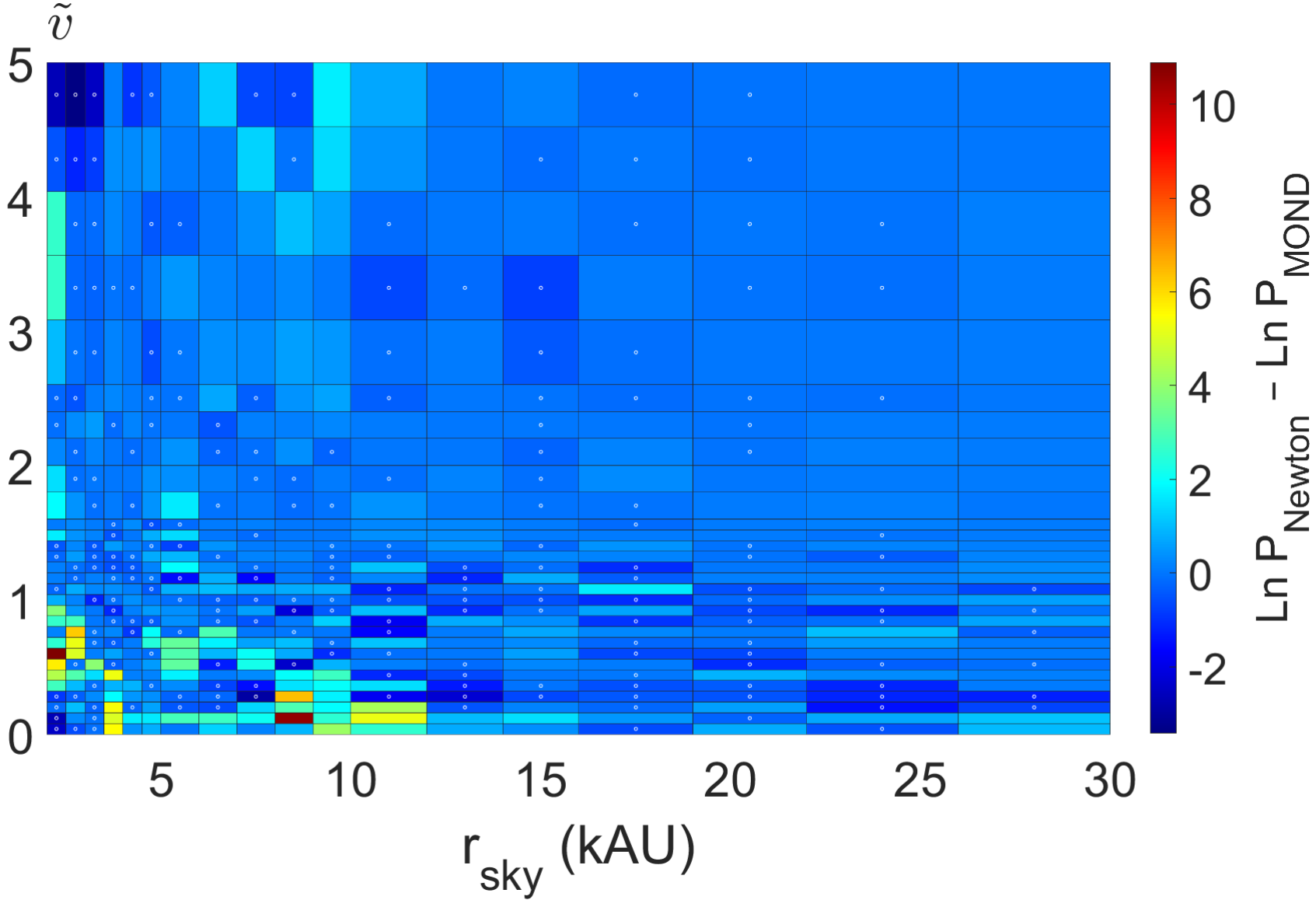}
    \caption{The relative performance of our best-fitting Newtonian and MOND models, shown here for every pixel. The binomial log-likelihood of each model is found by comparison with the observed number of WBs in each pixel (Figure~\ref{Observed_photo_tiled}). Open white circles indicate pixels for which MOND outperforms Newtonian gravity. The combined difference in log-likelihood at each $r_{\rm{sky}}$ summed across all $\widetilde{v}$ (and vice versa) is shown in Figure~\ref{Ln_P_Newton_MOND_summed}.}
    \label{Ln_P_Newton_MOND}
\end{figure}

In Figure~\ref{Ln_P_Newton_MOND_summed}, we presented the difference in log-likelihood between our best-fitting Newtonian and Milgromian models. The panels in this figure showed the results for each $r_{\rm{sky}}$ but with all $\widetilde{v}$ pixels summed over, and vice versa. To give a better understanding of which pixels work better in which theory, we show the full 2D distribution of $\Delta \ln P$ between these models in Figure~\ref{Ln_P_Newton_MOND}. Since $\Delta \ln P$ is close to zero in several pixels, we clarify which model does better by adding an open white circle at the centre of a pixel if MOND fits it better. Out of the 540 pixels used in our analysis (Table~\ref{Pixellation_decision}), only 221 pixels (41\%) prefer MOND while the remaining 59\% prefer Newtonian gravity. However, we also need to consider that the pixels which prefer MOND only do so to a rather small extent, while sometimes pixels which prefer Newtonian gravity do so to a very large extent.

\section{The best fit model with reduced CB contamination}
\label{Best_fit_fCB03}

The only alteration to our nominal analysis which gives an appreciable preference for $\alpha_{\rm{grav}} > 0$ is the one in which we fix $f_{\rm{CB}} = 0.3$ (Table~\ref{MCMC_inferences}). We explore this model in more detail to understand if it plausibly fits the WB data with a gravity law closer to Milgromian.

\begin{figure*}
    \centering
    \includegraphics[width=0.48\textwidth]{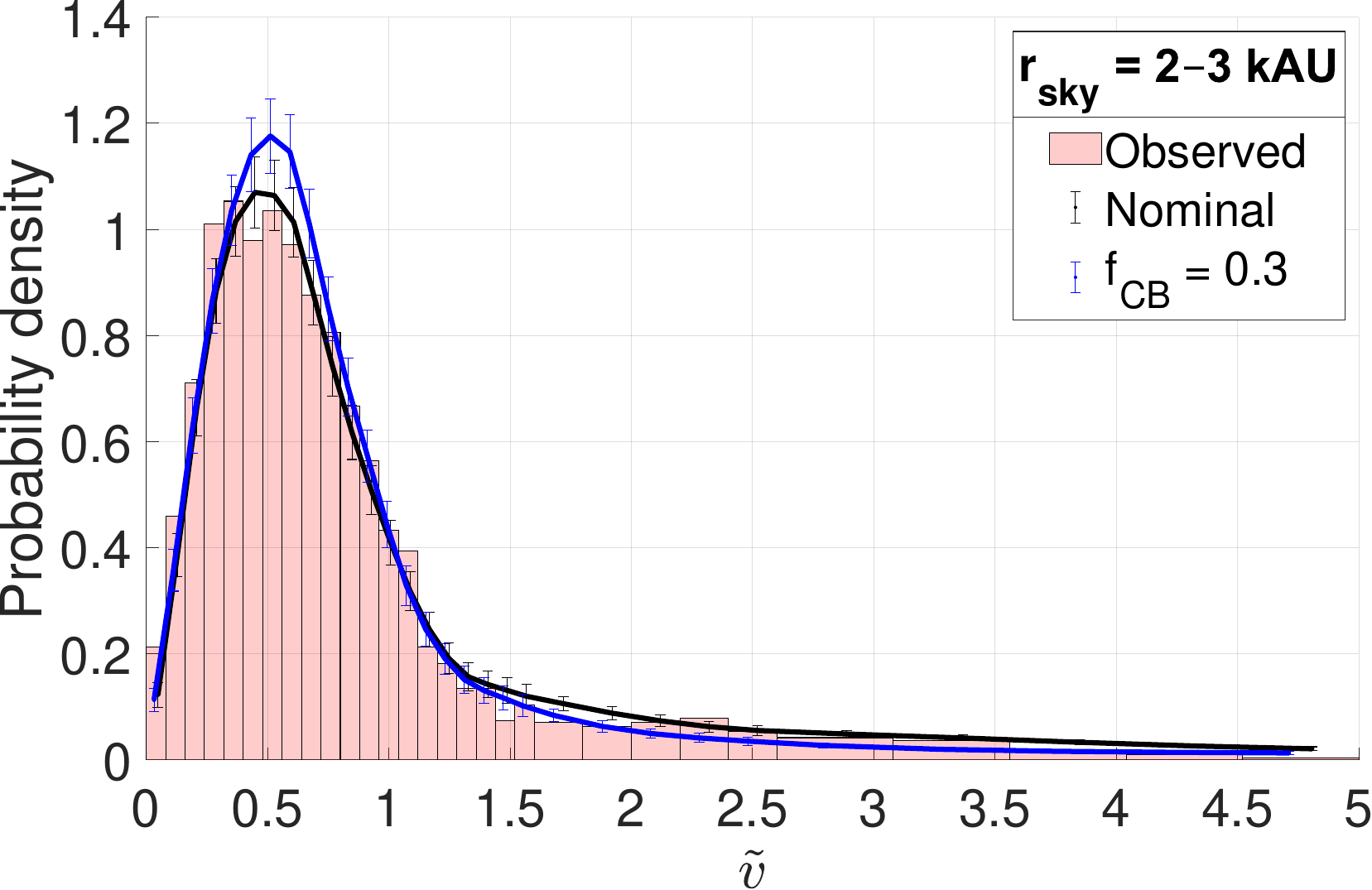}
    \hfill
    \includegraphics[width=0.48\textwidth]{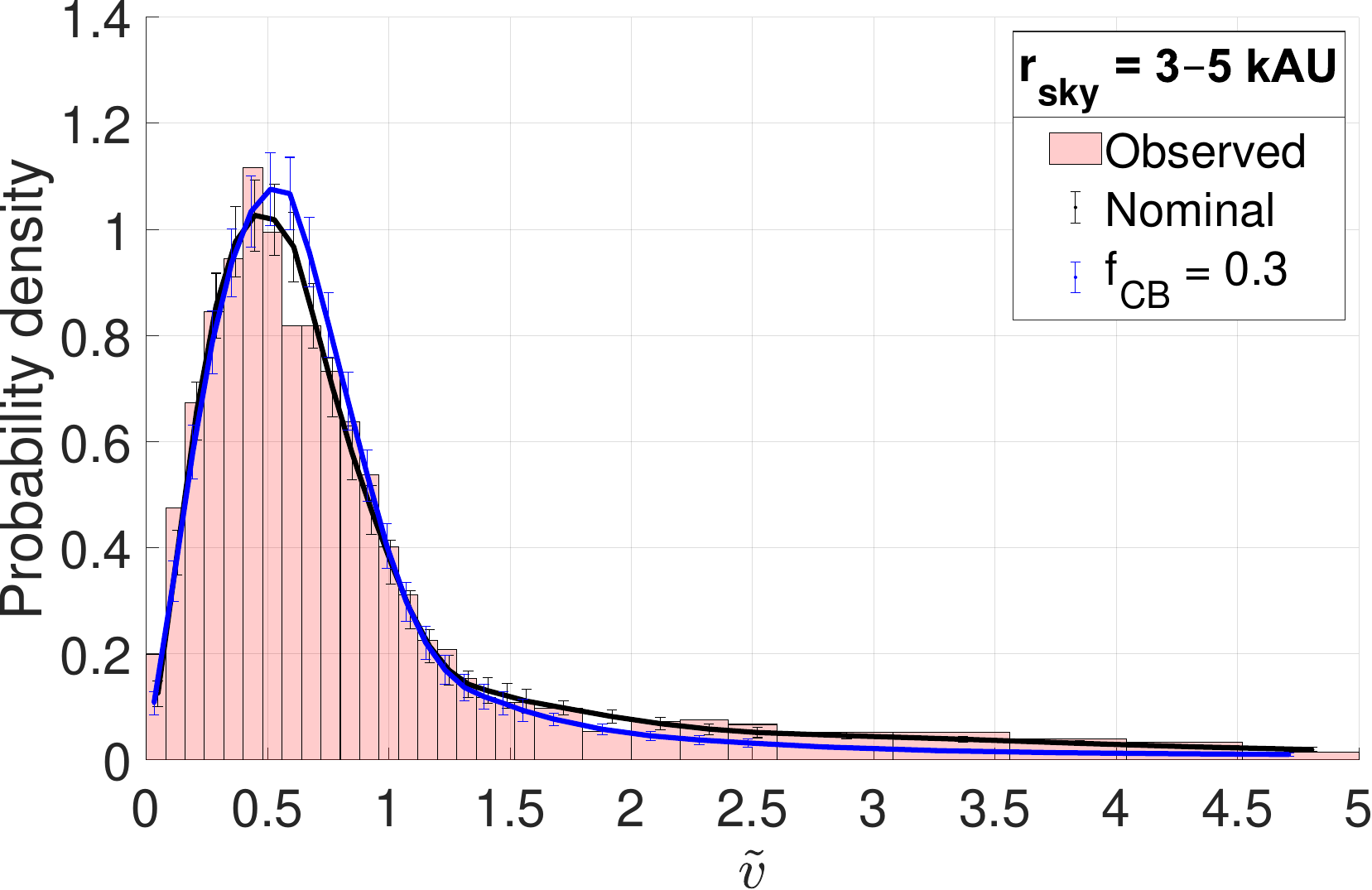}
    \includegraphics[width=0.48\textwidth]{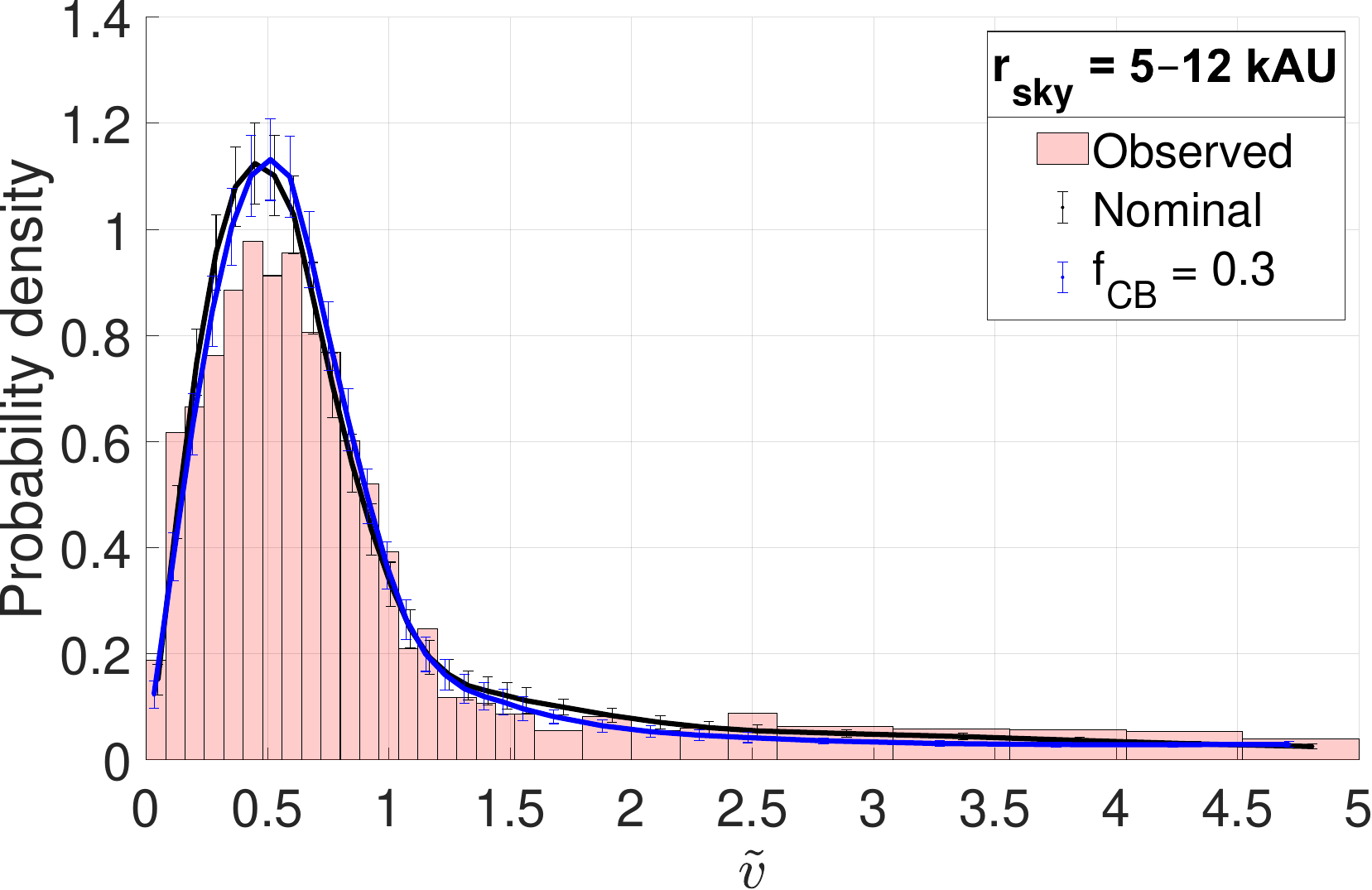}
    \hfill
    \includegraphics[width=0.48\textwidth]{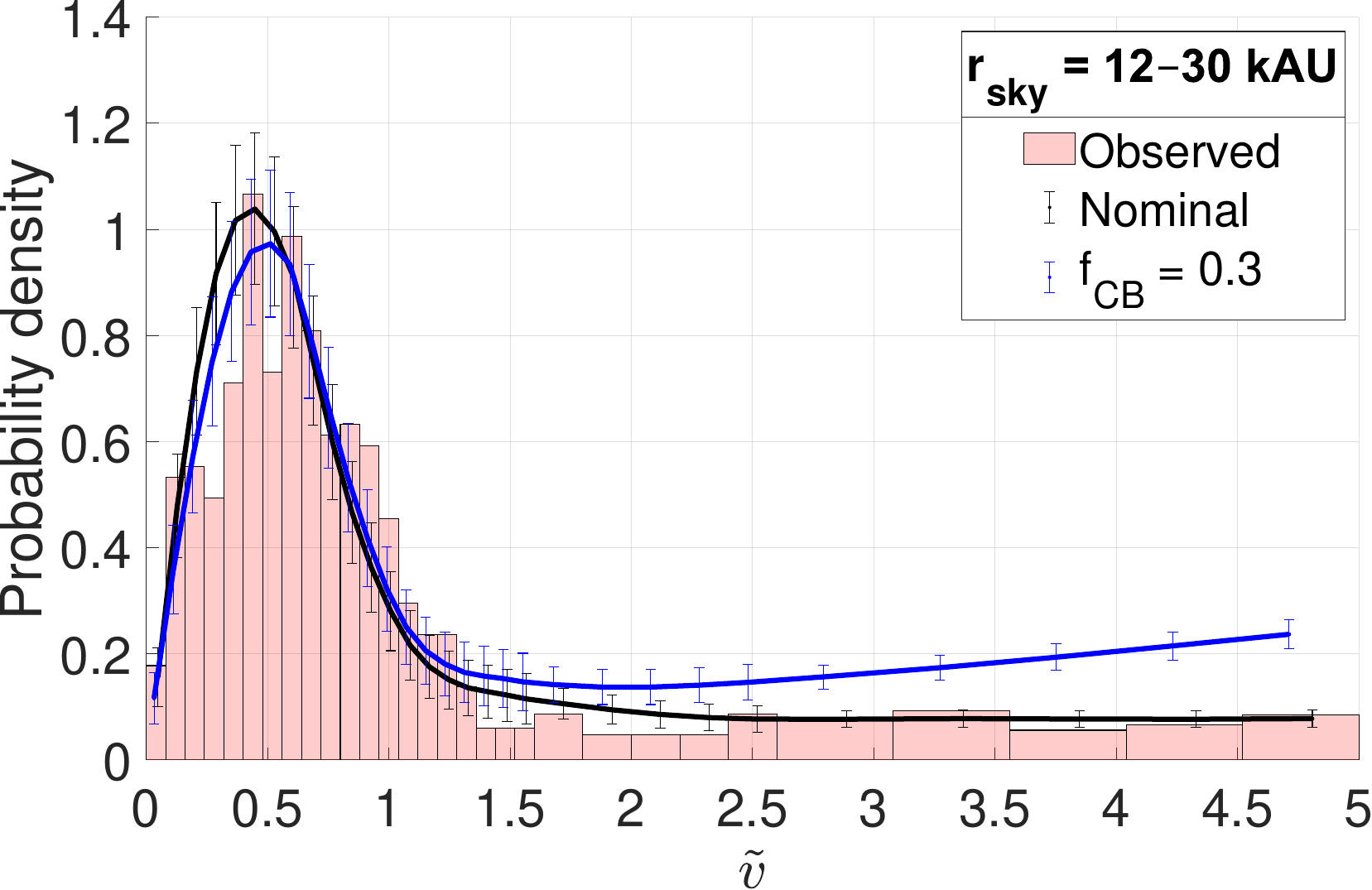}
    \caption{Comparison between the observed $\widetilde{v}$ distribution in four different $r_{\rm{sky}}$ ranges (solid red bars in different panels) with the prediction of our best model from our nominal analysis (black lines with error bars) and from our revised analysis with fixed $f_{\rm{CB}} = 0.3$ (blue lines with error bars). This revised analysis is the only one to appreciably shift the inferred gravity law towards MOND (Figure~\ref{Triangle_fCB}). The fit is much poorer in this case. The results shown here are normalized based on the observed WB distribution, helping to highlight that the $f_{\rm{CB}} = 0.3$ model predicts too many WBs at high $r_{\rm{sky}}$.}
    \label{Photo_best_nominal_fCB03}
\end{figure*}

In Figure~\ref{Photo_best_nominal_fCB03}, we compare the observed $\widetilde{v}$ distribution to the prediction of the best model with our nominal assumptions and with $f_{\rm{CB}} = 0.3$, considering both the final result of the gradient ascent and the whole MCMC chain in each case. It is clear that fixing $f_{\rm{CB}}$ to such a low value causes a catastrophic disagreement with nearly all aspects of the observations. The lack of sufficient CB contamination causes the peak of the $\widetilde{v}$ distribution to be more pronounced than in the \text{Gaia} number counts at very low $r_{\rm{sky}}$, though we see that both models struggle somewhat in the peak region at intermediate $r_{\rm{sky}}$. The most important differences concern the tail of the distribution, which cannot be adequately fit with so few CBs. The analysis has to try and fit the extended tail with substantially more LOS contamination, quadrupling the inferred $f_{\rm{LOS}}$. This may help somewhat at low $r_{\rm{sky}}$, but it leads to a rapidly rising $\widetilde{v}$ distribution at high $r_{\rm{sky}}$, in strong disagreement with the observations (bottom right panel).

These very serious issues cause the overall fit to be poorer than in our best-fitting nominal analysis by $\Delta \ln P = 413$, which implies that the latter is preferred at a confidence equivalent to $\sqrt{2 \Delta \ln P} = 29\sigma$ for a 1D Gaussian. Clearly, assuming a substantially lower likelihood of CB contamination is not a viable proposition. In any case, this only shifts the inferred gravity law towards MOND by a small amount, with Newtonian gravity still strongly preferred despite facing just over $3\sigma$ tension (Figure~\ref{Triangle_fCB}). It is therefore extremely difficult to reconcile our WB results with MOND as usually understood given other constraints.

\section{The artificial velocity limit in the \texorpdfstring{\citet{CHAE_2023}}{Chae (2023)} analysis}
\label{vtilde_limit_Kyu_impact}

In Section~\ref{Other_WBT_results}, we discussed several problems with the WB analysis of \citet{CHAE_2023}. One additional problem which may be noteworthy is that the sample is based on the WB catalogue of \citet{Badry_2021}. Each WB is assigned a likelihood of being genuine based on several factors, one of the most important being the magnitude of $\bm{v}_{\rm{sky}}$ (see their equation~7).\footnote{This is estimated using the difference in proper motions without allowing for the systemic RV, which should be a reasonable approximation given the small angular sizes of WBs.} Unlike our limit that $\widetilde{v} < 5$, their limit to $v_{\rm{sky}}$ does not easily translate into a limit on $\widetilde{v}$ because there is no allowance for more massive WBs having a faster Newtonian $v_c$. Those authors require that
\begin{eqnarray}
    v_{\rm{sky}} ~<~ 2.1 \sqrt{\frac{\rm{kAU}}{r_{\rm{sky}}}} \, \text{km/s} \, ,
\end{eqnarray}
which in terms of $\widetilde{v}$ corresponds to
\begin{eqnarray}
    \widetilde{v} ~<~ \sqrt{\frac{5 \, M_\odot}{M}} \, .
    \label{Kyu_vtilde_limit_equation}
\end{eqnarray}

\begin{figure}
    \centering
    \includegraphics[width=0.47\textwidth]{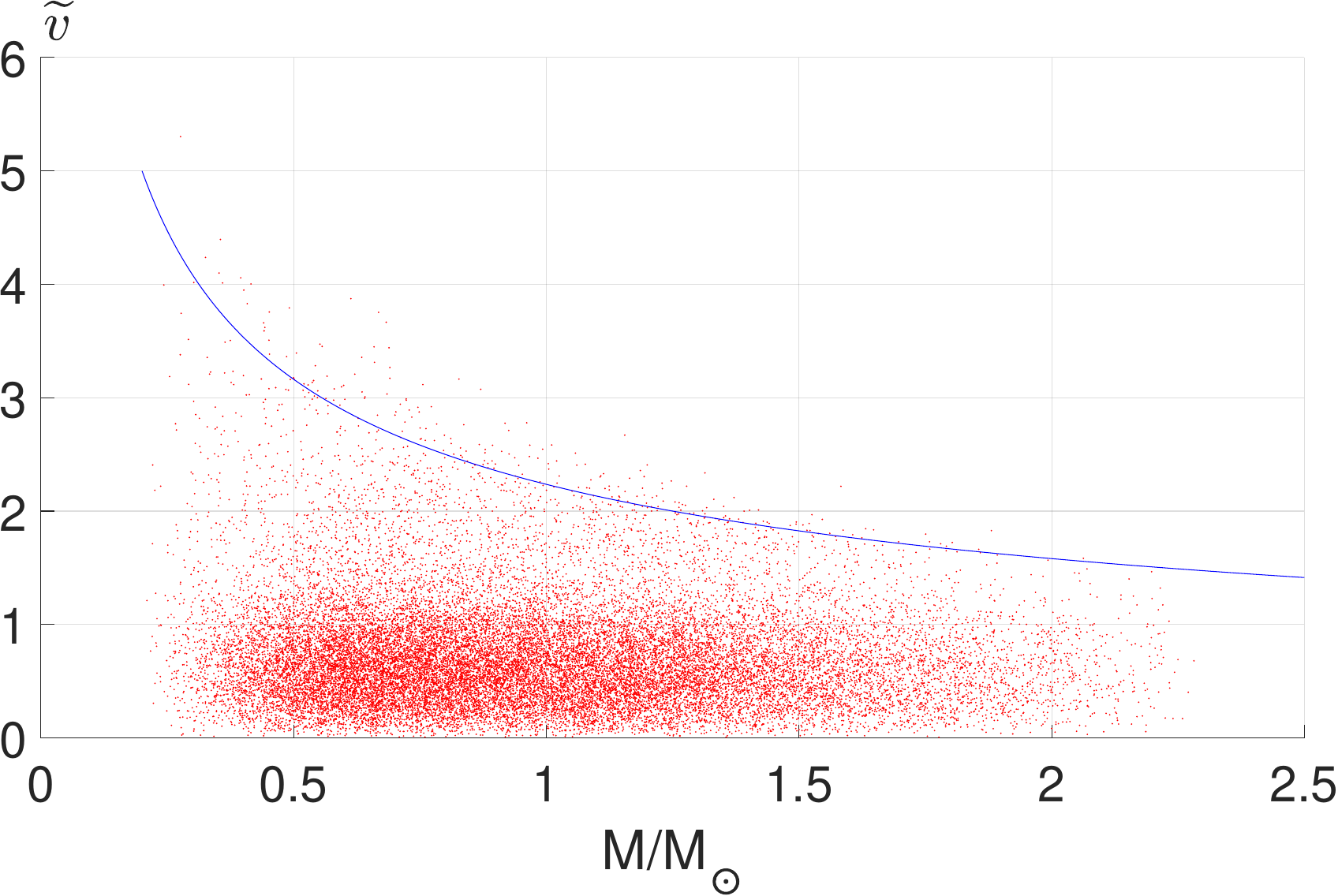}
    \caption{The red dots show the distribution of WB mass and $\widetilde{v}$ for the nominal sample of \citet{CHAE_2023}. Notice how the vast majority of WBs lie below the solid blue line (Equation~\ref{Kyu_vtilde_limit_equation}) due to selection effects \citep[equation~7 of][]{Badry_2021}.}
    \label{Kyu_vtilde_limit}
\end{figure}

We illustrate this cutoff in Figure~\ref{Kyu_vtilde_limit}, where we plot $\widetilde{v}$ as a function of $M$ for the 26615 WBs in the nominal sample of \citet{CHAE_2023}. Nearly all of these WBs lie below the blue line showing the limit given by Equation~\ref{Kyu_vtilde_limit_equation}. This is not a completely strict limit because each WB is required to be consistent with this bound within the $2\sigma$ observational uncertainty on $v_{\rm{sky}}$ \citep[see section~2 of][]{Badry_2021}. Even so, it should be clear that a WB sample selected in this way is not ideal for the WBT.\footnote{K. El-Badry, private communication.}

\begin{figure}
    \centering
    \includegraphics[width=0.47\textwidth]{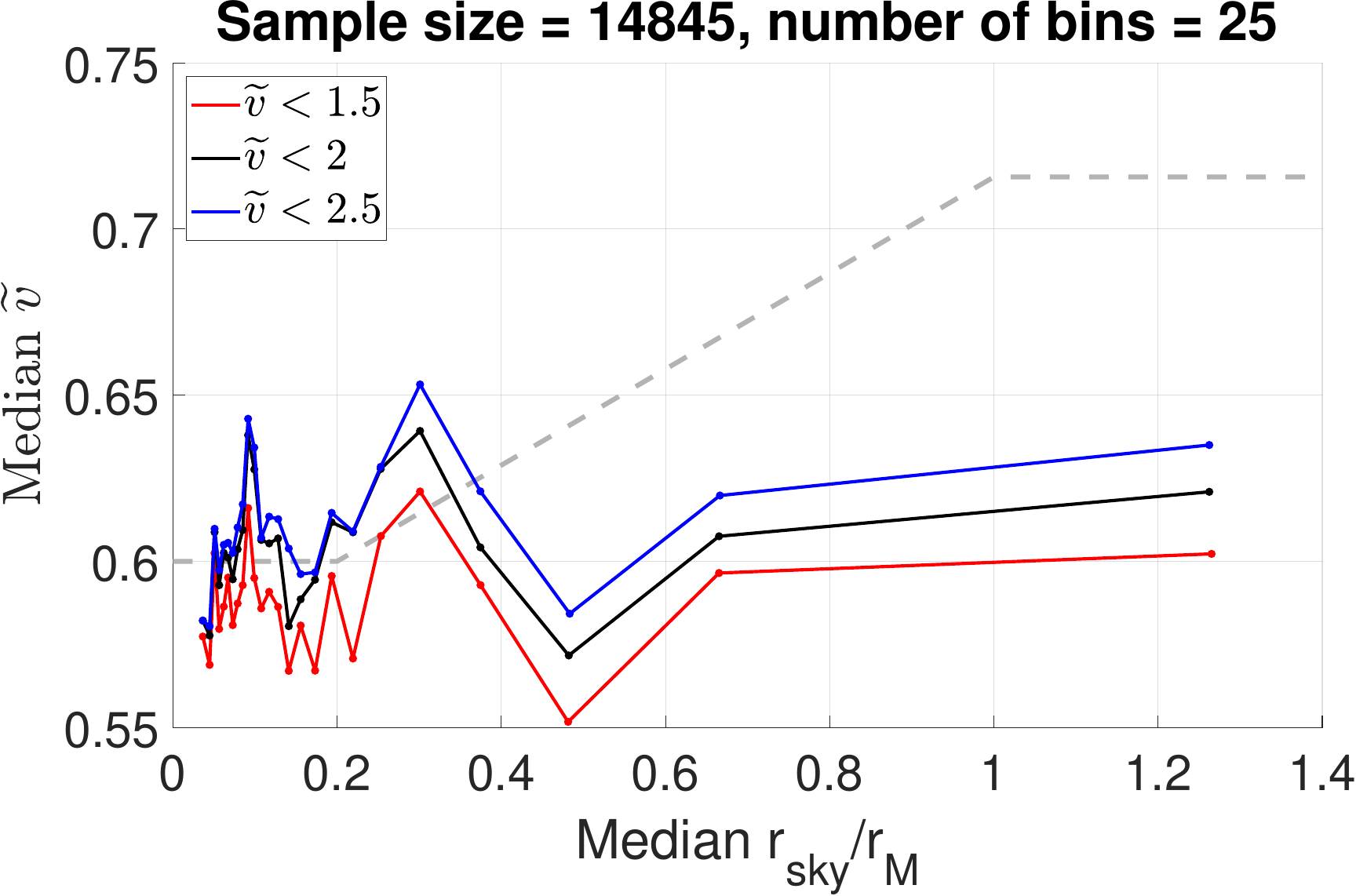}
    \caption{Similar to Figure~\ref{Kyu_medians_v01}, but restricted further to only WBs with $M < 1.25 \, M_\odot$. This ensures that the artificial limit to the $\widetilde{v}$ distribution in the \citet{CHAE_2023} sample (blue line in Figure~\ref{Kyu_vtilde_limit}) does not influence the results for $\widetilde{v} < 2$. The mass limit has little impact on the results, which continue to show a flat trend once we impose that $\sigma \left( \widetilde{v} \right) < 0.1 \max \left( 1, \widetilde{v}/2 \right)$, as done here.}
    \label{Kyu_medians_v01_M1p25}
\end{figure}

To check if our results in Section~\ref{Other_WBT_results} remain reliable despite this complication, we need to impose an upper limit on the mass so that results remain reliable up to some maximal $\widetilde{v}$. This is equivalent to drawing a rectangular cut towards the bottom left of Figure~\ref{Kyu_vtilde_limit}. To retain most of the WB sample while having reliable results beyond the main peak of the $\widetilde{v}$ distribution, we impose a limit of $M < 1.25 \, M_\odot$. With this mass limit, regardless of the mass of a WB, there is no truncation to the $\widetilde{v}$ distribution for $\widetilde{v} < 2$. We then use Figure~\ref{Kyu_medians_v01_M1p25} to show the median $\widetilde{v}$ as a function of $r_{\rm{sky}}/r_{_M}$ for the case where the sample is further limited to only those WBs where the estimated $\sigma \left( \widetilde{v} \right) < 0.1 \max \left( 1, \widetilde{v}/2 \right)$, as done in Figure~\ref{Kyu_medians_v01}. Both figures show very similar results, indicating that the $\widetilde{v}$ limit imposed by Equation~\ref{Kyu_vtilde_limit_equation} has little impact on the flat trend of the median $\widetilde{v}$ with respect to our proxy for the internal WB acceleration. This is presumably because the main peak of the $\widetilde{v}$ distribution lies at $\widetilde{v} < 1$, which is not much affected by this cut (Figure~\ref{Kyu_vtilde_limit}). Even so, having a mass-dependent limit to $\widetilde{v}$ could in principle bias the WBT because a WB with a lower mass also has a lower acceleration at the same $r_{\rm{sky}}$. Since the main idea of the WBT is that some typical measure of $\widetilde{v} \propto \sqrt{g/g_{_N}}$, one should be cautious about a sample selection that allows lower mass systems to reach a higher $\widetilde{v}$ than higher mass systems.

\end{appendix}

\bsp
\label{lastpage}
\end{document}